\documentclass[twocolumn]{aastex62}
\received{}
\revised{}
\accepted{}

\shorttitle{Evolution of 3-D shape of galaxies at $z<1$}
\shortauthors{Satoh et al.}

\begin{document}

\title{Evolution of 3-dimensional Shape of Passively Evolving and Star-forming
 Galaxies at $z<1$}

\author{Yuki K. Satoh}
\affiliation{Graduate School of Science and Engineering, Ehime University}
\affiliation{Department of Earth and Space Science, Graduate School of Science, 
Osaka University, 1-1 Machikaneyama-cho, Toyonaka, 
Osaka 560-0043, Japan; {\url{satohyk@iral.ess.sci.osaka-u.ac.jp}}}

\author{Masaru Kajisawa}
\affil{Graduate School of Science and Engineering, Ehime University, Bunkyo-cho, Matsuyama 790-8577, Japan; {\url{kajisawa@cosmos.phys.sci.ehime-u.ac.jp}}}
\affil{Research Center for Space and Cosmic Evolution, Ehime University}

\author{Kazuharu G. Himoto}
\affil{Graduate School of Science and Engineering, Ehime University, Bunkyo-cho, Matsuyama 790-8577, Japan}



\begin{abstract}
Using the HST/ACS $I_{\rm F814W}$-band data,  
we investigated distribution of apparent axial ratios of
$\sim21000$ galaxies with $M_{V}<-20$ at 
$0.2<z<1.0$ in the COSMOS field as a function of stellar mass, specific star 
formation rate (sSFR), and redshift.
We statistically estimated intrinsic 3-dimensional shapes of these galaxies
 by fitting the axial-ratio distribution with 
triaxial ellipsoid models characterized by 
face-on (middle-to-long) and edge-on (short-to-long) axial ratios 
$B/A$ and $C/A$. 
We found that the transition from thin disk to thick spheroid occurs at 
$\Delta$MS $\sim-1$ dex, i.e., 10 times lower sSFR than 
that of the main sequence for galaxies with 
$M_{\rm star} = 10^{10}$--$10^{11} M_{\odot}$ at $0.2<z<1.0$.
Furthermore, the intrinsic thickness ($C/A$) of passively evolving galaxies 
with $M_{\rm star}=10^{10}$--$10^{11}M_{\odot}$ significantly 
decreases with time from 
 $C/A \sim 0.40$ -- 0.50 at $z\sim 0.8$ to $C/A\sim0.33$ -- 0.37 at $z\sim0.4$, 
while those galaxies with $M_{\rm star}>10^{11}M_{\odot}$ have $C/A\sim0.5$ 
irrespective of redshift. 
On the other hand, star-forming galaxies on the main sequence 
with $10^{9.5}$--$10^{11}M_{\odot}$ show no significant 
evolution in their shape at $0.2<z<1.0$,   
but their thickness depends on stellar mass;  
more massive star-forming galaxies tend to 
have lower $C/A$ (thinner shape) than low-mass ones.
These results suggest that some fraction of star-forming 
galaxies with a thin disk, which started to appear around $z\sim1$, 
quench their star formation without violent morphological change,  
and these newly added quiescent galaxies with a relatively thin shape cause 
the significant evolution in the axial-ratio distribution of 
 passively evolving galaxies with $M_{\rm star}<10^{11}M_{\odot}$ at $z<1$. 
\end{abstract}



\section{Introduction} \label{sec:intro}

One of the most striking feature of galaxies is the variety of their morphology.
Since \cite{hub26}, it is known that the morphology of galaxies is closely 
correlated with their physical properties such as luminosity, 
stellar mass, color, star formation rate (SFR), gas contents, 
and so on (e.g., \citealp{rob94}; \citealp{bel12}; \citealp{blu19}).
Elliptical galaxies show spheroidal shapes with smooth light distributions 
and mainly contain old stars with little star formation.
On the other hand, spiral galaxies have a flat stellar disk with 
characteristic spiral arms and form new stars from a thin gas disk.
S0 galaxies have intermediate properties between ellipticals and spirals 
and show a flat disk with smooth light distributions and little star formation. 
Many studies have proposed mechanisms to form such galaxies with the intermediate 
properties, but which process(es) dominates in the formation of S0 galaxies
 is still unclear (e.g., \citealp{lar80}; \citealp{agu12}). 
Recent observational studies with the integral field spectroscopy suggest 
that early-type galaxies are well classified into slow and fast rotators by 
their stellar kinematics 
and that the classification by the specific angular momentum of the stellar 
component can be more fundamental for understanding of these galaxies 
than the E/S0 classification (e.g., \citealp{ems11}).

In the point of view of star formation histories, 
galaxies are well divided into two populations, namely, 
star-forming galaxies and passively evolving galaxies with little 
star formation at $z\lesssim 3$ (e.g., \citealp{bel04}; \citealp{wil09}; 
\citealp{whi11}). 
The star formation activity of star-forming galaxies with similar stellar masses 
is rather uniform, and these galaxies form a tight correlation between 
SFR and stellar mass, namely, the main sequence of star-forming galaxies 
(e.g., \citealp{noe07}; \citealp{elb07}).
Some fraction of star-forming galaxies are expected to stop their star formation 
by some mechanisms and then evolve into the passively evolving population 
(e.g., \citealp{fab07}; \citealp{pen10}).
Because of the uniformity of star-forming galaxies,  
such quenching of star formation is considered to be the most important process
in star formation histories of galaxies.
There are many proposed physical mechanisms for the quenching, namely, 
galactic wind expelling gas from the galaxy by supernova feedback,   
gas heating by AGN feedback, shock heating of gas infalling into dark matter 
halos, gravitational stabilization of gas disk by the bulge, 
environmental effects such as ram-pressure stripping, harassment, 
strangulation/starvation, 
and so on (e.g., \citealp{dek86}; \citealp{fab12}; \citealp{bir03}; 
\citealp{mar09}; \citealp{aba99}; \citealp{moo96}; \citealp{bal00}). 
Since some of the quenching mechanisms also affect the morphology/shape 
of the galaxies with various ways, 
the correlation between the morphology and star formation activity 
 could reflect such physical processes.

The shape of the stellar body of galaxies is closely related with the 
kinematics of their structures and is expected to reflect 
their formation and evolution processes.
Many previous theoretical studies suggested that the shape and structure 
of galaxies are affected by various physical processes such as 
the gas accretion to the galaxies in dark matter halos, 
gravitational instability in gas disks, galaxy merger/interaction, 
stellar/AGN feedback, and so on (e.g., \citealp{hop09}; 
\citealp{sal09}; \citealp{ose10}; 
\citealp{sal12}; \citealp{fia15}; \citealp{rod17}; \citealp{elb18}).
For example, it is considered that 
rotationally supported flat disks are 
formed through a gradual accretion of gas with a rather stable spin axis 
from a quasi-hydrostatic hot corona, which has been 
heated to around the virial temperature by shock in infalling to the halo
(e.g., \citealp{whi91}). 
On the other hand, a direct accretion of cold gas from distinct filaments 
with misaligned spin directions to the galaxy is expected to lead to 
 more thick and spheroidal-like structures (e.g., \citealp{dek09a}).
While dry major mergers between gas-poor galaxies are considered to result in 
 the remnants with a spheroidal shape (e.g., \citealp{bar88}), 
 gas-rich mergers may form those with a significant disk component 
(e.g., \citealp{spr05}). 
The secular evolution by the bar or spiral arms 
and the gas inflow to the center of the galaxy by 
the galaxy interaction/minor merger may cause bulge growth without 
a destruction of the thin disk (e.g., \citealp{kor04}; \citealp{gue13}). 
Thus the distribution of the intrinsic shape of galaxies and its dependence 
on other physical properties such as stellar mass, SFR, and redshift 
can provide us important clues to understand how galaxies formed and evolved 
through such physical processes.

Although it is difficult to measure the intrinsic shape of a galaxy individually, 
one can statistically estimate the 3-dimensional shapes for a sample of 
galaxies from 
the distribution of the apparent axial ratio projected on the celestial sphere. 
Several pioneering works investigated the distribution of the apparent axial
ratio of nearby galaxies, and confirmed that elliptical galaxies have relatively 
thick and spheroidal shapes, while spiral (and S0) galaxies have flat and thin disk 
shapes (e.g., \citealp{san70}; \citealp{bin80}; \citealp{bin81}; \citealp{lam92}). 
A large sample of galaxies at $z \lesssim 0.1$ drawn from the SDSS survey enables to 
investigate the intrinsic shape with high statistical accuracy and its dependence on 
other physical properties such as luminosity, stellar mass, and surface brightness
profile 
(\citealp{ryd04}; \citealp{vin05}; \citealp{pad08}; \citealp{van09}).
\cite{ryd04} and \cite{pad08} confirmed that spiral galaxies have a flat disk 
with an edge-on axial ratio of 0.2 -- 0.25 and that 
their face-on axial ratio is clearly 
smaller than unity, which suggests that their disks are not 
circular shape on the face-on view.
\cite{pad08} and \cite{vin05} reported that luminous early-type galaxies 
have round intrinsic shapes, while those faint galaxies show a flatter distribution 
of the apparent axial ratio, which indicates relatively thin disk shapes. 
\cite{van09} found that the axial-ratio distribution of passively 
evolving galaxies abruptly changes around $M_{\rm star} \sim 10^{11} M_{\odot}$, 
and suggested that those galaxies with $M_{\rm star} < 10^{11} M_{\odot}$ have 
 disk-like intrinsic shapes, while massive ones with $M_{\rm star} > 10^{11} M_{\odot}$ 
have spheroidal shapes, which can be formed preferentially by major mergers.

Using high-resolution imaging data taken with Hubble Space Telescope 
(HST) in the GOODS, COSMOS, and CANDELS surveys, previous studies 
carried out the similar analyses for star-forming and passively evolving 
galaxies at high redshifts, mainly $z\sim 2$--3 
to study the evolution of their 3-dimensional shapes 
(e.g., \citealp{rav06}; \citealp{yum11}; \citealp{yum12}; \citealp{hol12}; 
\citealp{law12}; \citealp{cha13}; 
\citealp{van14a}; \citealp{tak15}; \citealp{zha19}; \citealp{hil19}).
\cite{rav06} and \cite{law12} investigated the axial-ratio distribution 
for rest-UV color-selected Lyman Break Galaxies (LBGs) and BX/BM galaxies 
at $z\sim$ 1.5--5, and found that they have more thick and prolate shapes 
with $B/A\sim0.7$ and $C/A\sim 0.3$, where $B/A$ is the face-on intrinsic 
axial ratio and $C/A$ is the edge-on axial ratio (i.e., $A>B>C$), than 
star-forming galaxies in the present universe. 
 \cite{rav06} also reported that both those galaxies with the exponential 
and $r^{1/4}$-like surface brightness profiles have the similar prolate shapes, 
which suggests that the surface brightness profile does 
not necessarily represent the 3-dimensional shape for these high-$z$ 
star-forming galaxies especially in the rest-UV wavelength.
\cite{yum11} and \cite{yum12} investigated the apparent axial ratio of 
star-forming BzK galaxies at $z\sim$ 1.4 -- 2.5 in the rest-frame UV and 
optical 
wavelengths, and found that these galaxies also have similar prolate 
3-dimensional shapes with $C/A \sim$ 0.26 -- 0.28 and $B/A\sim$ 0.6 -- 0.8.
\cite{tak15} reported that the intrinsic shapes of star-forming galaxies 
on the main sequence evolve from the prolate shapes at $z\sim2$ to more oblate 
(disky) shapes with $B/A\sim 0.9$ at $z\sim 0.7$.
\cite{van14a} and \cite{zha19} also found that the prolate shapes are seen  
preferentially in star-forming galaxies with smaller stellar mass at higher redshift.
While most star-forming galaxies show thin disk shapes at $z<1$, 
those with $M_{\rm star} < 10^{10} M_{\odot}$ tend to have the prolate shapes
at $z \gtrsim$ 1 -- 1.5.
On the other hand, \cite{hol12}, \cite{cha13}, and \cite{hil19} 
studied the intrinsic shapes for passively evolving galaxies at high redshifts.
\cite{hol12} and \cite{hil19} found that the distribution of the apparent 
axial ratio for those galaxies at $z\sim 0.7$ significantly changes at 
$M_{\rm star} \sim 10^{11} M_{\odot}$; 
massive quiescent galaxies with $M_{\rm star} > 10^{11} M_{\odot}$ have thick and 
spheroidal shapes, while those with $10^{10}$--$10^{11} M_{\odot}$ show thin disk-like 
shapes, which is similar with those galaxies at $z\sim 0$ as mentioned above. 
\cite{cha13} and \cite{hil19} also reported that massive galaxies 
with $M_{\rm star} > 10^{11} M_{\odot}$ show more thin and oblate shapes at $z>1$ 
than those at $z<1$, and the shapes of those at $z\sim$2--3 
are similar with massive star-forming galaxies at the same redshift.

Although many previous studies have investigated the evolution of the 
intrinsic shape 
of galaxies, the sample sizes in these studies were relatively small,  
especially at $z<1$, and therefore detailed studies with high statistical 
accuracy for galaxies at $z<1$ have been difficult to carry out.
However, since the morphologies similar with the present-day Hubble sequence 
have started to appear around $z\sim1$ (e.g., \citealp{abr01};
\citealp{kaj01}; \citealp{con05}), 
it is important to investigate the 3-dimensional shape of galaxies 
at $z\lesssim 1$
 as a function of stellar mass, star formation activity, and redshift 
in order to understand how the galaxy morphology and its correlation with 
other physical properties of galaxies in the present universe formed.
In this paper, we measure the apparent axial ratios of 
$\sim 21000$ galaxies with $M_{V}<-20$ at $0.2<z<1.0$ with the HST/ACS data 
over 1.65 deg$^2$ region in the COSMOS field to estimate the 3-dimensional 
shapes 
as a function of stellar mass, specific SFR ($= SFR/M_{\rm star}$, 
hereafter sSFR), and redshift. 
The large sample of galaxies allows us to study the evolution of the intrinsic 
shape of galaxies with high statistical accuracy and its dependence 
on stellar mass and sSFR in detail.
Section \ref{sec:sample} describes the sample selection, 
and Section \ref{sec:analysis} describes the methods to measure the 
apparent axial ratios of sample galaxies and estimate the 3-dimensional shapes.
In section \ref{sec:result}, we present the distribution of the apparent axial ratio
and the estimated intrinsic shape as a function of stellar mass, sSFR, 
and redshift, and examine possible biases in our analysis. 
We discuss our results and their implications in Section \ref{sec:discus}, 
and summarize the results in Section \ref{sec:summary}.
Throughout this paper, magnitudes are given in the AB system.
 We adopt a flat universe with $\Omega_{\rm matter}=0.3$, $\Omega_{\Lambda}=0.7$, 
and $H_{0}=70$ km s$^{-1}$ Mpc$^{-1}$.

\section{Sample} \label{sec:sample}

In this study, we used a sample of galaxies with $M_{V}<-20$ at 
$0.2 < z_{\rm phot} < 1.0$ in the 1.65 deg$^2$ COSMOS HST/ACS field 
drawn from the COSMOS photometric redshift 
catalog (\citealp{ilb09}; \citealp{ilb13}). 
We chose the absolute magnitude limit of $M_{V}<-20$ to 
secure enough accuracy in measurements of their axial ratio even at 
$z\sim1$.
\begin{figure}
\epsscale{1.1}
\plotone{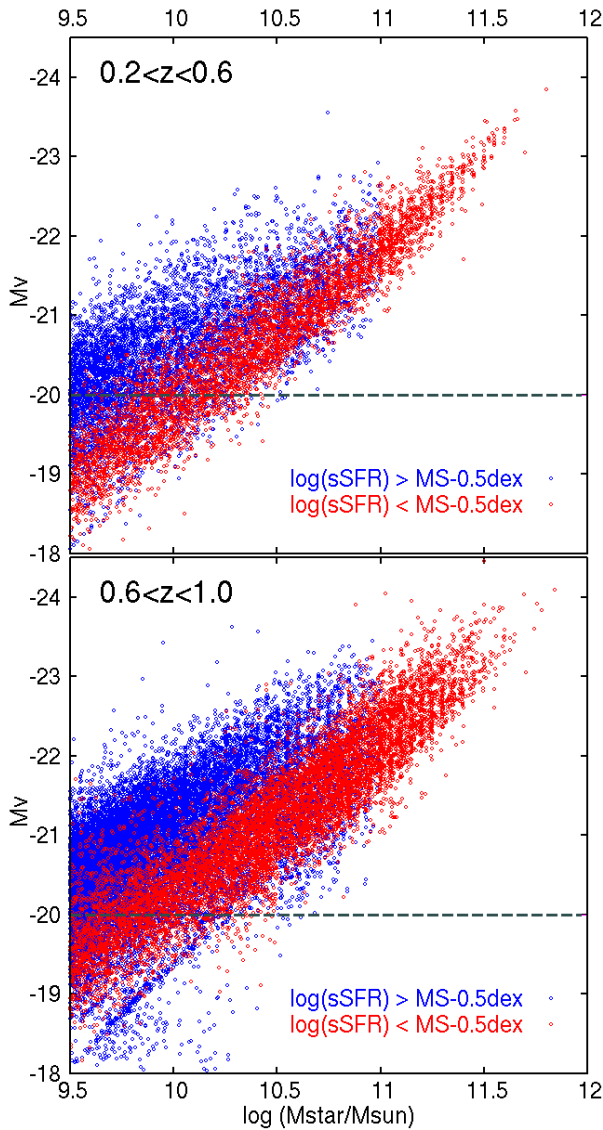}
\caption{The rest-frame $V$-band absolute magnitude vs. stellar mass for galaxies 
at $0.2<z<0.6$ (top) and $0.6<z<1.0$ (bottom) in the COSMOS field.
Blue symbols show star-forming galaxies with $\Delta$MS $ > -0.5$ dex, 
and red ones represent the other galaxies with lower sSFRs of $\Delta$MS $ < -0.5$ dex.
 The dashed line indicates the absolute magnitude limit of $M_{V}<-20$ for our sample.
 \label{fig:msmv}}
\end{figure}
The sample galaxies were detected on the Subaru/Suprime-Cam $i'$-band data
(\citealp{tan07}; \citealp{cap07}),  
and their photometric redshifts were estimated from the multi-band photometric 
data 
from UV to MIR, which include GALEX FUV and NUV, CFHT $u*$ and $i*$, 
Subaru $B$, $V$, $g'$, $r'$, $i'$, $z'$, and the 12 intermediate ($IA$) bands 
\citep{tan15}, VISTA $Y$, $J$, $H$, $K$, Spitzer IRAC1, 2, 3, 4 bands. 
The accuracy of the photometric 
redshift is very high with a small fraction of the catastrophic 
failures especially for galaxies at $z\lesssim 1.2$, 
where the $IA$ bands sample the Balmer break \citep{ilb13}.   
\cite{ilb13} also estimated the stellar mass and SFR by fitting the same 
multi-band photometry with the GALAXEV population synthesis library 
\citep{bru03}. In the SED fitting, they used  
exponentially decaying star formation histories and Calzetti extinction law 
\citep{cal00}, and assumed \cite{cha03}'s Initial Mass Function.

We checked the completeness for galaxies with $M_{V}<-20$ as a function of 
stellar mass (Figure \ref{fig:msmv}), and limited our sample to those with 
$M_{\rm star} > 10^{9.5} M_{\odot}$ for star-forming galaxies on the main sequence
(see below) and those with $M_{\rm star} > 10^{10} M_{\odot}$ for the other 
galaxies with lower sSFRs. 
We note that 
$\sim 35$ \% ($\sim$ 20 \%) of galaxies with $10^{9.5}$--$10^{10} M_{\odot}$ 
at $0.2<z<0.6$ ($0.6<z<1.0$) for the main-sequence galaxies are missed by the 
absolute magnitude limit of $M_{V}<-20$, 
 while $\sim 20$ \% ($\sim $ 5--6 \%) of the other galaxies with 
$10^{10}$--$10^{10.5} M_{\odot}$ at $0.2<z<0.6$ ($0.6<z<1.0$) are missed by
the same limit.
We examine how the incompleteness affects our results in Section \ref{subsec:incom}.  
Finally, we selected total 21294 galaxies 
(13132 main-sequence galaxies with $M_{\rm star} >10^{9.5}M_{\odot}$ 
and 8162 galaxies with a lower sSFR and $M_{\rm star} > 10^{10}M_{\odot}$). 

\begin{figure}
\epsscale{1.15}
\plotone{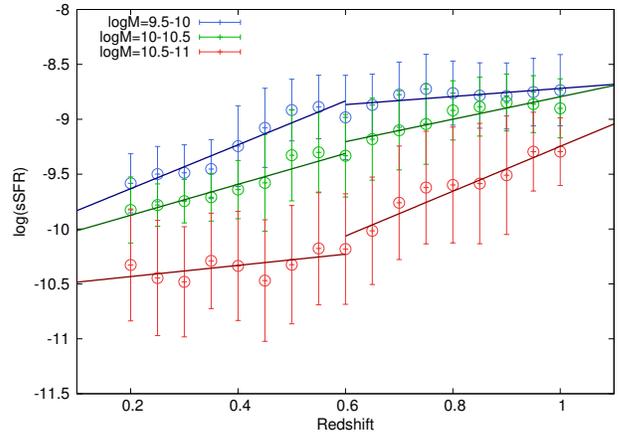}
\caption{The clipping mean sSFRs for star-forming galaxies 
with different stellar masses 
in our sample as a function of redshift.
The mean values are calculated with 2$\sigma$ clipping and additional exclusion
of those with a sSFR more than 10 times lower than the mean.
The different colors of symbols show galaxies with the different stellar masses.
The solid lines represent the linear fitting results for those galaxies 
at $0.2<z<0.6$ and $0.6<z<1.0$.  
\label{fig:ms}}
\end{figure}

\begin{deluxetable}{lrl} 
\tablecaption{fitting results of the mean sSFR of the main sequence \label{tab:mseq}}
\tablehead{
\colhead {redshift} & \colhead {stellar mass} & \colhead{fitted sSFR} 
}
\startdata
$z=0.2$--0.6 & $\log{M_{\rm star}}=9.5$--10 & $\log{(sSFR)}=1.998z-10.032$\\
 & 10--10.5 & $\log{(sSFR)}=1.408z-10.156$\\
 & 10.5--11 & $\log{(sSFR)}=0.510z-10.535$\\
$z=0.6$--1.0 & $\log{M_{\rm star}}=9.5$--10 & $\log{(sSFR)}=0.369z-9.088$\\
 & 10--10.5 & $\log{(sSFR)}=1.024z-9.819$\\
 & 10.5--11 & $\log{(sSFR)}=2.043z-11.290$ \\
\enddata
\end{deluxetable}

In Section 4, we divide our sample by sSFR and investigate the distribution 
of the axial ratio as a function of sSFR. 
In addition to sSFR itself, we also use differences in $\log{(sSFR)}$ from the main 
sequence of star-forming galaxies, namely, $\Delta$MS.  
In order to define the main sequence, we calculated the clipping mean of sSFR 
for galaxies in each redshift bin with a width of $\Delta z = 0.1$ 
(Figure \ref{fig:ms}). In the calculation, we used 2$\sigma$ clipping with 
additionally excluding objects with sSFR more than an order of magnitude 
lower than the mean. 
Since the sSFR of the main sequence of star-forming galaxies depends on 
stellar mass especially at $z\lesssim 1$ (e.g., \citealp{kaj10}; 
\citealp{ilb15}; 
\citealp{pop19}), we separately estimated the mean sSFR for galaxies 
with different stellar masses. Figure \ref{fig:ms} shows the different 
evolutionary trends of the mean sSFR for star-forming galaxies with different
masses. In order to take the evolution of the main sequence into account, 
we fitted the logarithm of the mean sSFR as a function of redshift with 
a linear line for galaxies at $0.2<z<0.6$ and $0.6<z<1.0$ separately.
The fitting results are summarized in Table \ref{tab:mseq}. 
Because there are few star-forming galaxies with 
$M_{\rm star} > 10^{11} M_{\odot}$, 
we did not define the main sequence for those with 
$M_{\rm star} > 10^{11} M_{\odot}$.
In the following, we use these equations to calculate the sSFR of the main 
sequence and $\Delta$MS ($ = \log{(sSFR)}-\log{(sSFR_{\rm MS})}$) for each object with 
$M_{\rm star} < 10^{11} M_{\odot}$.

\section{Analysis} \label{sec:analysis}

\subsection{Measurement of axial ratio} \label{sec:measure}

We measured apparent axial ratios of the sample galaxies on publicly available  
COSMOS HST/ACS $I_{\rm F814W}$-band data version 2.0 \citep{koe07} using 
the SExtractor software version 2.5.0 \citep{ber96}.
The pixel scale of the data is 0.03 arcsec/pixel and 
the FWHM of PSF is $\sim 0.1$ arcsec.
At first, we cut out a $12^{\prime\prime} \times 12^{\prime\prime}$ 
$I_{\rm F814W}$-band image 
centered on the position of each sample galaxy detected on the Subaru 
$i'$-band data. We made the SEGMENTATION image for the $i'$-band data 
with SExtractor, and used it to mask pixels belonging to any other 
$i'$-selected objects on the $12^{\prime\prime} \times 12^{\prime\prime}$ 
$I_{\rm F814W}$-band image. 
We then ran SExtractor on the masked $I_{\rm F814W}$-band images to 
measure apparent axial ratios of the sample galaxies.    
A detection threshold of 1.3 times the local background root mean square over 
12 connected pixels was used. 
In order to avoid the over-deblending due to the dust extinction, in particular, 
 the dust lane in edge-on disk galaxies, 
we chose no deblending (DETECTED\_MINCONT$=$1) and required the position 
of the detected object on the $I_{\rm F814W}$-band image 
to coincide that of the target object selected on the $i'$-band image 
within 0.6 arcsec, which roughly corresponds to the FWHM of the PSF 
of the $i'$-band data. 

\begin{figure}
\epsscale{1.1}
\plotone{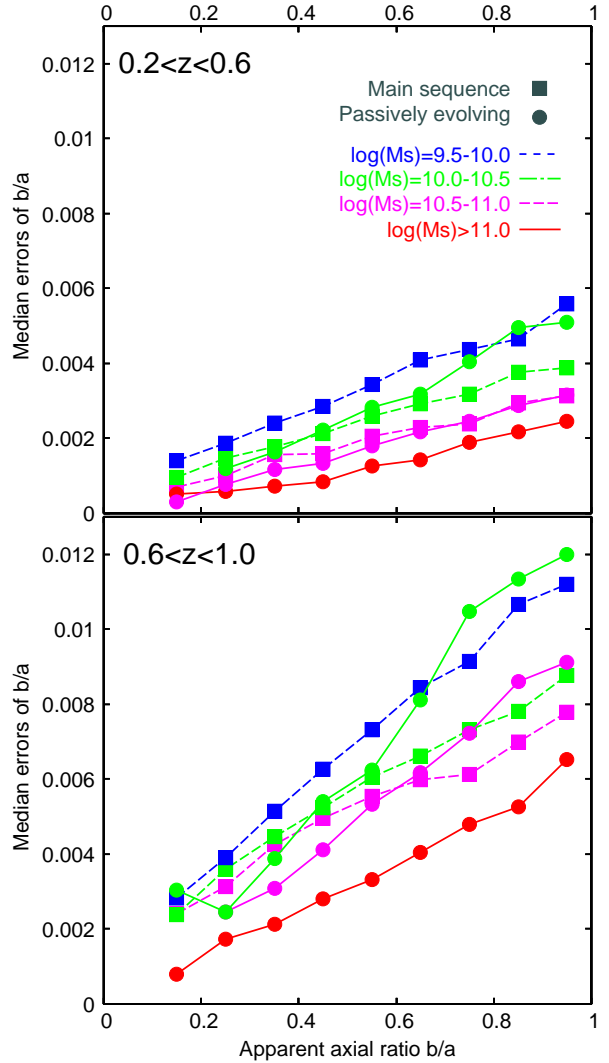}
\caption{ 
The median errors of the measured apparent axial ratio for our subsamples 
at $0.2<z<0.6$ (top) and $0.6<z<1.0$ (bottom) as a function of axial ratio.
Squares show star-forming galaxies on the main sequence with 
$\Delta$MS $ = -0.5$ -- $+0.5$ dex, and circles show 
passively evolving galaxies with $\Delta$MS $ < -1.5$ dex.
The different colors of the symbols represent those galaxies with different 
stellar masses.
\label{fig:arerr}}
\end{figure}

SExtractor computed the second order moments along the major and minor axes 
of the object as follows:
\begin{equation}
a^2 = \frac{\overline{x^2} + \overline{y^2}}{2}+\sqrt{(\frac{\overline{x^2}-\overline{y^2}}{2})^2+\overline{xy}^2}
\end{equation}
\begin{equation}
b^2 = \frac{\overline{x^2} + \overline{y^2}}{2}-\sqrt{(\frac{\overline{x^2}-\overline{y^2}}{2})^2+\overline{xy}^2},
\end{equation}
where $\overline{x^2}$, $\overline{y^2}$, and $\overline{xy}$ are the second 
order moments of the object in the $x$-$y$ coordinate, i.e., 
\begin{equation}
\overline{x^2} = \frac{\sum_{i \in S} I_{i}(x_{i}-\overline{x})^2}{\sum_{i \in S} I_{i}}
\end{equation}
\begin{equation}
\overline{y^2} = \frac{\sum_{i \in S} I_{i}(y_{i}-\overline{y})^2}{\sum_{i \in S} I_{i}}
\end{equation}
\begin{equation}
\overline{xy} = \frac{\sum_{i \in S} I_{i}(x_{i}-\overline{x})(y_{i}-\overline{y})}{\sum_{i \in S} I_{i}}.
\end{equation}
Using the second order moments along the major and minor axes, 
we calculated $b/a$ as the axial ratio of the object.
Since an ellipse with semi-major and minor radii of $3\times a$ and $3\times b$ 
usually includes most of the flux for galaxies with a normal surface brightness 
profile, we excluded a galaxy with $b$ smaller than 1.3 pixel 
(i.e., $3\times b < $ 0.1 
arcsec), which is too small to reliably measure the apparent axial ratio 
 with the spatial resolution of the HST data.
There is only one object with such a small size in our sample,  
and it is excluded from the analysis. 

While those estimated from the surface brightness fitting under the 
assumption of 
some parametric forms of the profile were used in previous studies (e.g., 
\citealp{pad08}; \citealp{hol12}), 
the second order moments allow us to directly estimate the axial ratio 
without assuming the surface brightness profile of the object. 
On the other hand, the measured second order moments can be contributed from 
several components if exist for example, bulge and disk in spiral galaxies, 
because we didn't carry out the decomposition of these components.
It is expected that a larger and/or brighter component tends to dominate 
the second order moments of the object in such cases.
We keep in mind these things in discussing our results.

We show examples of sample galaxies with different measured values 
of the axial ratio in Appendix \ref{sec: montage}. 
Figure \ref{fig:arerr} shows median values of the statistical errors 
in the axial ratios 
as a function of axial ratio itself for the sample galaxies with different 
stellar masses, sSFRs, and redshifts. 
Although the uncertainty clearly increases with increasing axial ratio, 
the errors are $\delta (b/a) \sim 0.005$ at $0.2<z<0.6$ and $\sim 0.012$
 at $0.6<z<1.0$ 
even for the least massive (faintest) galaxies in our sample.
These are sufficiently small compared to a bin width of 0.1 
for the axial-ratio distribution we used in this study, 
and the measurement errors do not significantly 
affect our results in the following sections.

\subsection{Estimate of intrinsic 3-dimensional shape} \label{subsec:intrinsic}

In order to infer the intrinsic 3-dimensional shape of our sample galaxies, 
we fitted the distribution of the apparent axial ratio with triaxial ellipsoid 
models, following \cite{ryd04}.
The shape of the triaxial model is characterized by 
 two parameters, namely, face-on axial ratio $B/A$ and 
edge-on axial ratio $C/A$, where $A$, $B$, and $C$ are radii in the major, 
middle and minor axes ($A > B > C$).
We assumed Gaussian distributions for both $B/A$ and $C/A$,  
and performed Monte Carlo simulations to estimate the distribution of 
the apparent axial ratio. 
We chose the Gaussian distribution for $B/A$ instead of a log-normal 
distribution for $\epsilon = 1-B/A$ used in \cite{ryd04}, 
because the models with a Gaussian distribution for $B/A$  
could reproduce the observed distributions better than those with the 
log-normal one, 
especially for star-forming (disky) galaxies.
We also examined models with a Gaussian distribution for the triaxiality 
($T=(A^2-B^2)/(A^2-C^2)$) used by \cite{cha13}, \cite{van14a}, 
and \cite{zha19} 
instead of $B/A$, and confirmed that the both models 
produced the similar results. We preferred those with Gaussian distributions
for both $B/A$ and $C/A$ because of ease to understand the fitting results.
In the simulation, we calculated an apparent axial ratio for a given 
combination 
of $B/A$ and $C/A$ assuming a random viewing angle, following \cite{bin85}.
Thus free parameters in the fitting are the mean and dispersion of $B/A$ 
and $C/A$, 
namely, $\mu_{B/A}$, $\sigma_{B/A}$, $\mu_{C/A}$, and $\sigma_{C/A}$. 
For each combination of these four parameters, we carried out 100000 
simulations 
to calculate the distribution of the apparent axial ratio. 
We used a bin width of 0.1 to represent the both observed and simulated 
distributions 
of the apparent axial ratio. 
In the simulation, we set a criterion of $\mu_{C/A} < \mu_{B/A} \le 1$ to ensure 
$A > B > C$, and we calculated the apparent axial ratio 
by swapping $B/A$ and $C/A$ values in the case of $B/A < C/A$, 
which could occurs when $\mu_{C/A}$ is nearly equal to $\mu_{B/A}$ or 
$\sigma_{B/A}$ and/or $\sigma_{C/A}$ are relatively large.  
We also set a lower limit of $C/A > 0.05$ to match the observed 
distribution
of the apparent axial ratio, where there is no object with an apparent 
axial ratio of $b/a < 0.1$ in our sample. 
We calculated statistical errors based on the square root of 
the number of sample galaxies in the bins except for bins with a very small 
number of 
objects, for which we adopted the upper and lower confidence limits given by 
\cite{geh86}. 
We simply used the grid search to find the best-fitting parameters 
with the minimum $\chi^2$ method and 
estimated their 68\% confidence ranges with the $\Delta \chi^2$ method.
In the fitting procedures, $\mu_{B/A}$ and $\mu_{C/A}$ range from 0.05  
to 1 with a step of 0.005, and $\sigma_{C/A}$ ranges from 
0.01 to 0.5 with a step 
of 0.01. We used three step widths for $\sigma_{B/A}$, namely, 
0.01 at $0.01<\sigma_{B/A}<0.5$, 0.02 at $0.5<\sigma_{B/A}<0.8$, and 0.04 
at $0.8<\sigma_{B/A}<1.0$. 
We present the fitting results for subsamples used in this study 
in Appendix \ref{sec:compmodel} and Table \ref{tab:fitevol}. 
The best-fit models are consistent with the observed distributions  
 within the errors for all the subsamples.

\section{Results} \label{sec:result}

\subsection{Dependence of axial ratio \& 3-D shape on sSFR} \label{subsec:ssfr}

\begin{figure*}
\epsscale{0.86}
\plotone{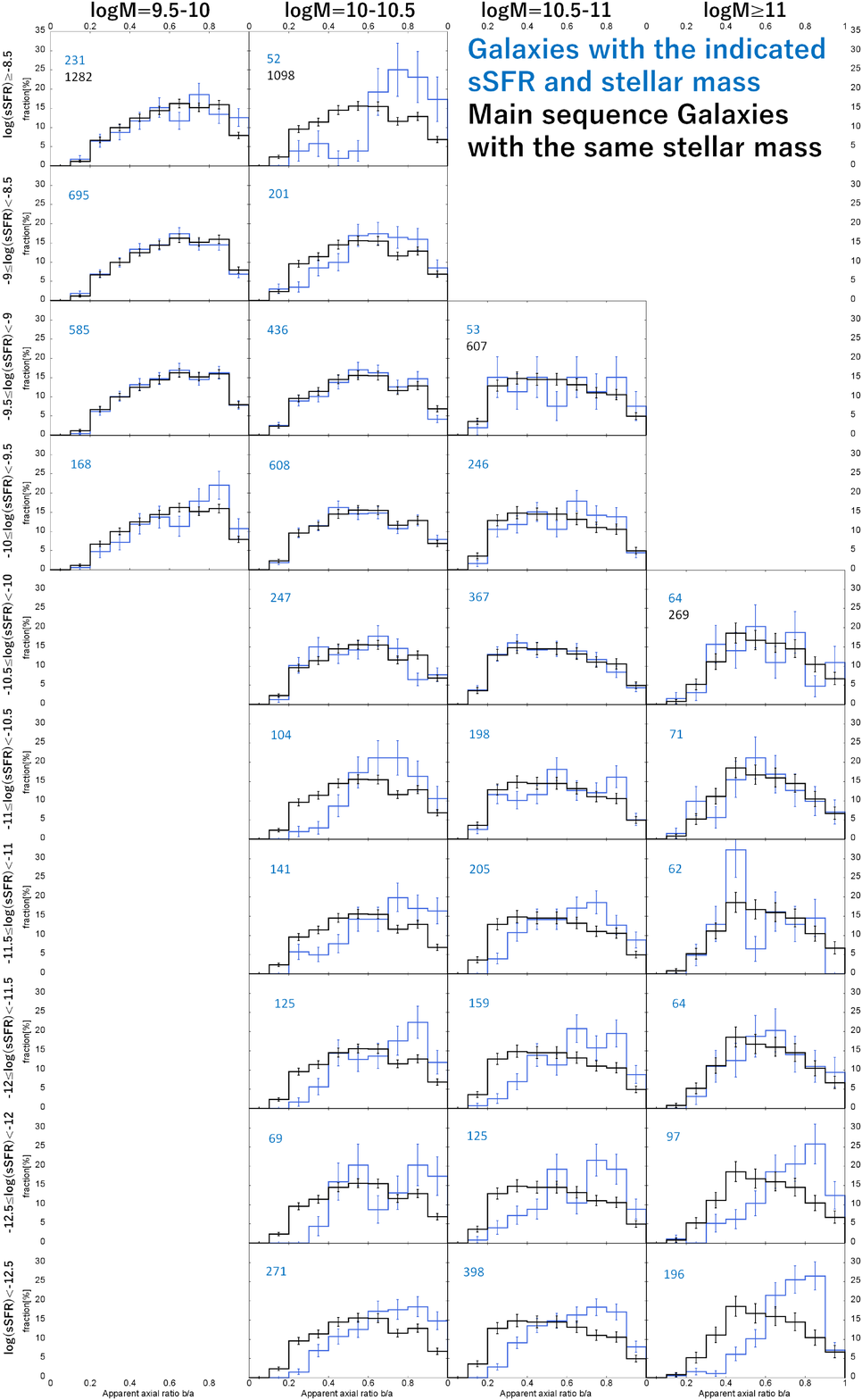}
\caption{ 
The distribution of the apparent axial ratio for subsamples at $0.2<z<0.6$ 
as a function of sSFR and stellar mass.
The sSFR decreases from top to bottom row, and the stellar mass increases 
from left to right column.
The error bars are based on the square root of the number of objects in 
the bin except for cases with a very small number of objects (see text). 
The distribution for star-forming galaxies on the main sequence with 
$\Delta$MS $ = -0.5$ -- $+0.5$ dex 
in the same mass range is also shown for reference, 
although that of those galaxies with 
$\log{(sSFR)} > -12$ are plotted for reference 
in the $M_{\rm star} > 10^{11} M_{\odot}$ bins.
The total number of objects in the subsample is also shown in each panel.
\label{fig:arlowz}}
\end{figure*}

\begin{figure*}
\epsscale{0.86}
\plotone{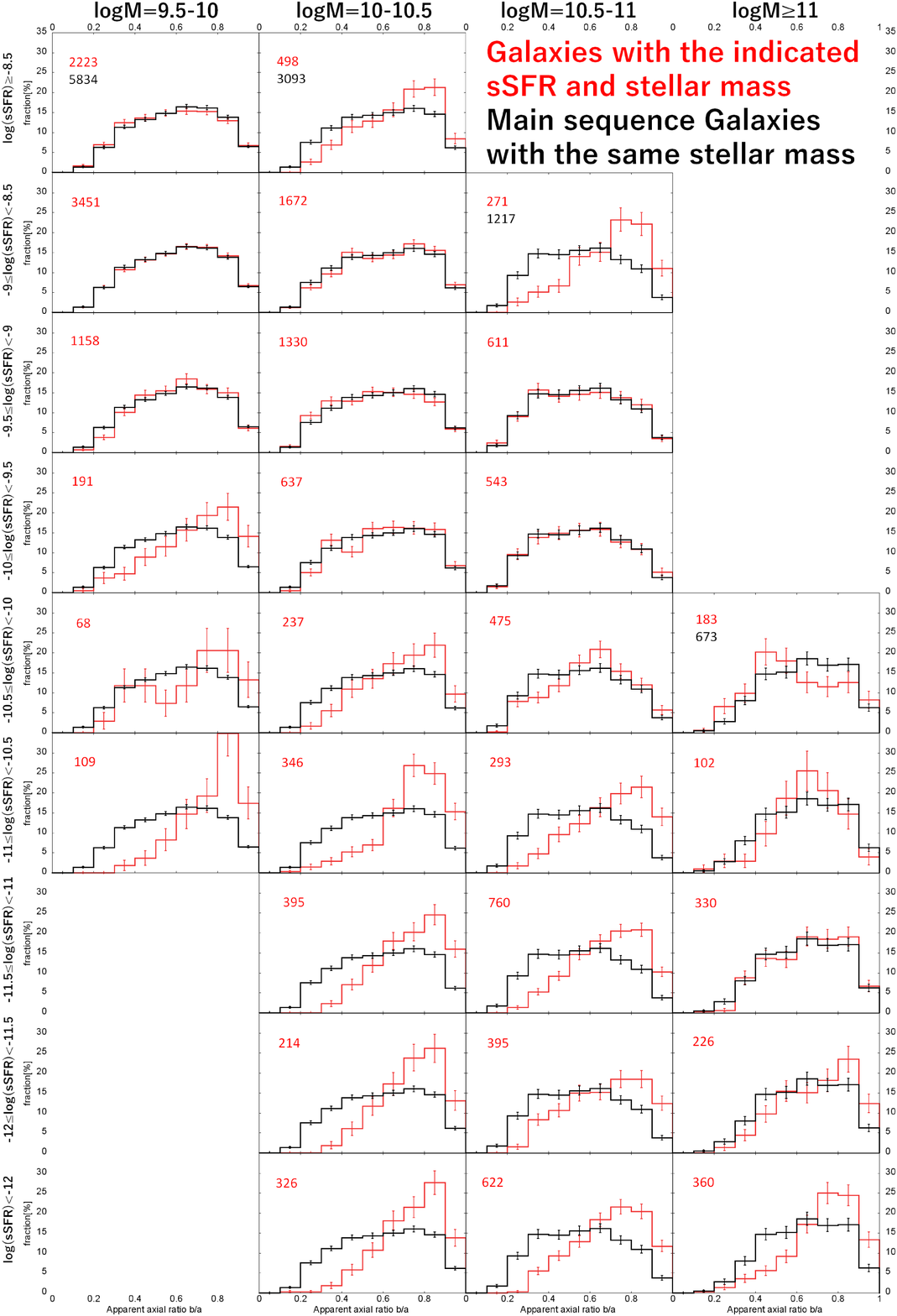}
\caption{ 
Same as Figure \ref{fig:arlowz}, but for galaxies at $0.6<z<1.0$.
In the $M_{\rm star} > 10^{11} M_{\odot}$ bins, that of those galaxies with 
$\log{(sSFR)} > -11.5$ is plotted for reference.
\label{fig:arhighz}}
\end{figure*}

\begin{figure*}
\epsscale{1.1}
\plotone{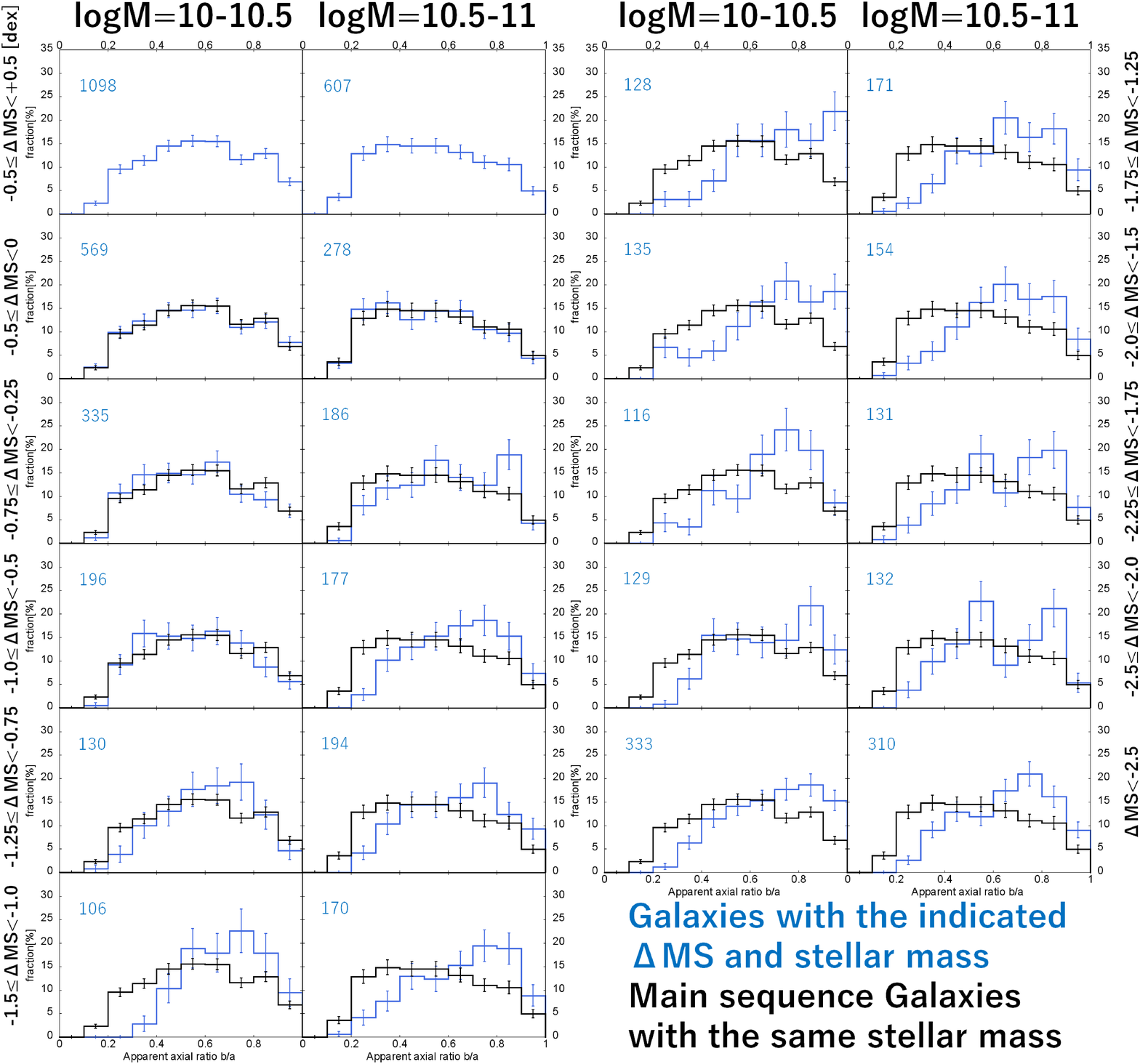}
\caption{ 
The distribution of the apparent axial ratio for subsamples at $0.2<z<0.6$ 
as a function of $\Delta$MS and stellar mass.
The $\Delta$MS decreases from top to bottom row, 
 and the panels in the left (right) column show galaxies with 
$M_{\rm star} = 10^{10}$--$10^{10.5} M_{\odot}$ ($10^{10.5}$--$10^{11} M_{\odot}$).
The distribution for the main-sequence galaxies with 
$\Delta$MS $ = -0.5$ -- $+0.5$ dex in the same mass range is also shown for reference.
The total number of objects in the subsample is also shown in each panel.
\label{fig:dmslowz}}
\end{figure*}

\begin{figure*}
\epsscale{1.1}
\plotone{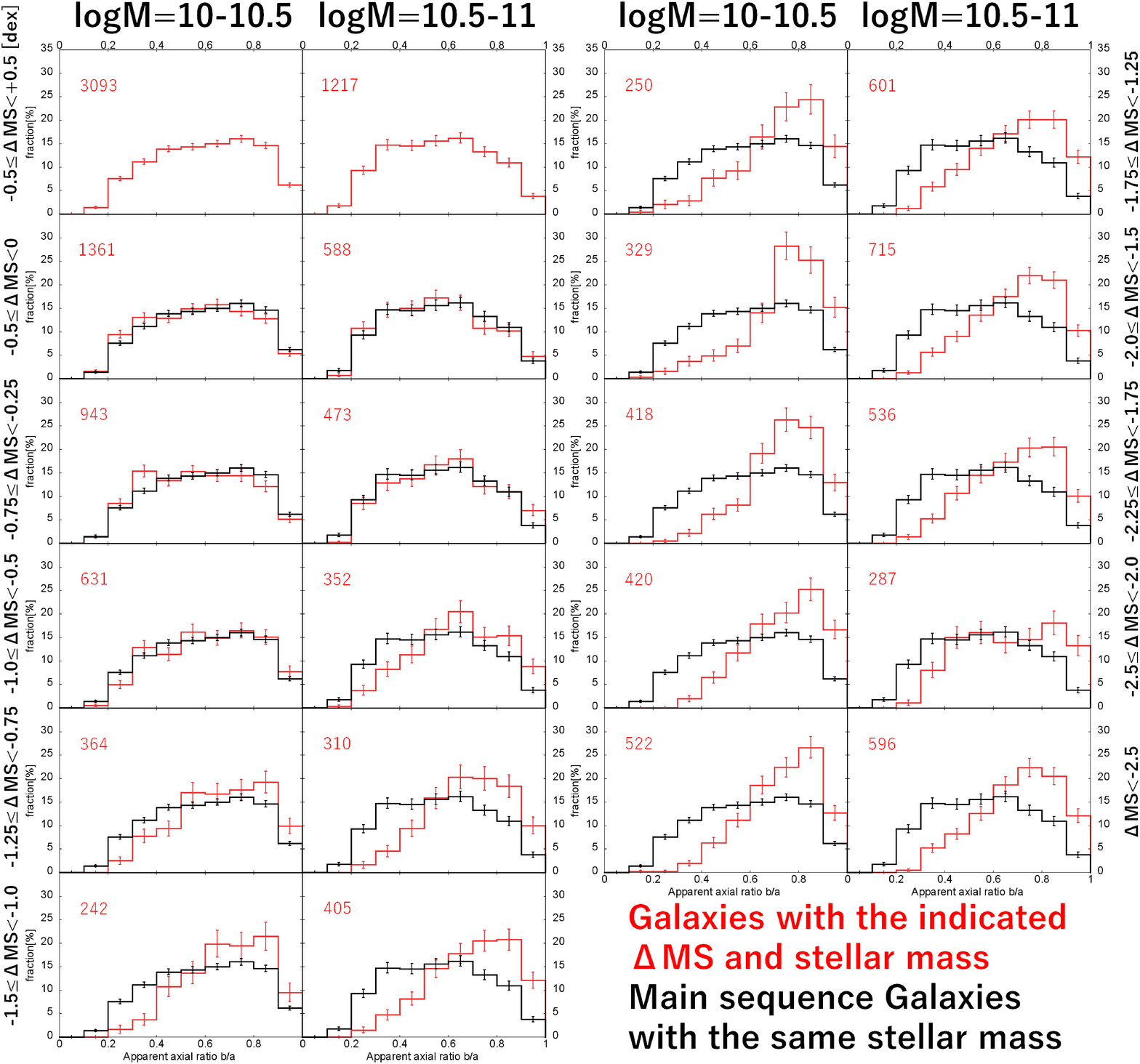}
\caption{ 
Same as Figure \ref{fig:dmslowz}, but for galaxies at $0.6<z<1.0$.
\label{fig:dmshighz}}
\end{figure*}

In order to investigate the 3-dimensional shape of galaxies as a function of 
redshift, stellar mass, and sSFR, we divided the sample galaxies into subsamples
with different properties and derived their distribution of the axial ratio 
separately. We basically divided by redshift into those at $0.2<z<0.6$ 
and $0.6<z<1.0$ to study the evolution, and by stellar mass into those with 
$M_{\rm star} = 10^{9.5}$--$10^{10} M_{\odot}$, $10^{10}$--$10^{10.5} M_{\odot}$, 
$10^{10.5}$--$10^{11} M_{\odot}$, and $M_{\rm star} > 10^{11} M_{\odot}$ to investigate 
the mass dependence. 
Figures \ref{fig:arlowz} and \ref{fig:arhighz} show the distribution of the 
axial ratio as a function of sSFR for galaxies at $0.2<z<0.6$ and $0.6<z<1.0$, 
respectively. 
We plot only those for the subsamples with sufficient number of objects 
($n > 50$) to ensure the statistical accuracy.
For $M_{\rm star} = 10^{9.5}$--$10^{10} M_{\odot}$, galaxies  
with a relatively low sSFR are missed from our sample due to the 
 incompleteness by the criterion of $M_{V} < -20$ 
as mentioned in Section \ref{sec:sample}.
On the other hand, the number of massive star-forming galaxies with $M_{\rm star} 
\gtrsim 10^{11} M_{\odot}$ is intrinsically small, 
and the results for these galaxies 
are also not shown in the figures.

\begin{figure*}[t]
\epsscale{1.1}
\plotone{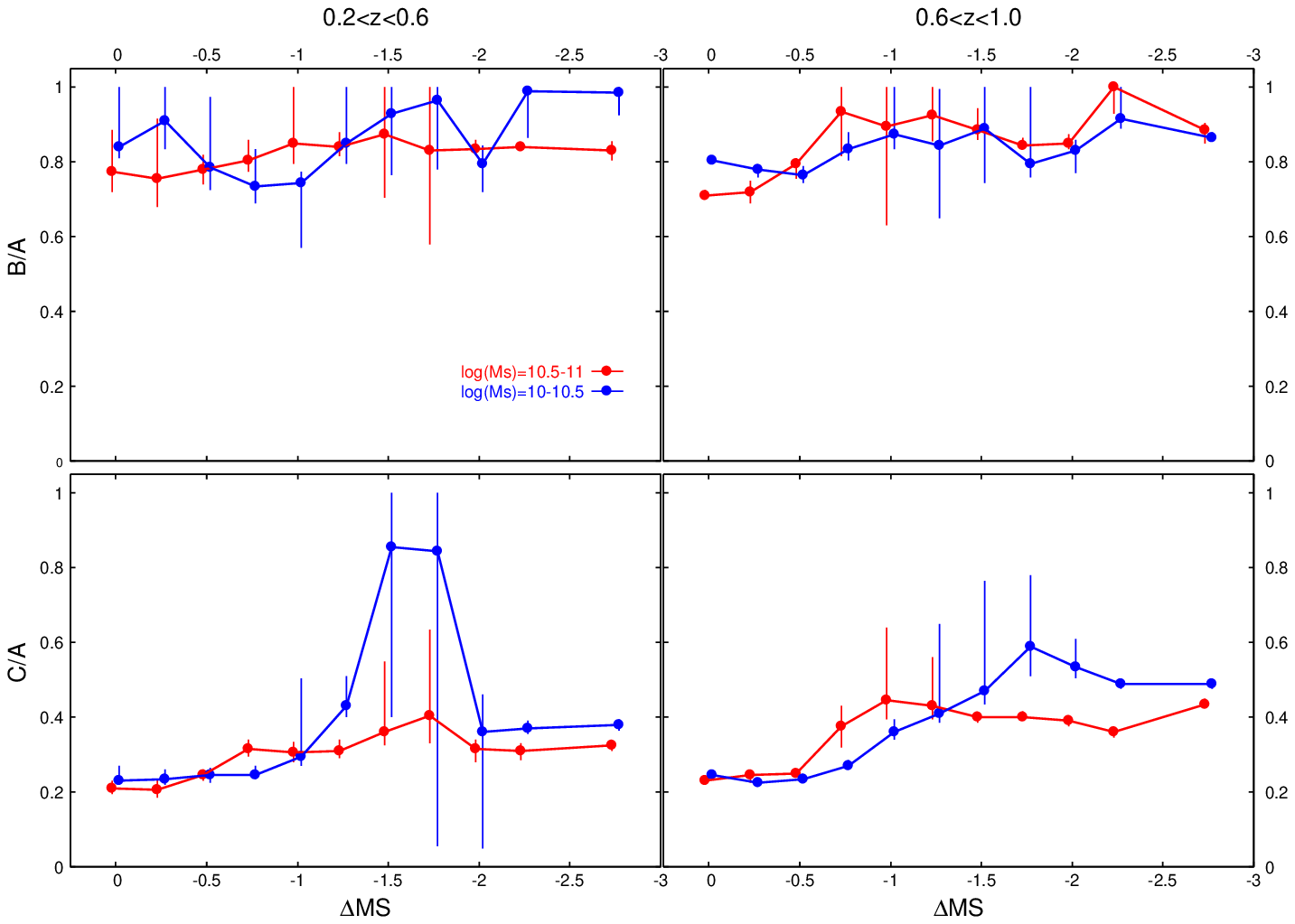}
\caption{ 
The best-fit mean values of the intrinsic face-on axial ratio $\mu_{B/A}$ 
(top panels) and edge-on axial ratio $\mu_{C/A}$ (bottom panels) as a function 
of $\Delta$MS for galaxies at $0.2<z<0.6$ (left) and $0.6<z<1.0$ (right).
Blue symbols show galaxies with $M_{\rm star} = 10^{10}$--$10^{10.5} M_{\odot}$, 
and red symbols show those with $10^{10.5}$--$10^{11} M_{\odot}$.
The error bars represent the 68\% confidence ranges of the $\mu_{B/A}$ and $\mu_{C/A}$.
\label{fig:intssfr}}
\end{figure*}

In Figures \ref{fig:arlowz} and \ref{fig:arhighz}, 
one can see the axial-ratio distribution changes with decreasing sSFR 
for most of the redshift and stellar mass bins.
We also plot the distribution for star-forming galaxies on the main sequence, 
i.e., those with $\Delta$MS $ = -0.5$ -- $+0.5$ dex in the same redshift 
and stellar mass bin for reference. 
For those with $M_{\rm star} > 10^{11} M_{\odot}$, we could not define 
the main sequence 
as mentioned in Section \ref{sec:sample} and plot that of galaxies with 
$sSFR > 10^{-12}$ yr$^{-1}$ at $0.2<z<0.6$ and those with 
$sSFR > 10^{-11.5}$ yr$^{^-1}$
 at $0.6<z<1.0$ in the same mass range for reference. 
The distribution for star-forming galaxies at a relatively high sSFR 
 tends to be flat with a plateau 
over $b/a = $ 0.3 -- 0.9. On the other hand, the fraction of galaxies 
with $b/a < 0.5$
decreases with decreasing sSFR, and the distributions at low sSFRs 
have a peak around 
$b/a \sim $ 0.8 -- 0.9. As shown in previous studies, 
the flat distribution indicates a relatively flat disk 3-dimensional shape, 
while that with a peak around $b/a \sim $ 0.8 -- 0.9 is expected for more thick 
spheroidal shape. Thus Figures \ref{fig:arlowz} and \ref{fig:arhighz} 
suggest that 
star-forming galaxies except for those with a extremely high sSFR are 
basically disk-dominated galaxies and passively evolving galaxies 
with a low sSFR 
have spheroidal morphology, which is consistent with the results 
in many previous 
studies at $z<1$ mentioned in Section \ref{sec:intro}.  

The sSFR at which the transition from the flat distribution to that with a peak 
around $b/a \sim $ 0.8 -- 0.9 occurs 
seems to depend on both redshift and stellar 
mass. The distributions for galaxies with 
$M_{\rm star} = 10^{10}$--$10^{10.5} M_{\odot}$ and 
$10^{10.5}$--$10^{11} M_{\odot}$ at $0.2<z<0.6$ significantly change around 
$sSFR \sim 10^{-11}$ yr$^{-1}$
 and $\sim 10^{-11.5}$ yr$^{-1}$, respectively. On the other hand, 
those for galaxies in the same mass ranges at $0.6<z<1.0$ change around 
$sSFR \sim 10^{-10.5}$ yr$^{-1}$ and $\sim 10^{-11}$ yr$^{-1}$, respectively. 
The transition sSFR is higher for less massive galaxies at higher redshift.
In order to investigate the relationship between the transition sSFR 
and the main 
sequence of star-forming galaxies, we divided our sample by $\Delta$MS 
defined in Section \ref{sec:sample} instead of 
sSFR itself and plot the distribution of the apparent axial ratio 
as a function of 
$\Delta$MS in Figures \ref{fig:dmslowz} and \ref{fig:dmshighz} for 
galaxies at $0.2<z<0.6$ and $0.6<z<1.0$, respectively.
We defined those with $-0.5 < \Delta$MS $ < 0.5$ as ``main-sequence'' galaxies 
and plot the distribution of these galaxies in each panel for reference. 
The other subsamples have a bin width of 0.5 dex in $\Delta$MS 
that is offsetted by 0.25 dex from the next ones.
In Figures \ref{fig:dmslowz} and \ref{fig:dmshighz}, the distribution for the 
main-sequence galaxies is flat, and the transition of the distribution
 occurs at $\Delta$MS $ \sim -1$ dex for the both redshift ranges.
The distribution of galaxies with $M_{\rm star} = 10^{10.5}$--$10^{11} M_{\odot}$ 
changes at 
$\Delta$MS $ \sim -0.75$ dex, which is slightly higher than that of galaxies 
with $10^{10}$--$10^{10.5} M_{\odot}$ in the 
both redshift bins.

\begin{figure*}[t]
\epsscale{1.15}
\plotone{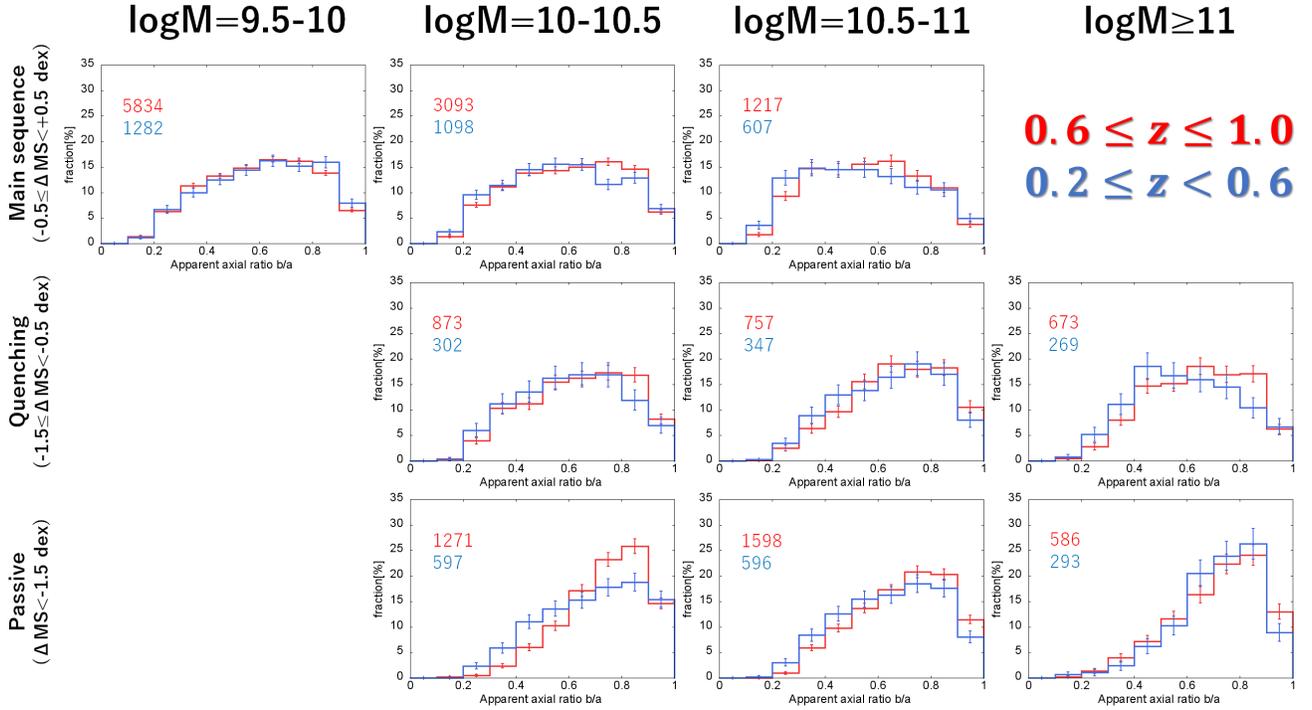}
\caption{ 
The evolution of the distribution of the apparent axial ratio as a function 
of $\Delta$MS and stellar mass.
Top, middle, and bottom panels show the main-sequence galaxies with 
$\Delta$MS $ = -0.5$ -- $+0.5$ dex, quenching galaxies with 
$\Delta$MS $ = -1.5$ -- $-0.5$ 
dex, and passively evolving galaxies with $\Delta$MS $ < -1.5$ dex, respectively.
The stellar mass increases from left to right panels. 
Blue lines show those galaxies at $0.2<z<0.6$ and red lines show those at $0.6<z<1.0$.
Note that massive galaxies with $M_{\rm star} > 10^{11} M_{\odot}$ are divided into 
quenching and passively evolving subsamples at a sSFR of $10^{-12}$ and $10^{-11.5}$ 
yr$^{-1}$ for those at $0.2<z<0.6$ and $0.6<z<1.0$, respectively, 
where the change of the 
axial-ratio distribution occurs (see text).
\label{fig:dms2evol}}
\end{figure*}

Using the Monte Carlo simulation described in Section \ref{subsec:intrinsic}, 
we then fitted the distribution of each subsample in Figures 
\ref{fig:dmslowz} and \ref{fig:dmshighz} with the triaxial ellipsoid models 
to estimate the intrinsic 3-dimensional shape.  
The comparisons with the best-fit models and the observed distributions 
for the subsamples are presented in Appendix \ref{sec:compmodel}.   
We show the estimated mean values of the intrinsic face-on axial ratio 
and edge-on axial ratio, namely, $\mu_{B/A}$ and $\mu_{C/A}$ 
as a function of $\Delta$MS in Figure \ref{fig:intssfr}. 
The face-on axial ratio $B/A$ seems to increase with decreasing $\Delta$MS 
from $B/A \sim $ 0.7 -- 0.85 at $\Delta$MS $ \sim 0$ to $B/A \sim $ 0.8 -- 0.95 
at $\Delta$MS $ \sim -2.5$
in the all redshift and stellar mass bins we investigated, but we cannot 
conclude it because the uncertainty in the estimates of $B/A$ is large. 
The $B/A$ of galaxies with $M_{\rm star} = 10^{10}$--$10^{10.5} M_{\odot}$ may be 
systematically higher than those with $10^{10.5}$--$10^{11} M_{\odot}$ 
at $0.2<z<0.6$ 
although the uncertainty is large, while those with the different mass ranges 
are consistent with each other within uncertainty at $0.6<z<1.0$.

On the other hand, the edge-on axial ratio $C/A$ (i.e., thickness) clearly 
increases with decreasing $\Delta$MS from $C/A \sim $ 0.2 -- 0.25 
at $\Delta$MS $ \sim 0$
 to $C/A \sim $ 0.3 -- 0.5 at $\Delta$MS $\lesssim -1.5$, 
although the uncertainty in some of bins at 
$-2 < \Delta$MS $ < -1$ is relatively large due to a small number of objects 
especially for less massive and lower redshift bins.
The transition from $C/A \sim $ 0.2 -- 0.25 to $\sim $ 0.3 -- 0.5 occurs around 
$\Delta$MS $ \sim -1$ dex in the all redshift and mass bins, 
which is consistent with 
the results in Figures \ref{fig:dmslowz} and \ref{fig:dmshighz}. 
One can also see the $C/A$ of galaxies with 
$M_{\rm star} = 10^{10}$--$10^{10.5} M_{\odot}$ 
changes at a slightly higher $\Delta$MS of $\sim -0.75$ dex than 
that of galaxies 
with $10^{10.5}$--$10^{11} M_{\odot}$ in the both redshift ranges.

We also note that the distributions of the apparent axial ratio of 
galaxies with 
$M_{\rm star} = 10^{10}$--$10^{10.5} M_{\odot}$ and $\Delta$MS 
$ \sim -2.0$ -- $-1.5$ 
at $0.2<z<0.6$ are fitted with models with a nearly spherical shape 
of $B/A \sim C/A \sim 1$ and their uncertainty of $\mu_{C/A}$ is 
extremely large. 
The small numbers of objects in these bins ($n \sim 130$) make the 
statistical uncertainty of the distribution relatively large, 
and the constraints on the intrinsic axial ratios are weak.
Furthermore, while the distributions have a broad peak around 
$b/a \sim 0.8$ -- 1.0,  
there are also a non-negligible fraction of galaxies with $b/a = 0.2$ -- 0.3.
Such distributions cannot be explained by the models with a 
narrow range of $C/A$, and only the models with a relatively 
large $\sigma_{C/A}$ enable to reproduce these distributions 
(see Appendix \ref{sec:largeerr} for details).
In such cases with a large $\sigma_{C/A}$, there are models with 
various $\mu_{C/A}$ values allowed by the observed distribution, 
which leads to the very large uncertainty of $\mu_{C/A}$ for these galaxies.
With the small sizes of these subsamples, we cannot conclude 
whether the non-negligible fractions of galaxies with $b/a = 0.2$ -- 0.3 
are caused by the statistical fluctuation or not.


\subsection{Evolution of axial ratio \& 3-D shape} \label{subsec:evol}

\begin{figure*}[t]
\epsscale{1.1}
\plotone{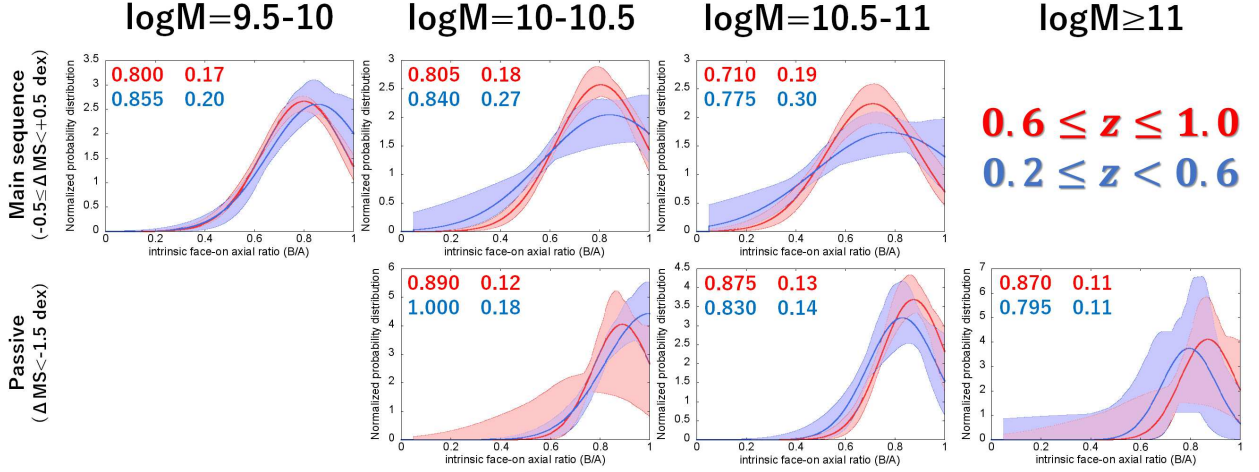}
\caption{ 
The probability distribution of the intrinsic face-on axial ratio $B/A$ 
calculated from the best-fit $\mu_{B/A}$ and $\sigma_{B/A}$ as a function of 
stellar mass for the main sequence (top) and passively evolving (bottom) 
subsamples. 
The stellar mass increases from left to right panels.
Blue lines show those galaxies at $0.2<z<0.6$ and red lines show 
those at $0.6<z<1.0$.
The shaded regions represent the 68\% confidence ranges calculated with 
the $\Delta \chi^{2}$ method. 
The best-fit $\mu_{B/A}$ and $\sigma_{B/A}$ values are also shown in each panel.
\label{fig:intabevol}}
\end{figure*}
\begin{figure*}
\epsscale{1.1}
\plotone{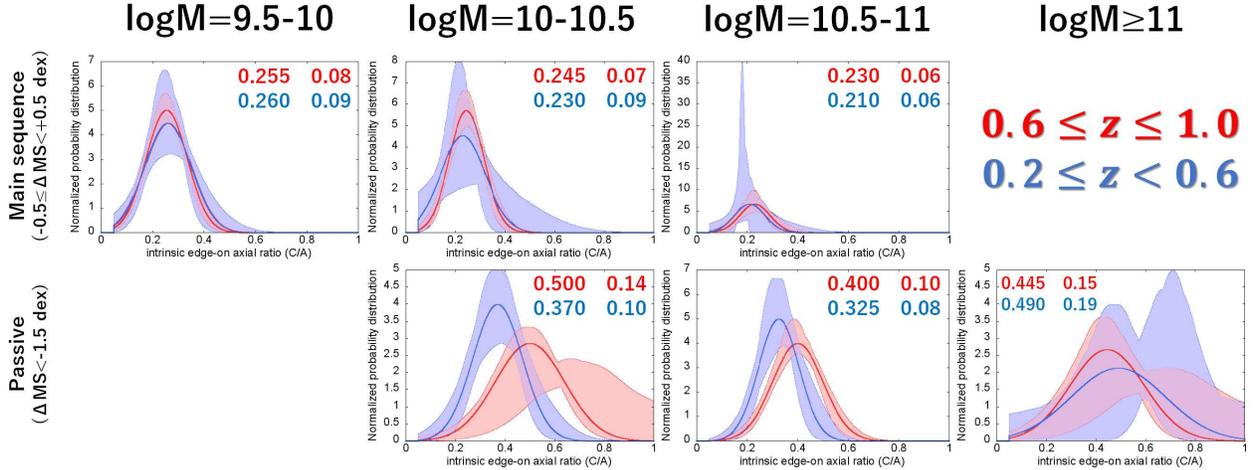}
\caption{ 
Same as Figure \ref{fig:intabevol}, but for the intrinsic edge-on axial ratio 
$C/A$ calculated from the best-fit $\mu_{C/A}$ and $\sigma_{C/A}$.
\label{fig:intacevol}}
\end{figure*}

\begin{deluxetable*}{rrccccc}[t] 
\tablecaption{the best-fit parameters of the triaxial ellipsoid models for 
the subsamples \label{tab:fitevol}}
\tablehead{
\colhead {stellar mass} & \colhead{$\Delta$MS} & \colhead{$\mu_{B/A}$} & \colhead{$\sigma_{B/A}$} & \colhead{$\mu_{C/A}$} & \colhead{$\sigma_{C/A}$} & \colhead{$\chi^2_{\rm min}$\tablenotemark{a}}
} 
\startdata
\multicolumn6c{$z=0.2$--0.6} \\
\hline
$\log{M_{\rm star}}=9.5$--10 & $\Delta$MS $ = -0.5$ -- $+0.5$ & $0.855^{+0.095}_{-0.055}$ & $0.20^{+0.07}_{-0.05}$ & $0.265^{+0.025}_{-0.020}$ & $0.09^{+0.04}_{-0.03}$ & 5.66 \\
 $\log{M_{\rm star}}=10$--10.5 & $\Delta$MS $ = -0.5$ -- $+0.5$ & $0.840^{+0.160}_{-0.070}$ & $0.27^{+0.25}_{-0.07}$ & $0.230^{+0.080}_{-0.025}$ & $0.09^{+0.11}_{-0.04}$ & 10.7 \\
 & $\Delta$MS $ = -1.5$ -- $-0.5$ & $0.780^{+0.220}_{-0.055}$ & $0.17^{+0.18}_{-0.05}$ & $0.275^{+0.065}_{-0.035}$ & $0.06^{+0.05}_{-0.05}$ & 1.27 \\
 & $\Delta$MS $ < -1.5$ & $1.000^{+0.000}_{-0.110}$ & $0.18^{+0.03}_{-0.07}$ & $0.370^{+0.035}_{-0.030}$ & $0.10^{+0.04}_{-0.02}$ & 1.61 \\
 $\log{M_{\rm star}}=10.5$--11 & $\Delta$MS $ = -0.5$ -- $+0.5$ & $0.775^{+0.225}_{-0.100}$ & $0.30^{+0.34}_{-0.09}$ & $0.210^{+0.06}_{-0.03}$ & $0.06^{+0.08}_{-0.05}$ & 1.07 \\
 & $\Delta$MS $ = -1.5$ -- $-0.5$ & $0.830^{+0.170}_{-0.050}$ & $0.14^{+0.13}_{-0.05}$ & $0.315^{+0.040}_{-0.035}$ & $0.08^{+0.04}_{-0.03}$ & 0.263 \\
 & $\Delta$MS $ < -1.5$ & $0.830^{+0.115}_{-0.035}$ & $0.14^{+0.10}_{-0.04}$ & $0.325^{+0.035}_{-0.020}$ & $0.08^{+0.02}_{-0.02}$ & 1.45 \\
 $\log{M_{\rm star}} > 11$ & $\log{(sSFR)}>-12.0$ & $0.835^{+0.165}_{-0.195}$ & $0.33^{+0.67}_{-0.17}$ & $0.335^{+0.125}_{-0.075}$ & $0.10^{+0.07}_{-0.05}$ & 0.847 \\
 & $\log{(sSFR)}<-12.0$ & $0.795^{+0.205}_{-0.130}$ & $0.11^{+0.89}_{-0.05}$ & $0.490^{+0.245}_{-0.055}$ & $0.19^{+0.31}_{-0.11}$ & 3.19 \\
\hline
\multicolumn6c{$z=0.6$--1.0} \\
\hline 
$\log{M_{\rm star}}=9.5$--10 & $\Delta$MS $ = -0.5$ -- $+0.5$ & $0.800^{+0.015}_{-0.020}$ & $0.17^{+0.02}_{-0.01}$ & $0.255^{+0.010}_{-0.010}$ & $0.08^{+0.01}_{-0.01}$ & 2.02 \\
 10--10.5 & $\Delta$MS $ = -0.5$ -- $+0.5$ & $0.805^{+0.020}_{-0.025}$ & $0.18^{+0.03}_{-0.03}$ & $0.245^{+0.010}_{-0.010}$ & $0.07^{+0.01}_{-0.01}$ & 11.4 \\
 & $\Delta$MS $ = -1.5$ -- $-0.5$ & $0.840^{+0.075}_{-0.045}$ & $0.15^{+0.07}_{-0.03}$ & $0.295^{+0.030}_{-0.015}$ & $0.07^{+0.03}_{-0.02}$ & 4.42 \\
 & $\Delta$MS$ < -1.5$ & $0.890^{+0.110}_{-0.170}$ & $0.12^{+0.26}_{-0.04}$ & $0.500^{+0.245}_{-0.025}$ & $0.14^{+0.15}_{-0.02}$ & 1.48 \\
 $\log{M_{\rm star}}=10.5$--11 & $\Delta$MS$ = -0.5$ -- $+0.5$ & $0.710^{+0.040}_{-0.020}$ & $0.19^{+0.05}_{-0.03}$ & $0.230^{+0.020}_{-0.020}$ & $0.06^{+0.02}_{-0.02}$ & 3.06 \\
 & $\Delta$MS $ = -1.5$ -- $-0.5$ & $0.915^{+0.085}_{-0.090}$ & $0.20^{+0.14}_{-0.08}$ & $0.385^{+0.080}_{-0.035}$ & $0.12^{+0.05}_{-0.03}$ & 5.92 \\
 & $\Delta$MS $ < -1.5$ & $0.875^{+0.030}_{-0.030}$ & $0.13^{+0.03}_{-0.03}$ & $0.400^{+0.015}_{-0.020}$ & $0.10^{+0.01}_{-0.02}$ & 5.06 \\
 $\log{M_{\rm star}} > 11$ & $\log{(sSFR)}>-11.5$ & $0.810^{+0.060}_{-0.050}$ & $0.16^{+0.09}_{-0.05}$ & $0.335^{+0.035}_{-0.030}$ & $0.09^{+0.04}_{-0.02}$ & 4.56 \\
 & $\log{(sSFR)}<-11.5$ & $0.870^{+0.130}_{-0.175}$ & $0.11^{+0.32}_{-0.04}$ & $0.445^{+0.285}_{-0.035}$ & $0.15^{+0.18}_{-0.04}$ & 0.490 \\
\enddata
\tablenotetext{a}{the minimum $\chi^2$ value in the fitting (9 degrees of freedom).}
\end{deluxetable*}

In order to investigate the evolution of the intrinsic shape of star-forming 
and passively evolving galaxies with high statistical accuracy, 
we here divided our sample by $\Delta$MS into three subsamples, namely, 
star-forming main-sequence galaxies with $-0.5 < \Delta$MS $ < 0.5$, 
quenching galaxies with $-1.5 < \Delta$MS $ < -0.5$, and passively evolving 
galaxies with $\Delta$MS $ < -1.5$. 
We chose these criteria in $\Delta$MS taking account of the $\Delta$MS 
dependence 
of the axial-ratio distribution and its transition found in 
the previous section.
The main-sequence galaxies have the relatively flat distribution 
of the apparent 
axial ratio, which suggests low-$C/A$ disky morphology, 
while passively evolving 
galaxies show higher-$C/A$ and thick spheroidal morphology.
Quenching galaxies are defined as that between these two populations.
For galaxies with $M_{\rm star} > 10^{11} M_{\odot}$, we cannot define 
the main sequence and therefore used constant $sSFR$ values as criteria.  
We divided these massive galaxies into quenching and passively evolving 
subsamples 
at $sSFR = 10^{-12}$ and $10^{-11.5}$ yr$^{-1}$ for those at $0.2<z<0.6$ 
and $0.6<z<1.0$,
 respectively. We also set these criteria for massive galaxies taking 
account of the transition of the axial-ratio distribution.

In Figure \ref{fig:dms2evol}, we compare the distributions of the 
apparent axial 
ratio for galaxies at $0.2<z<0.6$ and $0.6<z<1.0$ in each stellar mass and 
$\Delta$MS bin. 
Those of the main-sequence galaxies in the two redshift bins 
are consistent with each other within uncertainty in the all mass ranges, 
although the fraction of galaxies with $b/a \sim $ 0.7 -- 0.8  is marginally 
higher at $0.6<z<1.0$ for galaxies with 
$M_{\rm star} = 10^{10}$--$10^{11} M_{\odot}$. 
These galaxies basically show the flat distribution.
On the other hand, passively evolving galaxies with 
$M_{\rm star} < 10^{11} M_{\odot}$
show a significant evolution. 
Those galaxies at $0.2<z<0.6$ have the flatter distribution 
with higher fraction 
of objects with $b/a < 0.5$ and lower peak around $b/a \sim $ 0.8 -- 0.9 than 
those at $0.6<z<1.0$. The evolution seems to be stronger for passively evolving
galaxies with $M_{\rm star} = 10^{10}$--$10^{10.5} M_{\odot}$ than those with 
$10^{10.5}$--$10^{11} M_{\odot}$. 
In contrast to these galaxies with $M_{\rm star} < 10^{11} M_{\odot}$, 
massive passively evolving galaxies with $M_{\rm star} > 10^{11} M_{\odot}$ show 
no significant evolution, 
although the uncertainty is relatively large due to a small number of 
these massive galaxies.
These massive galaxies have more peaky distribution skewed toward 
high $b/a$ value 
than less massive ones with $10^{10.5}$--$10^{11} M_{\odot}$ 
in the both redshift ranges.

Quenching galaxies show marginal differences between those at $0.2<z<0.6$ and 
$0.6<z<1.0$, although the uncertainty is large. 
The subsamples at $0.2<z<0.6$ tend to have the flatter distribution or that 
skewed toward lower $b/a$ value, but this could depend on the choice of 
the criteria in $\Delta$MS for these galaxies. 
On the other hand, the evolutionary trends of the main-sequence and 
passively evolving populations mentioned above  
are not changed by the choice of the criteria.

\begin{figure*}[t]
\epsscale{1.15}
\plotone{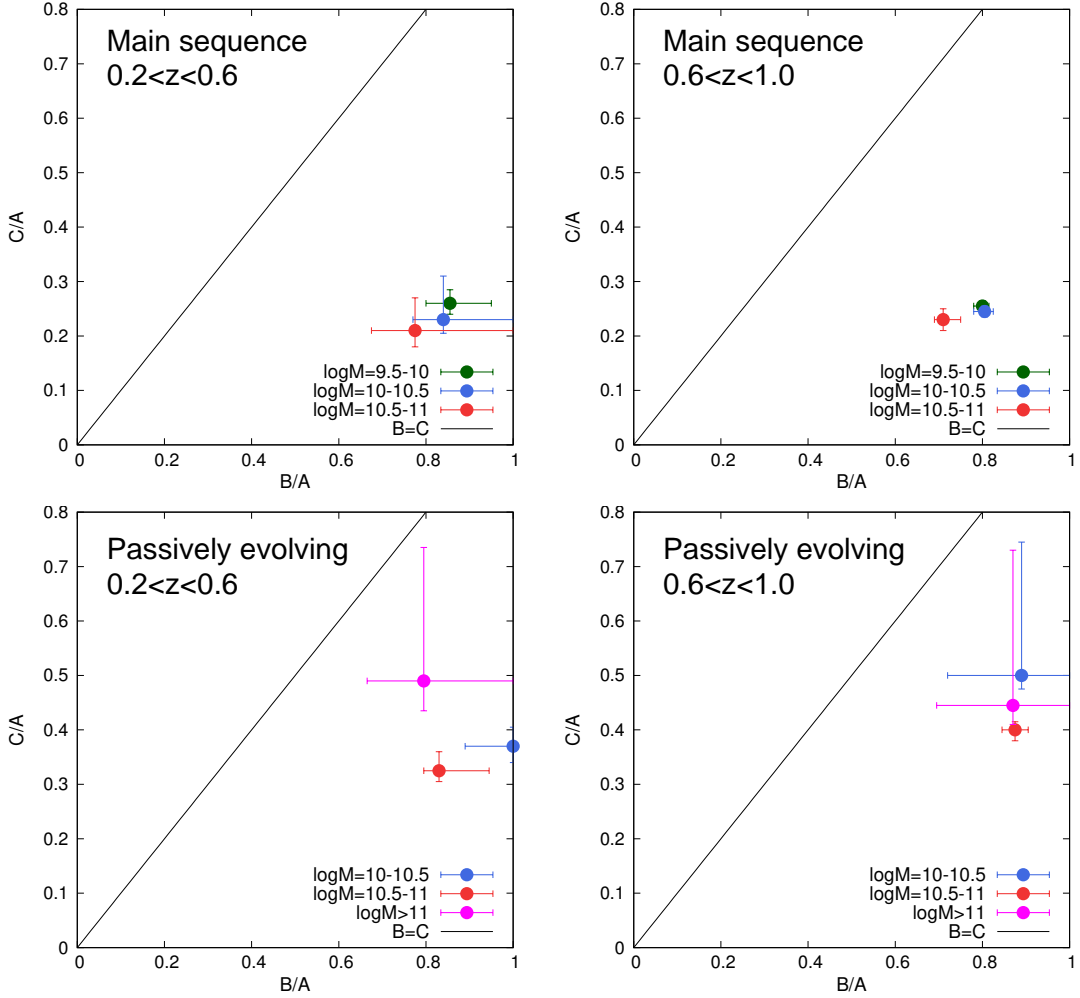}
\caption{ 
The best-fit values of the mean intrinsic face-on axial ratio $\mu_{B/A}$ and 
edge-on axial ratio $\mu_{C/A}$ for the main-sequence galaxies (top panels) 
and passively evolving galaxies (bottom panels) at $0.2<z<0.6$ (left) 
and $0.6<z<1.0$ (right). 
The different colors show galaxies with different stellar masses.
The error bars represent the 68\% confidence ranges of the $\mu_{B/A}$ and $\mu_{C/A}$.
The diagonal line indicates the criterion of $B/A > C/A$ 
set from the definition of $A>B>C$. 
\label{fig:intabcevol}}
\end{figure*}

We estimated the intrinsic 3-dimensional shape from the distribution of 
the apparent axial ratio for the main sequence and passively evolving populations  
with the Monte Carlo simulations, and show the results 
for the face-on axial ratio $B/A$ and edge-on axial ratio $C/A$ in 
Figures \ref{fig:intabevol} and \ref{fig:intacevol}, respectively.
In the figures, we plot the probability 
distributions of $B/A$ and $C/A$ 
and their uncertainty estimated from the best-fit parameters 
and their errors, and 
compare those for galaxies at $0.2<z<0.6$ and $0.6<z<1.0$. 
Note that these are not the likelihood functions of $\mu_{B/A}$ and $\mu_{C/A}$, 
but the Gaussian probability distributions calculated from the best-fit values 
of $\mu_{B/A}$, $\sigma_{B/A}$, $\mu_{C/A}$, and $\sigma_{C/A}$, and their 
confidence ranges.
Since each Gaussian distribution is normalized so that the integration 
over $B/A$ or $C/A$ becomes unity, the confidence range of the probability 
would be high if a relatively small value of $\sigma_{B/A}$ or $\sigma_{C/A}$, 
which corresponds to a narrow Gaussian distribution, is allowed for a certain 
value of $\mu_{B/A}$ or $\mu_{C/A}$. 
The fitting results are also summarized in Table \ref{tab:fitevol}.

The main-sequence galaxies show no evolution in the edge-on axial ratio $C/A$, 
while the face-on axial ratio $B/A$ of these galaxies show marginal changes 
at $M_{\rm star} > 10^{10} M_{\odot}$. 
The mean values of $B/A$ of those galaxies with 
$M_{\rm star} = 10^{10}$--$10^{10.5} M_{\odot}$ 
($10^{10.5}$--$10^{11} M_{\odot}$) evolve from $\mu_{B/A} = 0.805$ (0.710) 
at $0.6<z<1.0$ 
to $\mu_{B/A} = 0.840$ (0.775) at $0.2<z<0.6$, while $\sigma_{B/A}$ 
also increases with time.
On the other hand, passively evolving galaxies with 
$M_{\rm star} < 10^{11} M_{\odot}$ 
 show a significant evolution in the edge-on axial ratio $C/A$. 
The mean values of $C/A$ of passively evolving galaxies with 
$M_{\rm star} = 10^{10}$--$10^{10.5} M_{\odot}$ ($10^{10.5}$--$10^{11} M_{\odot}$) 
decreases 
with time from $\mu_{C/A}=0.500$ (0.400) at $0.6<z<1.0$ to 
$\mu_{C/A}=0.370$ (0.325) 
at $0.2<z<0.6$. Passively evolving galaxies with $M_{\rm star} > 10^{11} M_{\odot}$ 
have relatively high $C/A$ values but show no significant evolution in both 
$B/A$ and $C/A$. 
We summarize the mean values of $B/A$ and $C/A$ for star-forming and passively 
evolving galaxies in Figure \ref{fig:intabcevol}. 
It is seen that $C/A$ of passively evolving galaxies with 
$M_{\rm star} < 10^{11} M_{\odot}$ decreases with time, 
while main-sequence galaxies show no significant evolution but only marginal 
changes in $B/A$.

\begin{figure*}[t]
\epsscale{1.1}
\plottwo{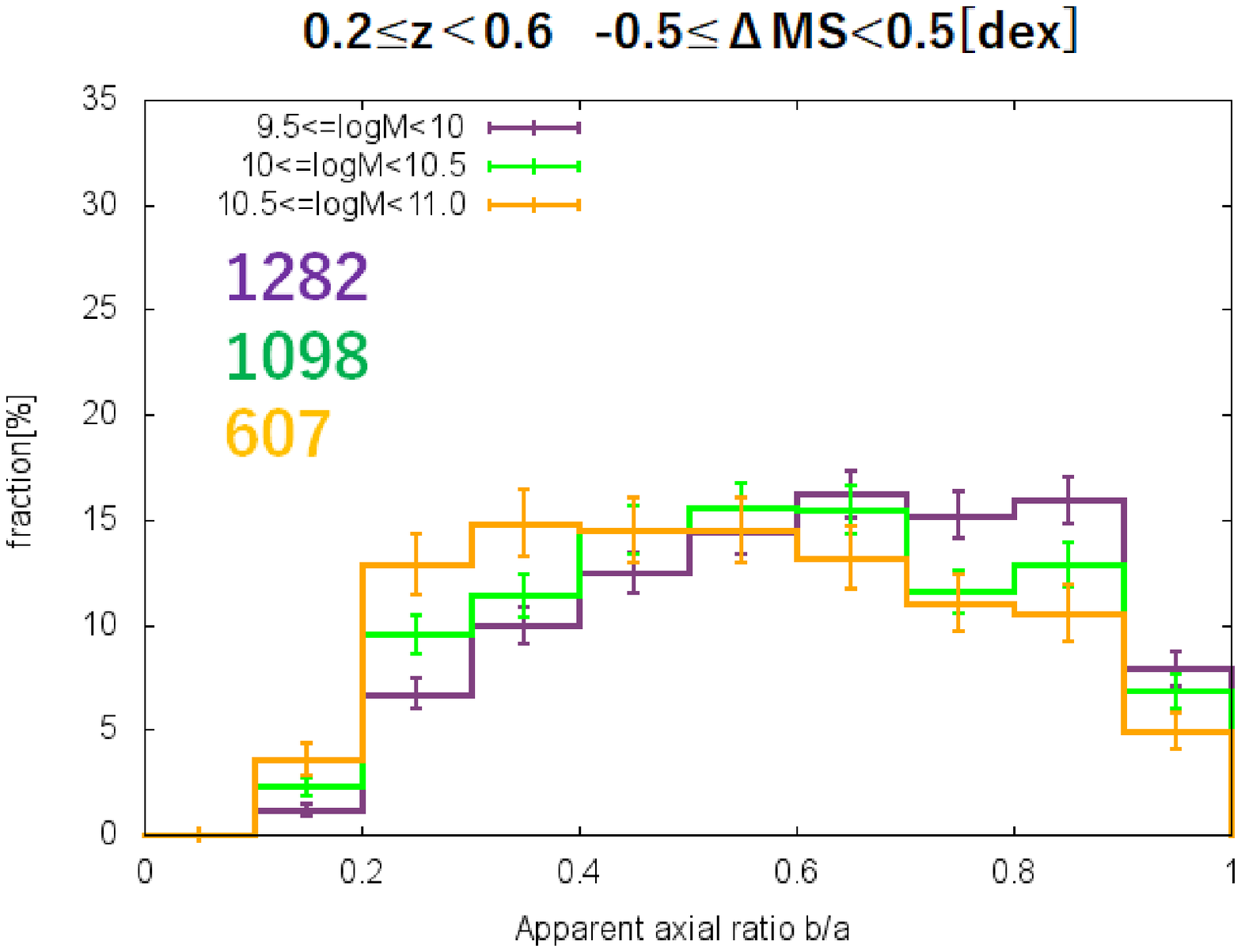}{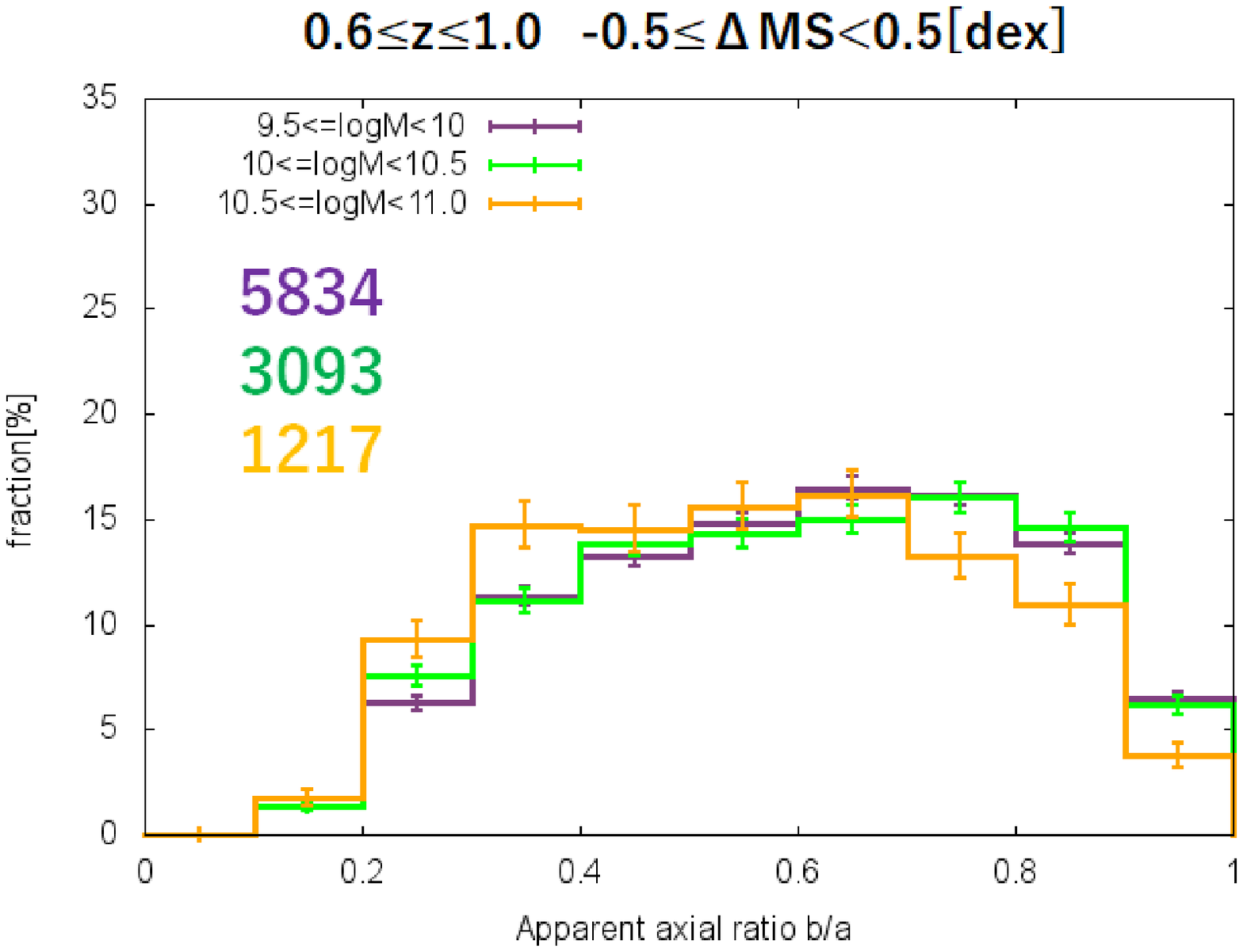}
\caption{ 
The mass dependence of the distribution of the apparent axial ratio 
for the main-sequence galaxies at $0.2<z<0.6$ (left) and $0.6<z<1.0$ (right).
The different colors of the lines show galaxies with different stellar masses.
The total numbers of objects in the subsamples are also shown in each panel.
\label{fig:msmass}}
\end{figure*}
\begin{figure*}
\epsscale{1.1}
\plottwo{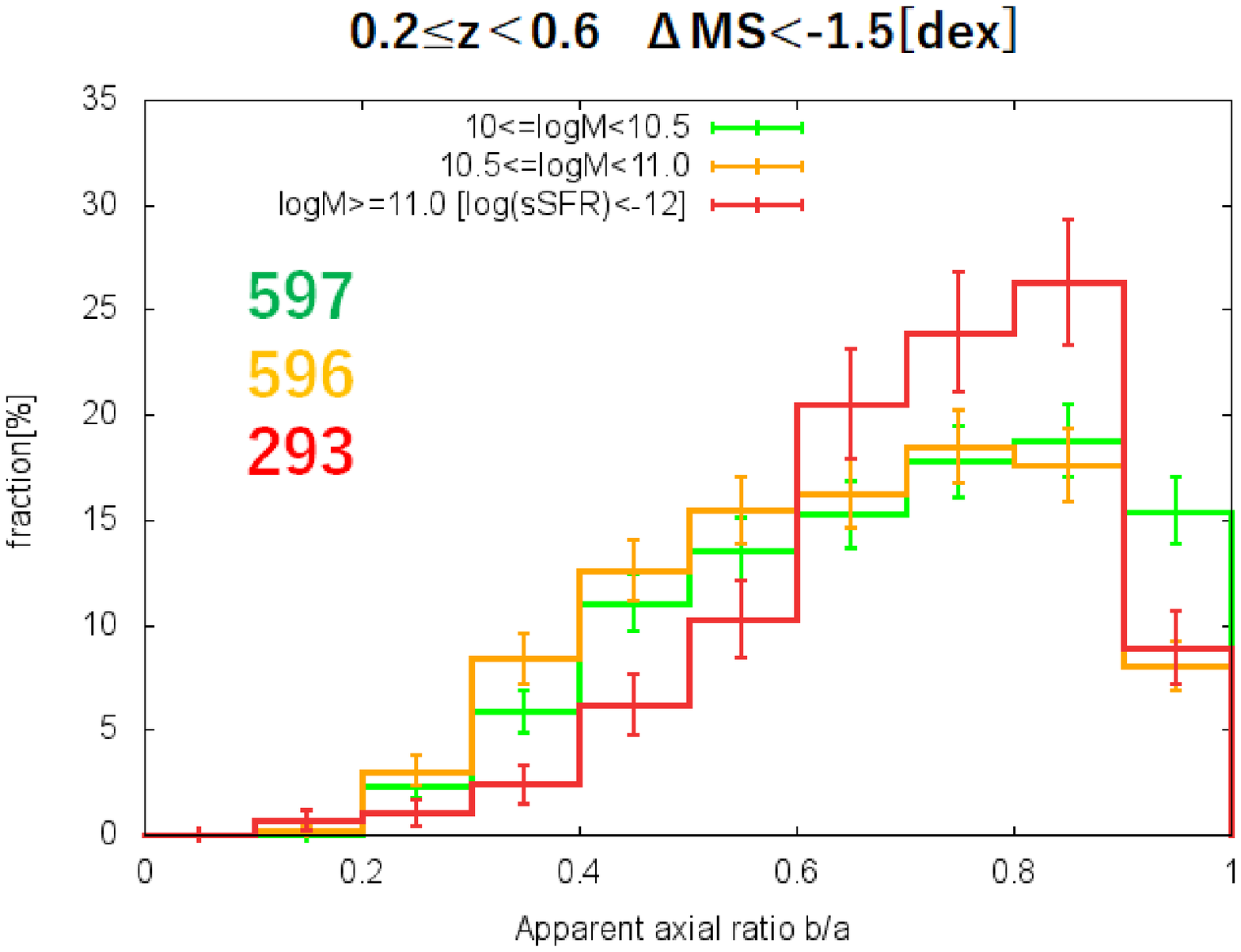}{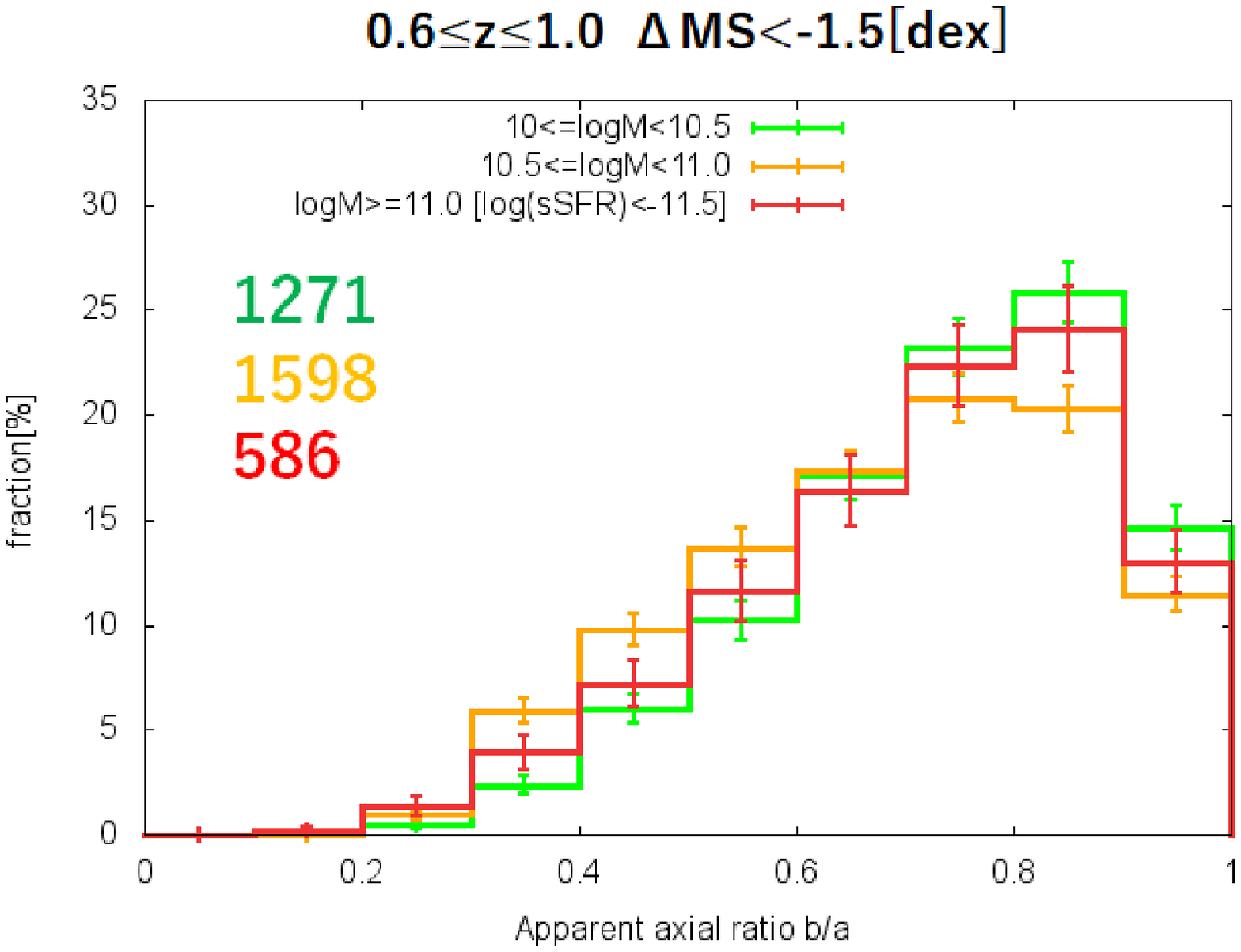}
\caption{ 
Same as Figure \ref{fig:msmass}, but for the passively evolving galaxies.
\label{fig:psmass}}
\end{figure*}

\subsection{Mass dependence of axial ratio \& 3-D shape} \label{subsec:mass}

In Figure \ref{fig:msmass}, we compare the distributions of the apparent 
axial ratio for the main-sequence subsamples with different stellar mass 
ranges to investigate the mass dependence of the 3-dimensional shape of 
star-forming galaxies.
While the distributions for all the main-sequence subsamples are relatively 
flat over $b/a \sim $ 0.2 -- 1.0, 
more massive galaxies tend to show lower values of $b/a$  
in the both redshift ranges. 
More massive main-sequence galaxies have a higher fraction of objects 
with $b/a<0.5$
and a lower fraction of those with $b/a>0.8$ than less massive ones.
The mass dependence of the distribution in the main-sequence galaxies reflects 
mass-dependent edge-on axial ratio $C/A$ (thickness) of these galaxies 
seen in Figure \ref{fig:intabcevol}.
Main-sequence galaxies with $M_{\rm star}=10^{9.5}$--$10^{10} M_{\odot}$, 
$10^{10}$--$10^{10.5} M_{\odot}$, and $10^{10.5}$--$10^{11} M_{\odot}$ at $0.2<z<0.6$ 
($0.6<z<1.0$) have the best-fit mean values of the edge-on axial ratio 
$\mu_{C/A} = $ 0.265, 0.230, and 0.210 (0.255, 0.245, and 0.230), respectively. 
The thickness of star-forming galaxies on the main sequence decreases 
with stellar mass in the both redshift ranges, although the uncertainty 
in these estimated values is not negligible, especially 
for more massive galaxies at lower redshift. 
It is also noted that those galaxies 
with $M_{\rm star}=10^{10.5}$--$10^{11} M_{\odot}$ 
show lower values of the face-on axial ratio $B/A$ than those less massive 
galaxies with $M_{\rm star} < 10^{10.5} M_{\odot}$ in the both redshift ranges.
The difference in $B/A$ may reflect the contribution from the bulge or bar 
structure, since more massive star-forming galaxies tend to show the high 
bulge fraction and/or strong bar (e.g., \citealp{blu19}; \citealp{cer17}).

\begin{figure*}[t]
\epsscale{1.1}
\plotone{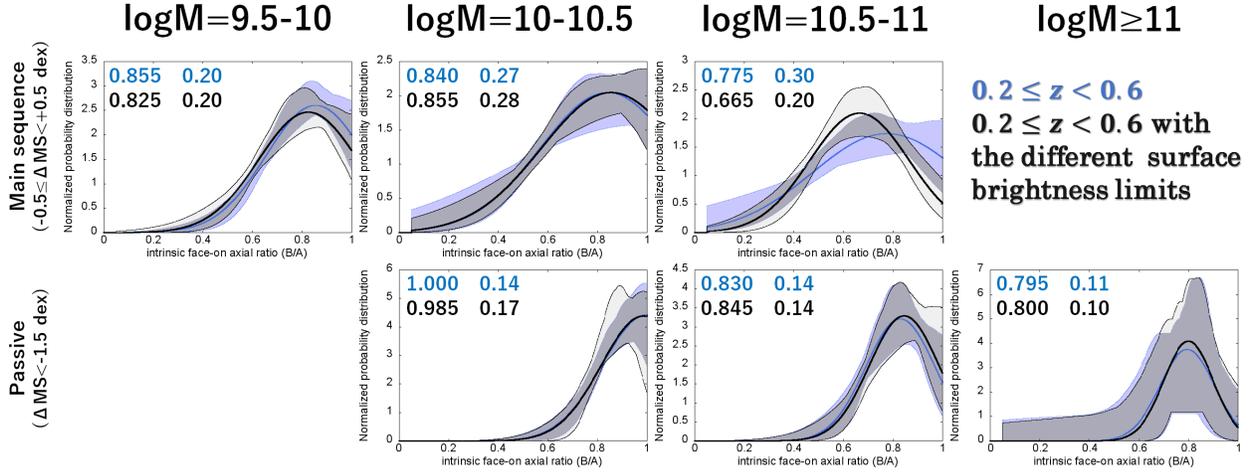}
\caption{ 
Same as Figure \ref{fig:intabevol}, but for galaxies at $0.2<z<0.6$ with 
the different surface brightness limits in the measurements of the apparent 
axial ratio.
While blue lines show the results in the original analysis, 
the black lines show those with the two times brighter surface brightness limit. 
\label{fig:intabdim}}
\end{figure*}

\begin{figure*}
\epsscale{1.1}
\plotone{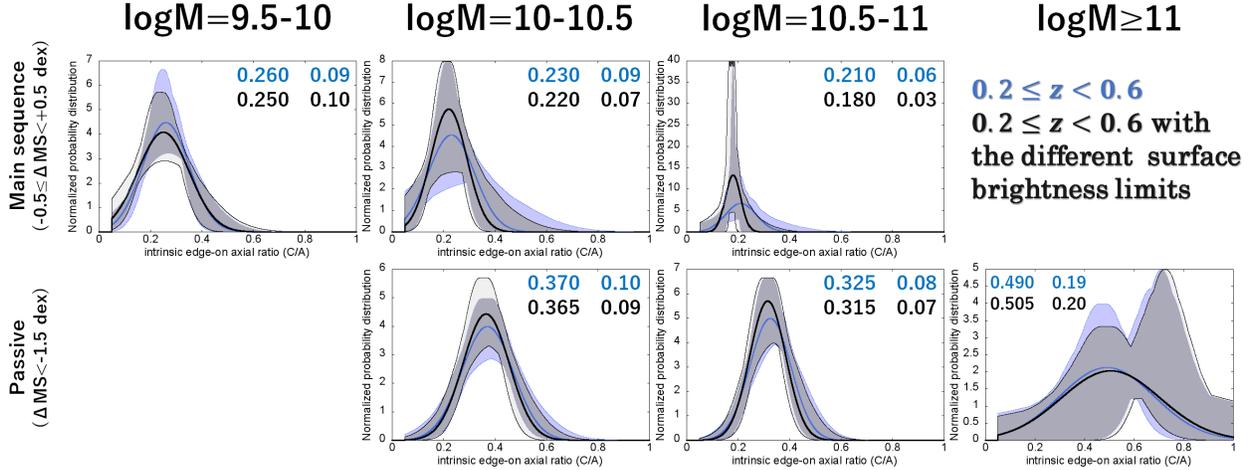}
\caption{ 
Same as Figure \ref{fig:intabdim}, but for the intrinsic edge-on axial ratio 
$C/A$.
\label{fig:intacdim}}
\end{figure*}

On the other hand, passively evolving 
galaxies show more complex dependence of the 
axial-ratio distribution on stellar mass.
Figure \ref{fig:psmass} shows the distribution of the apparent axial ratio 
for passively evolving galaxies with the different mass ranges in each redshift 
range.
In the both redshift ranges, those galaxies with $M_{\rm star} = 10^{10.5}$--$10^{11} M_{\odot}$
 show more flatter distribution and a higher fraction of objects with $b/a<0.5$ 
than the subsamples with $M_{\rm star} > 10^{11} M_{\odot}$ and 
$M_{\rm star} = 10^{10}$--$10^{10.5} M_{\odot}$.
The distribution of those with $M_{\rm star} = 10^{10}$--$10^{10.5} M_{\odot}$ 
is clearly flatter than massive galaxies with $M_{\rm star} > 10^{11} M_{\odot}$ 
at $0.2<z<0.6$, while the distribution of those low-mass galaxies is 
similar with that of massive galaxies at $0.6<z<1.0$.
In Figure \ref{fig:intabcevol}, passively evolving galaxies with 
$M_{\rm star} = 10^{10.5}$--$10^{11} M_{\odot}$ similarly show lower $\mu_{C/A}$ 
values than those with $M_{\rm star} > 10^{11} M_{\odot}$ and 
$M_{\rm star} = 10^{10}$--$10^{10.5} M_{\odot}$ in the both redshift ranges.
The $\mu_{C/A}$ value of those with $10^{10}$--$10^{10.5} M_{\odot}$ is lower 
than the most massive galaxies at $0.2<z<0.6$, while they show the similar 
$\mu_{C/A}$ values at $0.6<z<1.0$. 
As seen in the previous section, the thickness of those with 
$10^{10}$--$10^{10.5} M_{\odot}$ and $10^{10.5}$--$10^{11} M_{\odot}$ clearly decreases 
with time from $\mu_{C/A}=0.500$ and 0.400 at $0.6<z<1.0$ to 0.370 and 0.325 
at $0.2<z<0.6$, respectively. 
Although the evolution in the thickness is stronger for lower mass galaxies, 
the $\mu_{C/A}$ value of those with $10^{10}$--$10^{10.5} M_{\odot}$ is still higher 
than those with $10^{10.5}$--$10^{11} M_{\odot}$ even at $0.2<z<0.6$.
The most massive galaxies show no significant evolution in their intrinsic shape 
and they have relatively high $\mu_{C/A}$ values of 0.45 -- 0.49.

\subsection{possible biases} \label{subsec:bias}

We here examine possible biases that could affect the results described in 
the previous sections, namely, the cosmological surface brightness dimming, 
morphological K-correction, incompleteness by the absolute magnitude limit, 
environmental effect, and size dependence of the axial-ratio distribution.

\subsubsection{Cosmological surface brightness dimming}

\begin{deluxetable*}{rcccccc} 
\tablecaption{the best-fit parameters for galaxies at $0.2<z<0.6$ with the  
two times brighter surface brightness limit  
 \label{tab:fitdim}}
\tablehead{
\colhead {stellar mass} & \colhead{$\Delta$MS} & \colhead{$\mu_{B/A}$} & \colhead{$\sigma_{B/A}$} & \colhead{$\mu_{C/A}$} & \colhead{$\sigma_{C/A}$} & \colhead{$\chi^2_{\rm min}$\tablenotemark{a}}
} 
\startdata
\multicolumn6c{$z=0.2$--0.6} \\
\hline
$\log{M_{\rm star}}=9.5$--10 & $\Delta$MS $ = -0.5$ -- $+0.5$ & $0.825^{+0.135}_{-0.045}$ & $0.20^{+0.10}_{-0.05}$ & $0.250^{+0.025}_{-0.025}$ & $0.10^{+0.05}_{-0.03}$ & 6.55 \\
 $\log{M_{\rm star}}=10$--10.5 & $\Delta$MS $ = -0.5$ -- $+0.5$ & $0.855^{+0.145}_{-0.085}$ & $0.28^{+0.17}_{-0.07}$ & $0.220^{+0.050}_{-0.020}$ & $0.07^{+0.09}_{-0.02}$ & 5.54 \\
 & $\Delta$MS $ = -1.5$ -- $-0.5$ & $0.780^{+0.095}_{-0.040}$ & $0.14^{+0.09}_{-0.04}$ & $0.255^{+0.030}_{-0.030}$ & $0.04^{+0.03}_{-0.03}$ & 2.09 \\
 & $\Delta$MS $ < -1.5$ & $0.985^{+0.015}_{-0.115}$ & $0.17^{+0.04}_{-0.09}$ & $0.365^{+0.030}_{-0.030}$ & $0.09^{+0.03}_{-0.02}$ & 1.11 \\
 $\log{M_{\rm star}}=10.5$--11 & $\Delta$MS $ = -0.5$ -- $+0.5$ & $0.665^{+0.050}_{-0.030}$ & $0.20^{+0.07}_{-0.04}$ & $0.180^{+0.030}_{-0.020}$ & $0.03^{+0.06}_{-0.02}$ & 3.54 \\
 & $\Delta$MS $ = -1.5$ -- $-0.5$ & $0.840^{+0.160}_{-0.040}$ & $0.14^{+0.16}_{-0.05}$ & $0.310^{+0.050}_{-0.035}$ & $0.08^{+0.04}_{-0.02}$ & 2.75 \\
 & $\Delta$MS $ < -1.5$ & $0.845^{+0.155}_{-0.040}$ & $0.14^{+0.12}_{-0.04}$ & $0.315^{+0.035}_{-0.020}$ & $0.07^{+0.03}_{-0.01}$ & 1.61 \\
 $\log{M_{\rm star}} > 11$ & $\log{(sSFR)}>-12.0$ & $0.755^{+0.245}_{-0.285}$ & $0.26^{+0.74}_{-0.12}$ & $0.330^{+0.185}_{-0.090}$ & $0.11^{+0.39}_{-0.06}$ & 2.39 \\
 & $\log{(sSFR)}<-12.0$ & $0.800^{+0.200}_{-0.110}$ & $0.10^{+0.78}_{-0.04}$ & $0.505^{+0.245}_{-0.060}$ & $0.20^{+0.30}_{-0.12}$ & 0.701 \\
\enddata
\tablenotetext{a}{the minimum $\chi^2$ value in the fitting (9 degrees of freedom).}
\end{deluxetable*}

As described in Section \ref{sec:measure}, we used the 
surface brightness threshold of 1.3 times the local background root mean square 
in the measurements of the apparent axial ratio on the $I_{\rm F814W}$-band data.
Since the surface brightness of objects decreases with increasing redshift by a 
factor of $(1+z)^{4}$ ($(1+z)^{3}$ in AB mag/arcsec$^2$) due to the cosmological 
expansion, the constant isophotal threshold in the measurements corresponds 
to the brighter intrinsic surface brightness limit for galaxies at higher redshifts.
Thus the apparent axial ratio of a galaxy at higher redshift tends to be measured 
in a brighter part of the object. 
This bias could affect our results about the evolution of the axial-ratio 
distribution in the previous sections.

In order to check the effects of the cosmological surface brightness dimming, 
we re-analyzed the sample galaxies at $0.2<z<0.6$ with a surface brightness threshold
two times brighter than that in the original analysis. 
We chose the factor of two considering the average dimming factor ratio of  
$(1+0.8)^3/(1+0.4)^3$ between our subsamples at $0.2<z<0.6$ and $0.6<z<1.0$.
Using the measurements with the brighter surface brightness threshold, 
we carried out the same analyses to calculate the distribution of the apparent 
axial ratio as a function of stellar mass and $\Delta$MS 
and estimate the 3-dimensional shape with the Monte Carlo simulations. 
In Figures \ref{fig:intabdim} and \ref{fig:intacdim}, we show 
the results of the intrinsic face-on axial ratio $B/A$ and edge-on axial ratio 
$C/A$ and compare them with those in the original analysis.
The fitting results are also summarized in Table \ref{tab:fitdim}.
The all results with the brighter threshold are consistent with the original ones 
within the errors. 
Although the edge-on axial ratio $C/A$ tends to be lower values by $\sim 0.01$ 
in the results with the brighter threshold, the differences are small and within 
the uncertainty. We conclude that the cosmological surface brightness dimming 
does not significantly affect our results. 
We also note that the effects of the brighter threshold on the results 
do not depend on stellar mass.
Thus, it is not the case that low-mass (faint) galaxies are preferentially affected 
by the surface brightness limit.
For example, the mass dependence of the $C/A$ for the main-sequence galaxies, 
i.e., thicker shapes for low-mass star-forming ones, is not changed by 
the brighter surface brightness threshold at all.
 
\begin{figure*}[t]
\epsscale{1.1}
\plotone{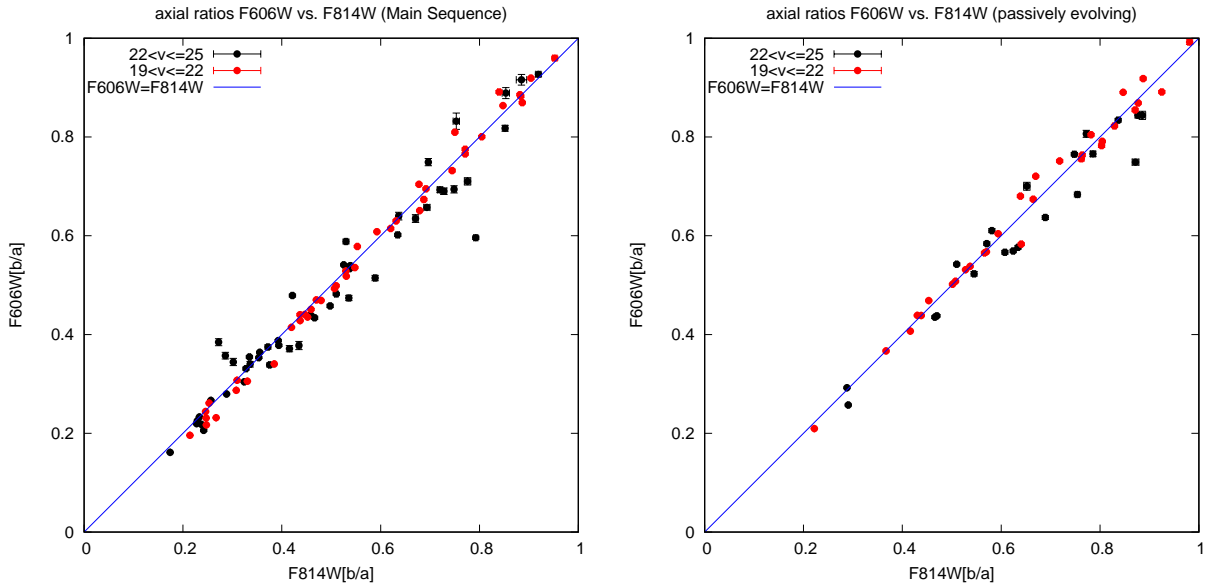}
\caption{ 
Comparison of the apparent axial ratios measured in the $V_{\rm F606W}$ and 
$I_{\rm F814W}$ bands for the main-sequence galaxies (left) and passively evolving
 galaxies (right) at $0.2<z<0.6$.
Red symbols show relatively bright galaxies with $V_{\rm F606W} = 19$ -- 22, 
while blue symbols show fainter ones with $V_{\rm F606W} = 22$ -- 25.
\label{fig:kcorr}}
\end{figure*}

\begin{deluxetable*}{lrccc} 
\tablecaption{Comparison of the axial ratios measured in the $V_{\rm F606W}$ and 
$I_{\rm F814W}$ bands \label{tab:arkcorr}}
\tablehead{
\colhead {type} & \colhead {$V_{\rm F606W}$ mag} & \colhead{$N_{\rm obj}$} & \colhead{$(b/a)_{\rm F6060W}-(b/a)_{\rm F814W}$\tablenotemark{a}} & \colhead{(error)$_{\Delta b/a}$\tablenotemark{b}} 
}
\startdata
Main sequence & $V_{\rm F606W} < 25$ & 92 & -0.006 $\pm$ 0.037 & 0.004 \\
 & $V_{\rm F606W}<22$ & 43 & -0.003 $\pm$ 0.020 & 0.002 \\
Passively evolving & $V_{\rm F606W} < 25$ & 51 & -0.007 $\pm$ 0.033 & 0.004 \\
 & $V_{\rm F606W}<22$ & 30 & -0.003 $\pm$ 0.022 & 0.003 \\
\enddata
\tablenotetext{a}{the mean value and standard deviation in the differences of 
$b/a$ measured in the $V_{\rm F606W}$ and $I_{\rm F814W}$ bands.}
\tablenotetext{b}{the mean measurement errors for the differences of 
$b/a$ measured in the $V_{\rm F606W}$ and $I_{\rm F814W}$ bands.}
\end{deluxetable*}

\subsubsection{Morphological K-correction}

We measured the apparent axial ratio on the $I_{\rm F814W}$-band data, 
which correspond to the rest-frame $V$ band for galaxies at $z\sim 0.4$ 
and the rest-frame $B$ band for those at $z\sim 0.8$.
Such differences in the rest-frame wavelength could cause some biases 
in the morphological analysis due to the color differences between bulge and disk, 
the blue star-forming regions/clumps, the dust extinction effect, and so on 
(e.g., \citealp{win02}; 
\citealp{hue09}; \citealp{wuy12}; \citealp{vik13}; \citealp{mur14};
\citealp{mag18}). 
In order to check the effects of the morphological K-correction, 
we used publicly available HST/ACS $V_{F606W}$-band data over a 0.05 deg$^2$ 
region in the COSMOS field from the CANDELS survey 
(\citealp{gro11}; \citealp{koe11}).
With the $V_{\rm F606W}$-band data, we can measure the apparent axial ratio of 
galaxies at $0.2<z<0.6$ in the rest-frame $B$ band, and investigate 
to what extent the difference in the rest-frame wavelength affects the measurements.  
There are 92 main-sequence and 51 passively evolving galaxies 
with $V_{\rm F606W}<25$ at $0.2<z<0.6$ in the region, and 
we measured the apparent axial ratio of these galaxies on the $V_{\rm F606W}$-band 
data with the same way. 
In Figure \ref{fig:kcorr}, we compare the apparent axial ratios $b/a$ measured 
on the $V_{\rm F606W}$-band data with those measured on the $I_{\rm F814W}$-band data.
The differences between the $V_{\rm F606W}$ and $I_{\rm F814W}$ bands are also 
summarized in Table \ref{tab:arkcorr}.
The apparent axial ratios measured on the $V_{\rm F606W}$ and $I_{\rm F814W}$-bands data 
agree well with each other for the both main-sequence and passively evolving 
populations.  
The average values of $(b/a)_{F606W} - (b/a)_{F814W}$ are -0.006 and -0.007 for 
main-sequence and passively evolving galaxies, respectively. 
These systematic offsets from zero are slightly larger than the averages of 
the measurement errors, but much smaller than the dispersion around the mean value.
When we use only bright subsamples with $V_{\rm F606W}<22$, the results do not 
significantly change, although the average offsets and measurement errors 
become slightly smaller.
Since these systematic offsets are much smaller than the bin width of 0.1 
in the distribution of $b/a$ we used, the morphological K-correction does not 
significantly affect the distribution of the apparent axial ratio.

 \subsubsection{Incompleteness} \label{subsec:incom}

\begin{figure}
\epsscale{1.15}
\plotone{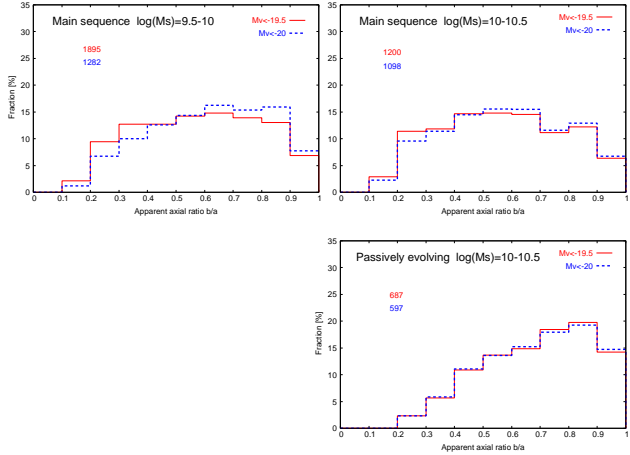}
\caption{
The distribution of the apparent axial ratio for galaxies with $M_{V} < -19.5$ 
at $0.2<z<0.6$. 
The top panels show those for the main-sequence galaxies with 
$M_{\rm star} = 10^{9.5}$--$10^{10} M_{\odot}$ (left) and 
$10^{10}$--$10^{10.5} M_{\odot}$ 
(right), while the bottom panel shows that for the passively evolving galaxies
with $10^{10}$--$10^{10.5} M_{\odot}$.
The distribution for galaxies with $M_{V}<-20$ and the same stellar mass, 
$\Delta$MS, and redshift ranges is also shown in each panel for reference.
The total numbers of objects in the subsamples are also shown in each panel.
\label{fig:ar_mvlim}}
\end{figure}

\begin{figure}
\epsscale{1.1}
\plotone{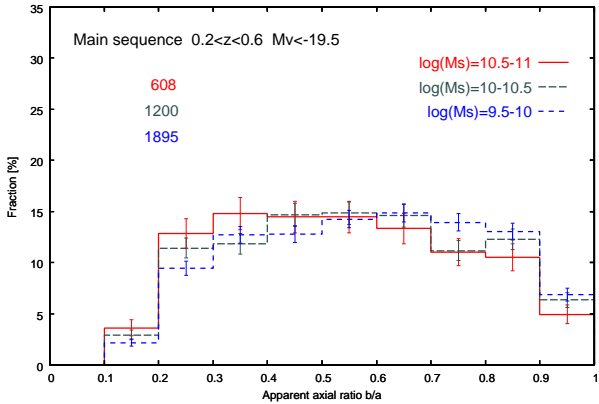}
\caption{
The mass dependence of the distribution of the apparent axial ratio 
for the main-sequence galaxies with $M_{V}<-19.5$ at $0.2<z<0.6$. 
The solid, long-dashed, and short-dashed lines represent those galaxies 
with $M_{\rm star} = 10^{10.5}$--$10^{11} M_{\odot}$, $10^{10}$--$10^{10.5} M_{\odot}$, 
and $10^{9.5}$--$10^{10} M_{\odot}$, respectively.
The total numbers of objects in the subsamples are also shown.
\label{fig:msmass_mvlim}}
\end{figure}

\begin{figure}
\epsscale{1.1}
\plotone{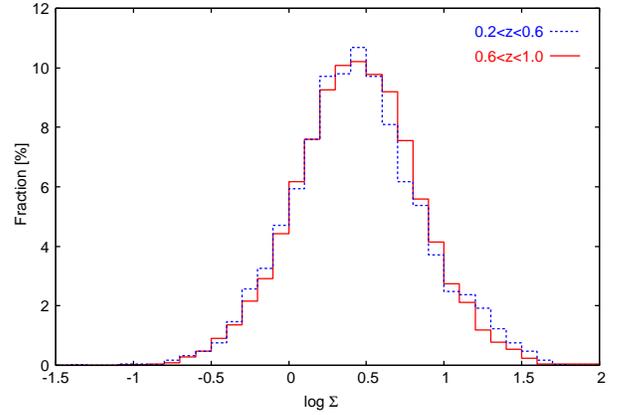}
\caption{ 
The distributions of the local surface density of our sample galaxies at 
$0.2<z<0.6$ (blue) and $0.6<z<1.0$ (red) estimated by \cite{dar17}. 
The local density $\Sigma$ is a unit of Mpc$^{-2}$ and calculated with 
the adaptive kernel smoothing and a redshift width of $\pm 1.5 \sigma_{\Delta z/(1+z)}$ 
(see text). 
\label{fig:envall}}
\end{figure}

\begin{figure*}
\epsscale{1.0}
\plotone{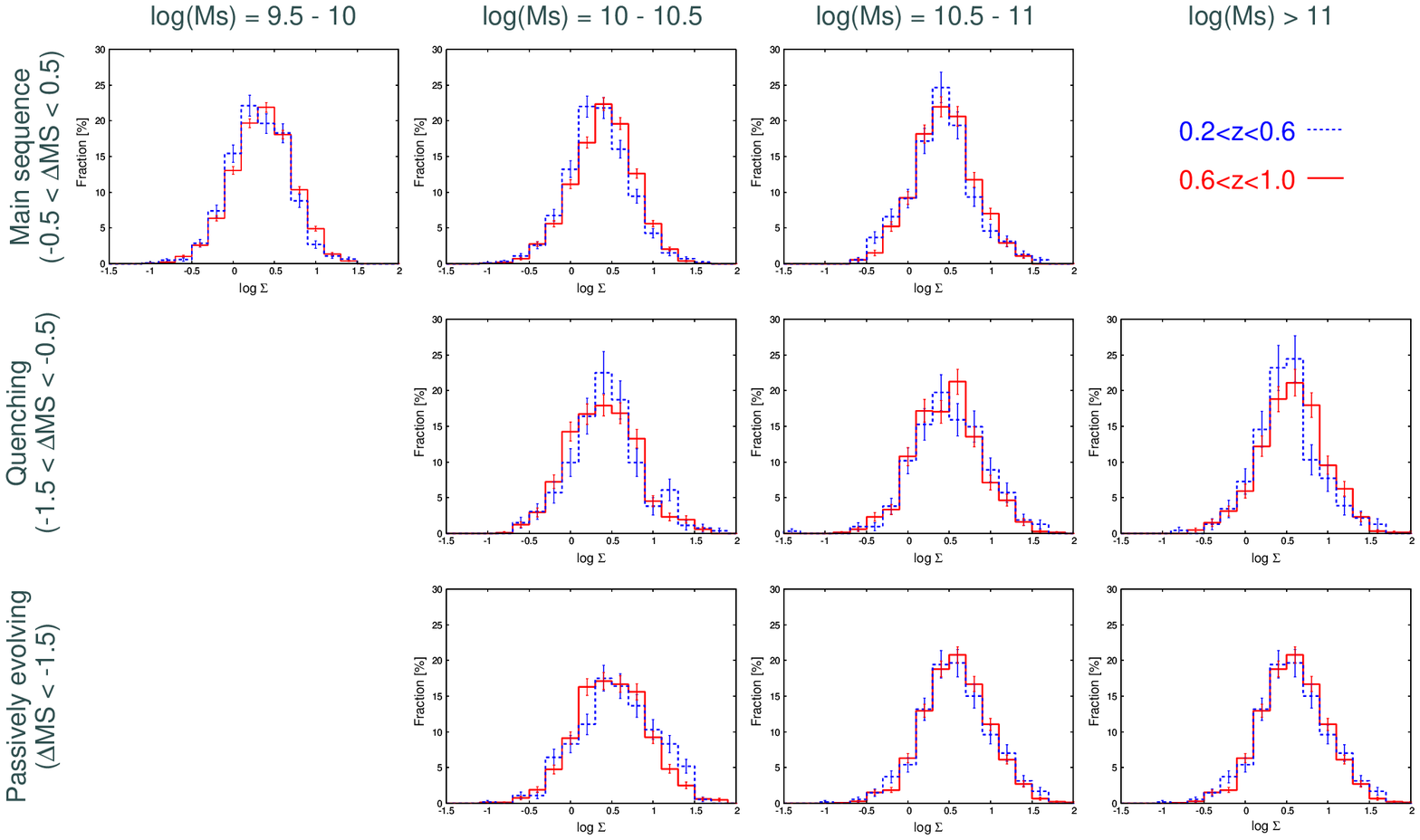}
\caption{ 
Same as Figure \ref{fig:envall}, but for subsamples with different stellar mass  
and $\Delta$MS ranges separately. 
 The configuration of the panels are the same as Figure \ref{fig:dms2evol}.
The error bars are based on the square root of the number of objects in 
the bin.
\label{fig:envdms}}
\end{figure*}

In Section \ref{sec:sample}, we noted that 
the absolute magnitude limit of $M_{V}<-20$ causes 
the incompleteness in the low-mass end of our sample 
especially for those at $0.2<z<0.6$.
We missed 
$\sim 35$ \% of the main-sequence galaxies with $10^{9.5}$--$10^{10} M_{\odot}$ 
at $0.2<z<0.6$ by the limit of $M_{V}<-20$, 
 while $\sim 20$ \% of the other galaxies with 
$10^{10}$--$10^{10.5} M_{\odot}$ at $0.2<z<0.6$ were missed by the same limit 
(Figure \ref{fig:msmv}).
In order to check how the incompleteness affects our results, 
we measured the apparent axial ratio of those faint galaxies 
with $-20 < M_{V} < -19.5$ at $0.2<z<0.6$ with the same manner. 
In Figure \ref{fig:ar_mvlim}, we show the distributions of the apparent axial ratio 
of galaxies with $M_{V}<-19.5$ for the main-sequence galaxies with 
$M_{\rm star} = 10^{9.5}$--$10^{10} M_{\odot}$ and $10^{10}$--$10^{10.5} M_{\odot}$, 
and for the passively evolving galaxies with $10^{10}$--$10^{10.5} M_{\odot}$, 
and compare them with the results for those with $M_{V}<-20$.
The distributions for the main-sequence galaxies with $M_{V}<-19.5$ 
are slightly skewed toward lower value of $b/a$ compared with those with $M_{V}<-20$
 in the both mass ranges. 
Those faint galaxies with $-20 < M_{V} < -19.5$ have systematically lower 
$b/a$, probably because edge-on star-forming galaxies are more affected by the 
dust extinction and are systematically fainter at a given stellar mass 
(e.g, \citealp{sha07}; \citealp{pad08}).
On the other hand, passively evolving galaxies with $10^{10}$--$10^{10.5} M_{\odot}$ 
show no significant difference in the distribution 
between those with $M_{V}<-19.5$ and $M_{V}<-20$.
The distribution of those faint galaxies with $-20 < M_{V} < -19.5$ is 
similar with that of brighter galaxies in the passively evolving population.

Figure \ref{fig:msmass_mvlim} shows the mass dependence of the axial-ratio 
distribution for the main-sequence galaxies with $M_{V}<-19.5$. 
While the distributions for those galaxies with $10^{9.5}$--$10^{10.5} M_{\odot}$ 
are skewed toward lower values as seen in Figure \ref{fig:ar_mvlim}, 
one can still see that more massive galaxies tend to have lower values of 
$b/a$. 
Note that the distribution for those galaxies with $10^{10.5}$--$10^{11} M_{\odot}$ 
is not affected by the choice of the magnitude limit, 
because there is only one faint galaxy with $M_{V} > -20$ in the subsample 
with $10^{10.5}$--$10^{11} M_{\odot}$.
Therefore we conclude that 
the distribution for the main-sequence 
galaxies really depends on stellar mass, 
although we probably overestimate the strength of the dependence 
to some extent due to the incompleteness effect.

\begin{figure*}[h!]
\epsscale{1.0}
\plotone{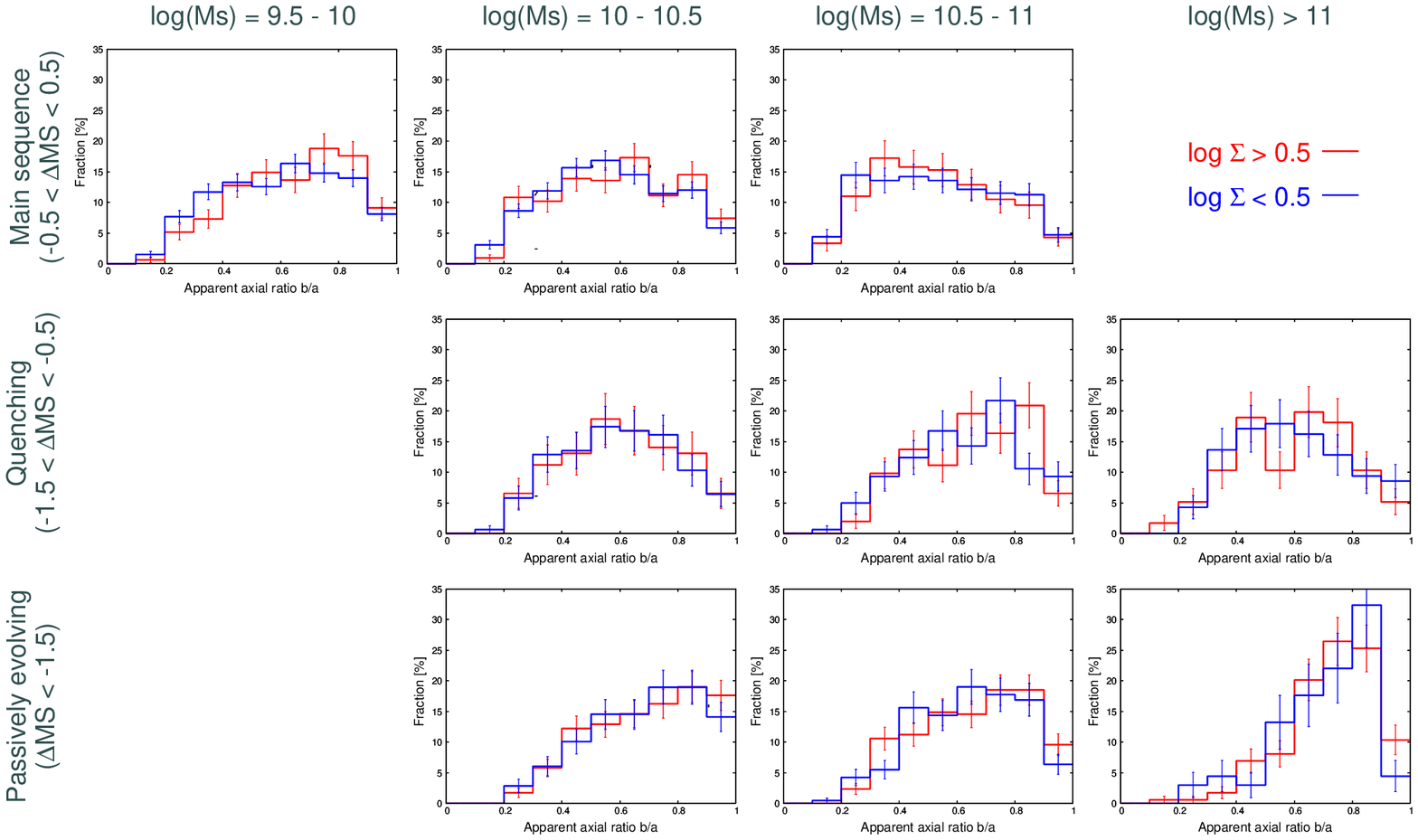}
\caption{
The distributions of the apparent axial ratio for galaxies at $0.2<z<0.6$ 
located in the different local densities as a function of stellar mass and 
$\Delta$MS.
The configuration of the panels are the same as Figure \ref{fig:dms2evol}.
The red lines show those galaxies in a high-density region with 
$\Sigma > 10^{0.5}$ Mpc$^{-2}$, while the blue lines represent those in 
a lower-density region with $\Sigma < 10^{0.5}$ Mpc$^{-2}$. 
\label{fig:arenvlowz}}
\end{figure*}
\begin{figure*}[t!]
\epsscale{1.0}
\plotone{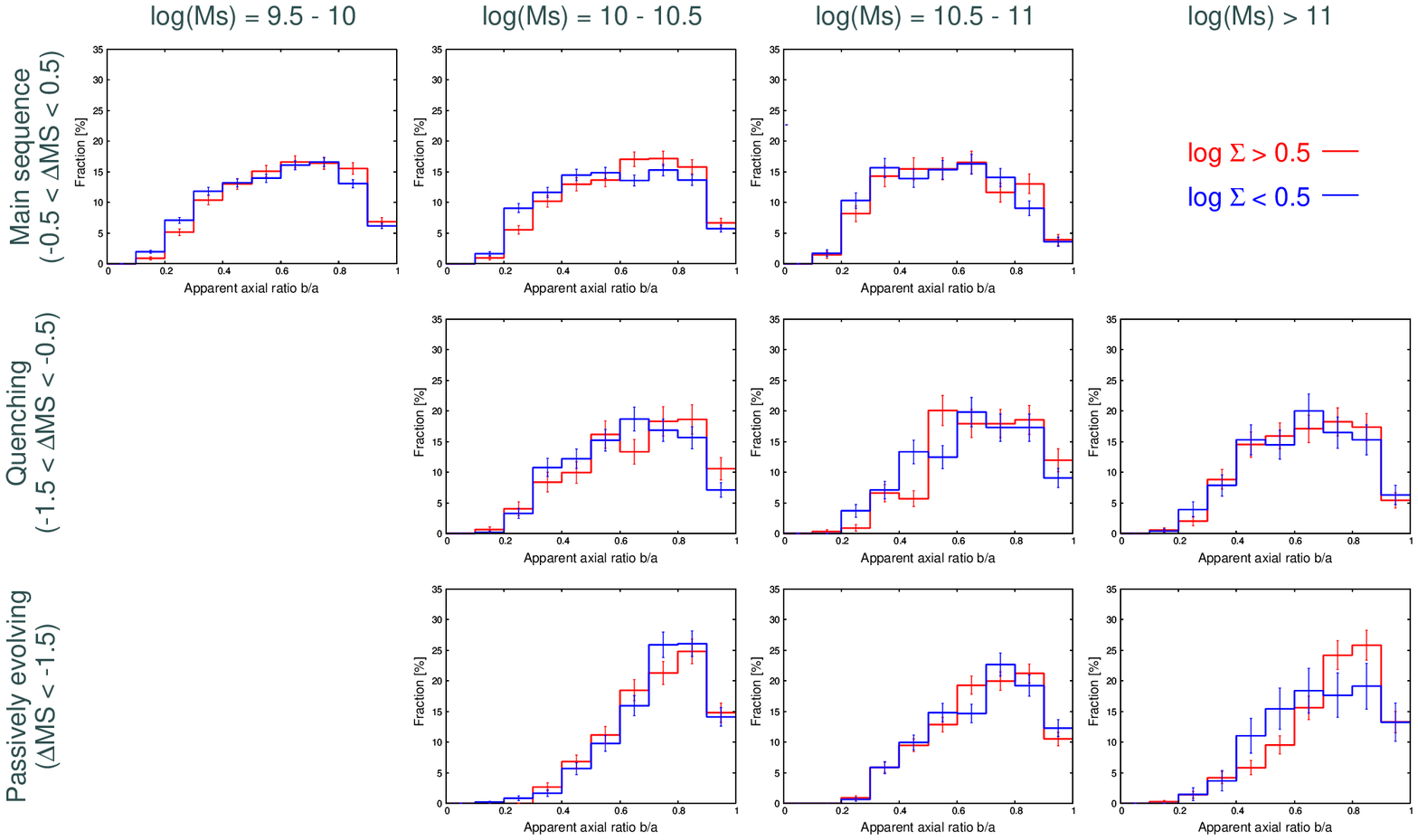}
\caption{ 
Same as Figure \ref{fig:arenvlowz}, but for galaxies at $0.6<z<1.0$.
\label{fig:arenvhighz}}
\end{figure*}

\subsubsection{Environments}

In order to ensure statistical accuracy, we divided our sample by redshift 
into those at $0.2<z<0.6$ and $0.6<z<1.0$, and have only two redshift bins.
Therefore differences of the environments between these two redshift bins 
could affect our results, although the co-moving survey volumes 
of these two bins 
are relatively large ($1.7 \times 10^{6}$ Mpc$^{3}$ for the $0.2<z<0.6$ bin 
and $4.3 \times 10^{6}$ Mpc$^{3}$ for the $0.6<z<1.0$ bin).
If the environments of these two redshift bins are significantly different 
from each other,  
we may mainly see the environmental dependence rather than 
the redshift evolution 
from the comparisons between these two bins.
In order to check this, we investigated the environments of sample galaxies 
in the two redshift bins by using the local surface number density of 
galaxies estimated with adaptive kernel smoothing by \cite{dar17}.
The local number density is calculated with a 2-dimensional Gaussian kernel 
with a width changing according to the density and the global width is selected 
to be 0.5 Mpc.
In the calculation,  
they used a redshift width of $\pm 1.5\sigma_{\Delta z/(1+z)}$, which roughly 
corresponds to be a comoving length of 550--600 Mpc over $0.2<z<1.0$ 
(Figure 2 of \citealp{dar17}),    
and the mean densities over the COSMOS field at $0.2<z<0.6$ and $0.6<z<1.0$ 
are similar (their Figure 3).
Since \cite{dar17} estimated the local number density for galaxies 
with $M_{\rm star} > 10^{9.6} M_{\odot}$  selected in the 
UltraVISTA field \citep{lai16}, 
a part of galaxies in our sample are not included in their catalog and unavailable 
in this analysis.
We matched total 19086 objects in our sample 
(4731 galaxies at $0.2<z<0.6$ and 14355 ones at $0.6<z<1.0$) 
with those in the catalog by \cite{dar17}.

Figure \ref{fig:envall} shows the distributions of the local surface number 
density of galaxies at $0.2<z<0.6$ and $0.6<z<1.0$. 
While the fraction of galaxies in a high-density region 
with $\Sigma > 10$ Mpc$^{-2}$
 at $0.6<z<1.0$ is systematically higher than those at $0.2<z<0.6$, 
the distributions of the local density are basically similar with each other.
In Figure \ref{fig:envdms}, we compared the distributions of the local density  
between the two redshift bins for the subsamples with different stellar mass 
and $\Delta$MS ranges separately. 
One can see that more massive galaxies with a lower $\Delta$MS tend to be located 
in higher-density regions.
While star-forming galaxies at $0.6<z<1.0$ seem to be located 
in slightly higher-density regions than those at $0.2<z<0.6$, 
there is no large difference 
in the distribution of the local density between the two redshift bins 
for all the subsamples. 
In order to check the environmental dependence of the axial-ratio distribution, 
we also divided the sample galaxies into those in high-density regions with 
$\Sigma > 10^{0.5}$ Mpc$^{-2}$ and those in lower-density regions, and 
compared the distributions of the apparent axial ratio between 
these two subsamples 
 at $0.2<z<0.6$ and $0.6<z<1.0$ in Figures \ref{fig:arenvlowz} and 
\ref{fig:arenvhighz}, respectively.
Although low-mass star-forming galaxies on the main sequence 
in the high-density regions tend to show 
slightly higher apparent axial ratios, which indicates thicker 
intrinsic shapes, than those
 in lower-density regions, the environmental dependence is not so strong in 
all the subsamples.
By comparing Figures \ref{fig:arenvlowz} and \ref{fig:arenvhighz}, 
we confirmed that the passively evolving galaxies with 
$M_{\rm star} < 10^{11} M_{\odot}$
 show the significant evolution in their axial-ratio distribution irrespective 
of environment. 
We conclude that the differences in the environment between the two 
redshift bins 
do not significantly affect our results about the evolution 
in the axial-ratio distribution.

\subsubsection{Size dependence}\label{subsec:size}

\begin{figure*}
\epsscale{1.1}
\plotone{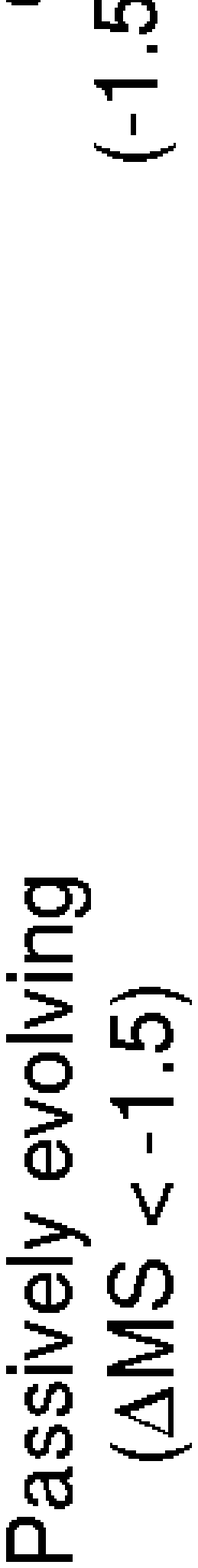}
\caption{ 
The axial-ratio distribution as a function of semi-major radius for 
sample galaxies at $0.2<z<0.6$. 
The blue line shows the median axial ratios in a given size range 
with a width of 0.1 dex. The histogram represents the size distribution 
of galaxies in each $M_{\rm star}$ and $\Delta$MS bin.
The short dashed curves in the upper left panel show the resolution limit 
of $b<1.3$ pixel for $z=0.2$ and $z=0.6$. 
\label{fig:arsizelowz}}
\end{figure*}

\begin{figure*}
\epsscale{1.1}
\plotone{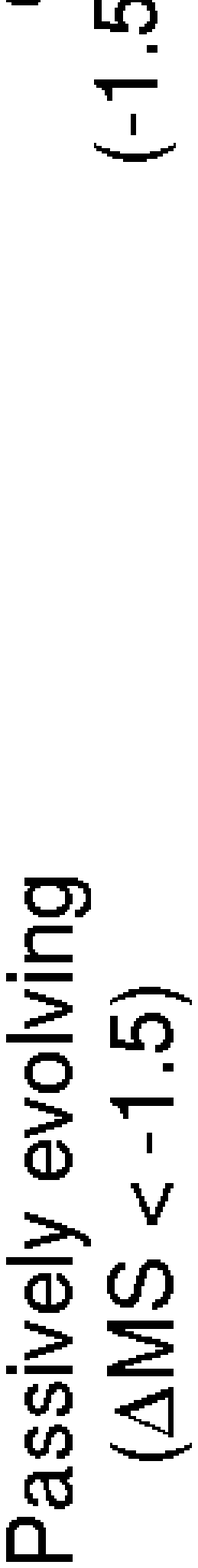}
\caption{ 
The same as Figure \ref{fig:arsizelowz}, but for galaxies at $0.6<z<1.0$.
\label{fig:arsizehighz}}
\end{figure*}

\cite{pad08} and \cite{zha19} reported that the axial-ratio distribution 
of galaxies depends on their sizes. The size dependence could bias 
our estimate of the 3-dimensional shape from the axial-ratio distribution.
Therefore we examine the axial-ratio distribution as a function of size 
in Figures \ref{fig:arsizelowz} and \ref{fig:arsizehighz}.
We used the semi-major radius $a$ calculated from the second order moment 
 described in Section \ref{sec:measure} as a size indicator.

The distributions of main-sequence galaxies in the Figures \ref{fig:arsizelowz} 
and \ref{fig:arsizehighz} show similar features with those seen 
in \cite{zha19} for star-forming galaxies at $0.5<z<1.0$, namely, 
a curved boundary at the lower left corner, a tail of galaxies at the 
lower right corner, and a deficiency of galaxies at the upper right 
corner. 
As the combination of these features, the median axial ratio of 
main-sequence galaxies decreases with increasing size.
In the upper left panels of the both figures, we also plot the resolution 
limit of $b < 1.3$ pixel described in Section \ref{sec:measure}.
The resolution limit does not affect the distribution of galaxies 
on the $b/a$ vs. $\log{(a)}$ plain.

The axial-ratio distribution for a given size tends to be flat over 
a wide range of size, and the range of the axial ratio is limited to 
be higher values at small radii. 
Although the curved boundary at the lower left corner could be caused by 
the prolate shape of galaxies \citep{zha19}, the prolate shape leads to 
a peak around a relatively low value of $b/a$, which is not seen in the 
observed distributions.  
Therefore, the observed curved boundary at the lower left corner indicates that 
the 3-dimensional shape of these galaxies is basically disk-like morphology 
and their thickness increases with decreasing size.
The tail at the lower right is probably caused by the dust extinction 
effect in edge-on galaxies as discussed in \cite{zha19}.
When a disk galaxy is viewed in a nearly edge-on view, the central part 
of galaxies tends to be heavily obscured due to a larger path-length, 
which leads to the overestimate of the second-order moment 
along the semi-major axis.
On the other hand, the apparent axial ratio $b/a$ is not significantly 
affected by the dust extinction, because the semi-minor radius along 
the height direction is similarly overestimated by the dust lane 
in the edge-on disk. 
As a result, edge-on disk galaxies whose sizes are overestimated make 
the tail at the lower right. 
The fact that the tail is negligible 
or less significant in passively evolving galaxies with small dust extinction 
supports this explanation.
The deficiency of galaxies at the upper right suggests that 
the face-on axial ratio $B/A$ is smaller than unity. 
The deficiency is more significant in massive main-sequence galaxies, 
which is consistent with the best-fit $\mu_{B/A}$ values of these galaxies 
seen in Figure \ref{fig:intabcevol} and Table \ref{tab:fitevol}.

The distribution of passively evolving galaxies similarly shows  
a curved boundary at the lower left and a deficiency at the upper right,  
but the distribution for a given size has a peak around $b/a \sim 0.8$, 
which suggests more thick spheroidal (oblate) shapes.
While the curved boundary at the lower left corner and the deficiency 
at the upper right are naturally expected for 
such thick spheroidal shapes \citep{zha19}, the concentration of small 
galaxies at high $b/a$ could reflect such compact galaxies have more 
spherical shapes than those with larger sizes.   

The features of the distribution on the $b/a$ vs. $\log{(a)}$ plain 
mentioned above 
are similar among the subsamples with different stellar masses for both 
main-sequence and passively evolving galaxies, while the size increases 
with increasing mass for the both populations.
These features and the size distributions suggest that 
the size dependence of the intrinsic shape does not significantly affect 
the overall axial-ratio distribution in each bin.
Therefore, we expect that our estimates for the intrinsic shape of 
galaxies with $M_{V}<-20$ at $z<1$ are not heavily biased 
by the size dependence, although those galaxies with smaller sizes 
tend to have thicker intrinsic shapes and (relatively rare) 
compact galaxies significantly below the mass-size relation 
could have systematically spherical shapes.

\section{Discussion} \label{sec:discus}

\subsection{3-dimensional shape transition at $\Delta$MS $\sim -1$ dex}

We investigated the distribution of the apparent axial ratio of galaxies 
with $M_{V}<-20$ at $0.2<z<1.0$ as a function of stellar mass and sSFR, 
and found that the distribution and estimated intrinsic shape change from 
the thin disk shape with $C/A \sim$ 0.2 -- 0.25 to the thicker spheroidal shape 
with $C/A \sim $ 0.3 -- 0.5 around $\Delta$MS $\sim -1$ dex irrespective of 
redshift.
We discuss possible origins of this transition around $\Delta$MS $ \sim -1$ dex
in the following. 

The major merger is a possible mechanism to cause both the morphological 
transition and the quenching of star formation.
Many previous studies with numerical simulations found that disks of 
merging galaxies 
could be destroyed and changed to the spheroidal remnants in the major mergers 
(e.g., \citealp{bar88}; \citealp{naa03}; \citealp{jes09}), 
while disks may survive in gas-rich major mergers 
(e.g., \citealp{spr05}; \citealp{gov09}).
It is also considered that intense starbursts are associated with major mergers 
through the gas inflow to the galaxy center and/or the gas 
compression/collapse in 
the dense tides or clouds, and then the merging remnants could cease 
star formation 
due to the gas exhaustion or blowout by the supernova feedback 
(e.g., \citealp{san96}; \citealp{tey10}; \citealp{ren14}; see also 
\citealp{spa16}).
In such major mergers, however, the timescales for the starbursts and 
the morphological transition seem to vary on a case-by-case basis.
For example, \cite{lot08} reported that the enhanced star formation 
continues after 
the morphological transition in their simulations. 
Thus the morphological transition to spheroidal shapes through major mergers 
does not necessarily occur when sSFR of the remnants corresponds to 
$\Delta$MS $\sim -1$ dex.
Furthermore, if the major merger is the main driver for the shape transition, 
it may be difficult to explain the observed evolution in the shape of 
passively evolving galaxies with $M_{\rm star}<10^{11} M_{\odot}$ 
as discussed in the next section.

The morphological quenching also can cause the decrease of sSFR through 
the stabilization of the gas disk against the fragmentation to star-forming 
clumps by the gravity of the dominant bulge \citep{mar09}. 
The bulge growth in star-forming galaxies can be stimulated by minor mergers 
and/or tidal interactions (e.g., \citealp{qu11}; \citealp{gue13}; 
\citealp{fia15}).
In this scenario, the morphological transition should precede the quenching of 
the star formation, and therefore there are bulge-dominated galaxies 
with a sSFR of the main sequence, i.e., $\Delta$MS $\sim 0$. 
\cite{mor18} investigated spatially resolved SFR and stellar mass 
for galaxies at $0.2<z<1.2$ with the multi-band HST data in the GOODS fields. 
They found that more massive star-forming galaxies on the main sequence 
are more bulge-dominated, and that those at the lower envelope of the main 
sequence show more quenched and dominant bulges, 
which suggests that the bulge growth proceeded on the main sequence.
Since the shape of galaxies changes into spheroidal-like 
before the star formation 
is quenched in this case, 
 an additional reason is needed to explain the shape transition around 
$\Delta$MS $ \sim -1$ dex. 
One possibility is that such bulge-dominated galaxies with $\Delta$MS $ \sim 0$ 
is minor in number in galaxies on the main sequence, because there are numerous 
normal disk-dominated star-forming galaxies. 
Since these normal disk galaxies are concentrated on the main sequence in sSFR, 
the bulge-dominated ones become significant in number as their sSFR decreases.
If the sSFR at which such quenching galaxies start to dominate corresponds to 
$\Delta$MS $ \sim -1$ dex irrespective of redshift, the shape transition around 
$\Delta$MS $ \sim -1$ dex can be explained.

\begin{figure}
\epsscale{1.1}
\plotone{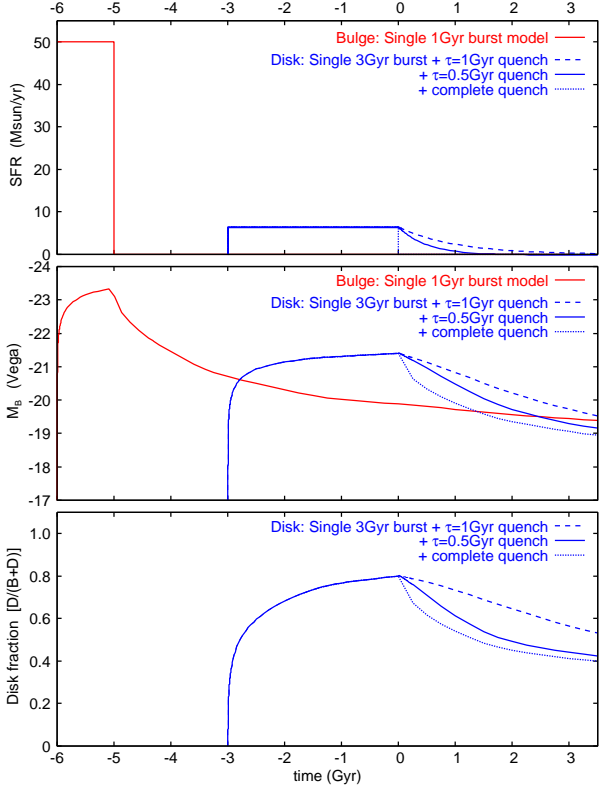}
\caption{ 
SFR (top), the rest-frame $B$-band absolute magnitude $M_{B}$ (middle), 
and the disk to total luminosity ratio (bottom) for the disk and 
bulge components 
in the toy model as a function of time ($t=0$ at the quenching of the disk).
The solid, dashed, and dotted lines show the different quenching timescales 
for the disk component (see text).
The model galaxy has a stellar mass of $3.8 \times 10^{10} M_{\odot}$ 
($M_{\rm bulge} = 2.6 \times 10^{10} M_{\odot}$ and 
$M_{\rm disk} = 1.2 \times 10^{10} M_{\odot}$)   
and a sSFR of $10^{-9.75}$ yr$^{-1}$ at the quenching of the disk.
\label{fig:sbmodel}}
\end{figure}

\begin{figure}
\epsscale{1.1}
\plotone{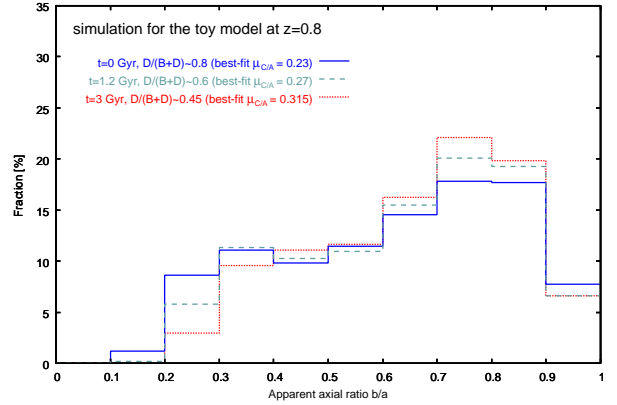}
\caption{ 
Distribution of the apparent axial ratio for the artificial objects 
with the different disk fractions.
The disk fractions correspond to the ages of the toy model 
with $\tau = 0.5$ Gyr in the legend.
The details of the artificial objects added to the $I_{\rm F814W}$-band images
are described in Appendix \ref{sec:monsim}.
The best-fit intrinsic thickness, $\mu_{C/A}$ for each disk fraction 
is also shown in the legend.
\label{fig:histsim}}
\end{figure}

On the other hand, the quenching of star formation in disks itself through, 
for example, the morphological quenching mentioned above or 
environmental effects such as the starvation or the ram-pressure 
stripping (e.g., \citealp{bal00}; \citealp{aba99}) 
could lead to the changes in the bulge/disk flux ratio. 
The luminosity of disks is expected to more rapidly decline just after the 
star formation stopped than that of the passively evolving bulge components.
In order to check this effect, we constructed a toy model that consists of 
a passively evolving bulge and a constantly star-forming and then 
quenching disk.
We used the GALAXEV population synthesis library to estimate the luminosity 
evolution of the bulge and disk with assumed star formation histories.
We assumed that the disk component continued a constant SFR for 3 Gyr and 
then quenched with timescales with $\tau=$ 1.0 Gyr, 0.5 Gyr, and 0 
(abruptly stopped), 
while the single 1 Gyr burst model with 6 Gyr age when the disk 
started to quench  
was adopted for the bulge component (the top panel of Figure \ref{fig:sbmodel}).
The model galaxy has $M_{\rm star} = 3.8\times10^{10} M_{\odot}$ 
($M_{\rm bulge} = 2.6\times 10^{10} M_{\odot}$ and 
$M_{\rm disk} = 1.2 \times 10^{10} M_{\odot}$),  
$sSFR = 10^{-9.75}$ yr$^{-1}$, and the disk to total luminosity 
ratio D/(B+D) $ = 0.8$ in the rest-frame $B$ band at the quenching of the disk.
Figure \ref{fig:sbmodel} shows the star formation histories and the rest-frame 
$B$-band absolute magnitudes of the bulge and disk components, and the 
disk to total luminosity ratio of the model as a function of time.
In the model, sSFR declines by 0.5 and 1 dex for 1.15 (0.58) and 2.3 (1.15) Gyr 
in the case with a quenching timescale of 1 (0.5) Gyr. 
The disk becomes fainter by $\sim 1$ mag during the sSFR declines
by 1 dex in the cases of $\tau =$ 1 and 0.5 Gyr, which leads to a decrease 
in the disk fraction from $\sim 0.8$ to $\sim 0.6$.
We then carried out a Monte Carlo simulation with the IRAF/ARTDATA package 
to examine the effect of the decrease of the disk fraction on 
the axial-ratio distribution.
We added artificial objects at $z=0.8$ with the disk and bulge components 
of the toy model to sky regions in the ACS $I_{\rm F814W}$-band images 
and measured the apparent axial ratios with SExtractor.
The magnitudes, sizes, and axial ratios of the disk and bulge components 
were adjusted to match with the observed main-sequence and passively 
evolving galaxies at $0.6<z<1.0$, respectively 
(see Appendix \ref{sec:monsim} for details).
We performed 10000 such simulations for the toy model with 
the disk fraction of 0.8, 0.6, and 0.45 and derived the axial-ratio 
distribution. 
Figure \ref{fig:histsim} shows the results of the simulation. 
The distribution of the apparent axial ratio for the model with 
D/(B+D) $ = 0.6$ are significantly different from that with D/(B+D) $ = 0.8$. 
We fitted these axial-ratio distributions with the triaxial ellipsoidal 
models as described in Section \ref{subsec:intrinsic} and obtained 
the intrinsic thickness of $\mu_{C/A}=0.23$ and 0.27 for D/(B+D) $ = 0.8$ and 
0.6, respectively. 
Therefore, we expect that the decrease of the disk fraction due to the 
quenching of star formation leads to a significant change in the 
axial-ratio distribution, although the observed change around 
$\Delta$MS $\sim -1$ dex in Figure \ref{fig:intssfr} 
may not be fully explained by this effect.

The slightly higher transition sSFR of $\Delta$MS $ \sim -0.75$ dex 
for galaxies with 
$M_{\rm star}=10^{10.5}$--$10^{11} M_{\odot}$ than those 
with $10^{10}$--$10^{10.5} M_{\odot}$ 
may also be explained by this effect, if more massive 
main-sequence galaxies tend to have the high bulge fraction 
as shown by \cite{mor18}. 
\cite {bre18} carried out the multi-component surface brightness fitting for 
galaxies with $M_{\rm star} = 10^{10.25}$--$10^{10.75} M_{\odot}$ at $z<0.2$ from the 
GAMA survey with the multi-band data, and found that most of green-valley 
galaxies
 show significant bulge and disk components. They also suggested that 
the migration from the blue cloud to the red sequence is caused 
by the disk fading.

\subsection{Shape evolution of passively evolving galaxies at $z<1$}

In Section \ref{subsec:evol}, we found that the distribution of the apparent 
axial ratio and intrinsic shape of passively evolving
 galaxies with $M_{\rm star} < 10^{11} M_{\odot}$ significantly evolve between 
$z\sim0.8$ and $z\sim0.4$; the edge-on axial ratio $C/A$ (thickness) 
decreases with time from  $\mu_{C/A}=0.500$ (0.400) at $0.6<z<1.0$ 
to $\mu_{C/A}=0.370$ (0.325) 
at $0.2<z<0.6$ for those with $M_{\rm star} = 10^{10}$--$10^{10.5} M_{\odot}$ 
($10^{10.5}$--$10^{11} M_{\odot}$). On the other hand, 
massive galaxies with $M_{\rm star} > 10^{11} M_{\odot}$ have a thick shape 
with $\mu_{C/A} = $ 0.45 -- 0.49 
and show no significant evolution in their shape.

\cite{van09} and \cite{hol12} measured the apparent axial ratio of 
quiescent galaxies at $z=$ 0.04 -- 0.08 with the SDSS data, and found that 
massive galaxies with $M_{\rm star} > 10^{11} M_{\odot}$ exclusively have 
$b/a > 0.6$, 
while those galaxies with $M_{\rm star} < 10^{11} M_{\odot}$ show more flatter 
distribution over $b/a = $ 0.2 -- 1.0. 
These results are consistent with those of passively evolving galaxies 
at $0.2<z<0.6$ in this study.
\cite{hol12} also investigated the axial-ratio distribution of quiescent galaxies 
at $z = $ 0.6 -- 0.8 with the GEMS and COSMOS data, and confirmed that 
these galaxies show the similar trends with the SDSS galaxies at $z\sim 0.06$, 
which is consistent with our results for those at $0.6<z<1.0$.
Although they found no significant evolution between $z \sim 0.7$ 
and $z \sim 0.06$ for quiescent galaxies with 
$M_{\rm star} < 10^{11} M_{\odot}$, their sample size for $z\sim0.7$ galaxies 
is $\sim$ 2--3 times smaller than our sample at $0.6<z<1.0$. 
The differences between passively evolving galaxies at $0.2<z<0.6$ and $0.6<z<1.0$ 
in the axial-ratio distribution in Figure \ref{fig:dms2evol} would be buried in noise 
especially for those galaxies with $M_{\rm star} = 10^{10.5}$--$10^{11} M_{\odot}$, 
 if we decrease the sample size by a factor of $\sim$ 2 -- 3.
\cite{hil19} studied the median values of the apparent axial ratio for 
quiescent and star-forming galaxies at $0.2<z<4.0$ with the CANDELS/3D-HST data.
They found that low-mass quiescent galaxies with $M_{\rm star} < 10^{10.5} M_{\odot}$ 
show a evolution of the median axial ratio from $\sim 0.68$ at $0.5<z<1.0$ to 
$\sim 0.63$ at $0.2<z<0.5$, while no significant evolution is seen for 
those with $M_{\rm star} > 10^{10.5} M_{\odot}$.
The result for those low-mass galaxies seems to be consistent with our result, 
although those with $M_{\rm star} = 10^{10.5}$--$10^{11} M_{\odot}$ show no significant 
evolution in \cite{hil19} probably due to a relatively small number of their samples
(total $\sim 800$ and 200 quiescent galaxies at $0.5<z<1.0$ and $0.2<z<0.5$ 
before divided by stellar mass).  
 
In our results, the thickness (edge-on axial ratio $C/A$) of 
passively evolving galaxies with $M_{\rm star} = 10^{10}$--$10^{11} M_{\odot}$ 
decreases with time on average. 
In general, it is difficult to make a stellar disk system thinner 
without forming new stars from a thin gas disk, because such a stellar disk 
system once formed tends to become thicker shape as time passes  
through minor mergers, galaxy interactions,  
and so on (e.g., \citealp{qui93}; \citealp{vil08}).
The minor merger/interaction could stimulate the bulge growth 
\citep{qu11}, which also leads to the thicker shape of those galaxies. 
Since the increase in the number density of passively evolving galaxies 
at $z\lesssim 1$ is much larger than that of star-forming galaxies 
(e.g., \citealp{bor06}; \citealp{ilb10}), most quiescent galaxies are 
expected not to experience a significant star formation after the 
star formation once stopped (e.g., \citealp{bel07}; but see also \citealp{man19}).
Thus passively evolving galaxies quenched in earlier epoch do not seem to evolve  
into thinner shape with time.

On the other hand, star-forming galaxies on the main sequence at $0.2<z<1.0$ 
have much thinner shapes with $\mu_{C/A}=$ 0.2 -- 0.25 
(Figure \ref{fig:intacevol} and Table \ref{tab:fitevol}).
Therefore, newly quenching galaxies from the main-sequence 
population could cause the evolution of the axial-ratio distribution for 
the passively evolving population, if their morphology does not violently 
change during the quenching. 
For example, while   
the disk fading and/or bulge growth mentioned in the previous section 
make the thickness of the newly quenching galaxies slightly larger than 
the typical value of main-sequence galaxies, 
such galaxies could be still sufficiently thinner than passively evolving 
ones quenched in earlier epoch.  
\cite{car13} investigated the number density evolution of passively evolving 
galaxies at $0.2<z<1.0$ in the COSMOS field as a function of galaxy size, 
and found that the fraction of those galaxies with a larger size increases with 
time. Since those galaxies with a larger size 
have slightly bluer rest-frame $U-V$ colors 
than those with a normal size, they suggested that newly quenching galaxies 
with a larger size mainly cause the evolution in the size distribution of 
passively evolving galaxies.
The scenario mentioned above seems to be consistent with the results by \cite{car13}, 
because star-forming galaxies tend to show larger sizes than passively evolving 
galaxies at a given stellar mass at $z\lesssim 3$ (e.g., \citealp{van14b}).  
Newly quenching galaxies from the main sequence are expected to have 
both a thinner shape and a larger size. 
In fact, from Figures \ref{fig:arsizelowz} and \ref{fig:arsizehighz}, 
we can see that (1) passively evolving galaxies tend to have smaller sizes 
than star-forming ones with the same mass and redshift, 
(2) passively evolving galaxies with larger sizes tend to show 
more extended axial-ratio distributions down to lower $b/a$ values, 
and (3) the sizes of passively evolving galaxies for a given mass 
significantly increase with time. 
These trends are consistent with the scenario.
In this scenario,  
the major merger, which tends to destroy the thin disk component, cannot be  
the main driver of the quenching of star formation, although a relatively small 
fraction of galaxies may quench through the major merger.

The transition of newly quenching galaxies with a thin shape 
from the main-sequence population 
could also explain 
the stellar mass dependence of the evolution in the axial-ratio distribution 
of passively evolving galaxies.
The evolution of the galaxy stellar mass function for quiescent galaxies 
in previous studies suggests that the number density increase with time 
at $z\lesssim 1$ in the passively evolving population is larger 
for less massive 
galaxies at $M_{\rm star} < 10^{11} M_{\odot}$ (e.g., \citealp{ilb10}; 
\citealp{ilb13}; \citealp{muz13}). 
Thus the expected fraction of galaxies newly quenched between $z\sim 0.8$ and 
$z\sim 0.4$ in the passively evolving population is higher for less massive 
galaxies. This can explain the result that passively evolving galaxies with 
$M_{\rm star} = 10^{10}$--$10^{10.5} M_{\odot}$ show the stronger evolution of the 
thickness from $\mu_{C/A} = 0.50$ at $0.6<z<1.0$ to 0.37 at $0.2<z<0.6$ 
than those with $10^{10.5}$--$10^{11} M_{\odot}$ (from $\mu_{C/A} =$ 0.40 to 0.33).

In this scenario, the relatively thick spheroidal shapes of 
 passively evolving galaxies with $M_{\rm star} < 10^{11} M_{\odot}$ 
at $0.6<z<1.0$ could be explained 
at least partly by a lack of the supply of quenching galaxies with a thin shape 
at higher redshifts, because star-forming galaxies at $z>1$ seem to have
 thicker shapes than those at $z<1$. 
Several studies have investigated the apparent axial ratios of star-forming 
galaxies at $z>1$ with the optical/NIR HST data to estimate their 
intrinsic shapes, and found that 
these galaxies at $z\sim$ 1.5--3 have thicker and more prolate (bar-like) 
shapes (\citealp{rav06}; \citealp{yum11}; \citealp{yum12}; \citealp{law12}; 
\citealp{van14a}; \citealp{tak15}; \citealp{zha19}).  
Ground-based observational studies with the integral field spectroscopy 
also found 
that star-forming galaxies with a rotationally supported disk 
at $z\sim 2$ show  
larger velocity dispersion than disk galaxies at lower redshifts, 
which suggests 
thicker shape of the disk (e.g., \citealp{for09}; \citealp{law12}; 
\citealp{wis15}). 
Therefore galaxies with a very thin shape seem to be rare at such high redshift 
even in the star-forming population,  
and it is difficult to form passively evolving galaxies with such a thin shape 
at $z>1$.  
The higher merger rate at higher redshifts (e.g., \citealp{mun17}; 
\citealp{dun19}) could also contribute to 
more thicker shapes of those high-redshift galaxies.

On the other hand, no significant evolution in the shape of 
massive passively evolving galaxies with 
$M_{\rm star} > 10^{11} M_{\odot}$ can be explained 
by a lack of such massive star-forming galaxies irrespective of shape.
The stellar mass function for star-forming galaxies 
indicates that massive star-forming galaxies with 
$M_{\rm star} > 10^{11} M_{\odot}$ 
are very rare at $z\lesssim 1$ (e.g., \citealp{ilb13}; \citealp{muz13}), 
and the expected number of such massive galaxies that newly quench the 
star formation without major mergers between $z\sim 0.8$ and $z\sim 0.4$ 
is very small. 
Instead, massive passively evolving galaxies with 
$M_{\rm star} > 10^{11} M_{\odot}$ are expected to be mainly formed 
through the major mergers  
of relatively massive objects with $M_{\rm star} \sim 10^{11} M_{\odot}$.
Since the quiescent fraction in galaxies with $M_{\rm star} \sim 10^{11} M_{\odot}$ 
becomes relatively high at $z\lesssim 1$ 
(e.g., \citealp{ilb10}; \citealp{kaj11}), 
such major mergers tend to be dry mergers or those with a relatively low 
gas-mass fraction, which leads to the spheroidal shapes of the remnants 
(e.g., \citealp{spr05}; \citealp{hop09}; \citealp{rod17}).
The lack of newly quenching galaxies with a thin shape and the formation 
through dry major mergers can explain the thick spheroidal shape and 
its no significant evolution at $z<1$ for passively evolving galaxies with 
$M_{\rm star} > 10^{11} M_{\odot}$ \citep{hol12}. 
The flatter shapes of massive quiescent galaxies with 
$M_{\rm star} > 10^{11} M_{\odot}$
 at $z>1$ reported by \cite{cha13} and \cite{hil19} could be explained 
similarly by their formation through the wet mergers,  
because the merger progenitors are expected to be more gas-rich 
at higher redshifts.

\subsection{Stellar mass dependence of thickness and its imprecations}

We found that the intrinsic edge-on axial ratio $C/A$ of main-sequence galaxies 
decreases with increasing stellar mass in the both redshift ranges 
(Figure \ref{fig:intabcevol} and Table \ref{tab:fitevol}). 
Recently, \cite{pil19} investigated the 3-dimensional shapes of stellar and 
gaseous components of star-forming 
galaxies in the high-resolution cosmological simulation, Illustris TNG50, 
and found that more massive star-forming galaxies show lower edge-on axial 
ratio $C/A$, i.e., thinner shapes of the stellar component than less massive 
galaxies at $0.5<z<4.0$. 
Their results for the thickness of those galaxies at $z\sim0.5$ 
as a function of stellar mass are qualitatively consistent with our results, 
although a direct comparison is difficult due to the different stellar mass 
and redshift binning between \cite{pil19} and this study.
In their simulation, the gas disks strongly evolve into thinner shapes 
as the $V_{\rm rot}/\sigma$ of the gas disk significantly decreases with time, 
and the star formation in the thin gas disk makes the stellar disk thinner.
The thickness of the gas disk at a given redshift tends to be smaller 
in more massive star-forming galaxies in all redshifts, which leads to 
the mass dependence of the thickness of the stellar disk.
On the other hand, \cite{sal12} found that the coherent alignment of 
the angular 
momentum of gas that has been accreting onto a galaxy over time is more 
important 
for the formation of the thin-disk morphology than the net spin or merger 
history 
of the dark matter halo the galaxy resides in the GIMIC cosmological simulation.
They also suggested that gas accretion from a quasi-hydrostatic hot corona, 
namely, ``hot-mode'' accretion preferentially forms a thin stellar disk,  
because such shock-heated gas in the halo is forced to homogenize its rotation 
properties before accreting onto the galaxy, which results in a rather gradual 
supply of gas with a relatively stable spin axis. 
In contrast with the hot-mode accretion, cold gas accretion, 
where gases from distinct filaments directly flow into the 
central galaxy with misaligned spins,  tends to disturb the gas kinematics and 
form a more thick spheroidal stellar component. 
The hot-mode accretion is expected to dominate preferentially 
in more massive dark matter 
halo with $M_{\rm halo} \gtrsim 10^{11.5} M_{\odot}$ from the previous 
theoretical studies (e.g., \citealp{bir03}; \citealp{ker05}).
It is suggested by the clustering and/or abundance matching analyses  
that more massive star-forming galaxies tend to be associated with more 
massive dark matter halos (e.g., \citealp{tin13}; \citealp{leg19}).
Therefore, the stellar mass dependence of the thickness of 
star-forming galaxies 
could be explained by the halo mass dependence of the contribution from the 
hot mode accretion in the gas supply to the galaxies.
In fact, \cite{leg19} found that the dark matter halo mass of 
star-forming galaxies at $0.2 \lesssim z \lesssim 1$ increases with 
stellar mass from $\sim 10^{11.3}$--$10^{11.7} M_{\odot}$ 
at $M_{\rm star}\sim 10^{9.5} M_{\odot}$ to $\sim 10^{12.3}-10^{12.7} M_{\odot}$ 
at $M_{\rm star} \sim 10^{11} M_{\odot}$, where the contribution from  
the hot mode accretion is expected to increase with halo mass.
More massive star-forming galaxies on the main sequence 
tend to be formed in massive halos dominated by the hot mode accretion, 
which may leads to their observed thinner shapes. 

This scenario could also explain the reason why star-forming galaxies with 
a relatively thin disk appeared around $z\sim 1$, because the hot mode 
accretion is expected not to dominate even in massive halos at $z \gtrsim 2$ 
due to gas supply through the cold gas stream (e.g., \citealp{ker09}; 
\citealp{dek09a}). 
Some of gas accreting to a dark matter halo is expected to 
penetrate surrounding hot gas in a form of filaments of dense and cold 
infalling gas and directly accrete onto the central galaxy 
at such high redshift, where the mass accretion rate and matter density 
tend to be high. 
Such direct gas supply through the filaments of cold gas could make the gas disk 
of the central galaxy more turbulent and gravitationally unstable, which leads to 
 thick and clumpy stellar disk and bulge formation/growth in some cases 
(e.g., \citealp{dek09b}; \citealp{cev10}; \citealp{dek14}).
Thus it seems to be difficult to form the thin stellar disks even in massive 
halos at $z \gtrsim 2$. 
After the hot-mode accretion starts to dominate at $z\sim2$ in relatively 
massive halos, 
the thin stellar disks may be gradually formed from thinner gas disks 
and appear around $z\sim1$. 
If some of these star-forming galaxies with a thin stellar disk quench 
and evolve into passively evolving galaxies without a violent morphological 
change as discussed in the previous section, 
it is understood that the fraction of passively evolving 
galaxies with a thin shape increases with time at $z<1$ rather than higher redshifts. 
Since more massive star-forming galaxies have a sufficient time 
to form a thin disk through the hot mode accretion in earlier epoch 
in this scenario \citep{nog19}, 
passively evolving galaxies with a thin shape may also 
be provided preferentially in more massive galaxy population at $z\sim1$.
This could explain our result that passively evolving galaxies 
with $M_{\rm star} = 10^{10.5}$--$10^{11} M_{\odot}$ already 
show a thinner shape than 
those with $10^{10}$--$10^{10.5} M_{\odot}$ at $0.6<z<1.0$. 
In fact, \cite{bez18} reported that $\sim 64$\% of quiescent galaxies with 
$M_{\rm star} \sim 10^{10.5}$--$10^{11} M_{\odot}$ at $0.6<z<1.0$ show a 
significant rotation, while those massive galaxies with 
$M_{\rm star} > 10^{11} M_{\odot}$
 show no or little rotation in the LEGA-C survey.
Such quiescent galaxies with a significant rotation 
might be recently quenched galaxies 
with a relatively thin disk that has grown through 
the hot mode gas accretion since $z\lesssim$ 2.

\section{Summary} \label{sec:summary}
With the COSMOS HST/ACS $I_{\rm F814W}$-band data over the 1.65 deg$^2$ region 
in the COSMOS field, we measured the apparent axial ratios of 
$\sim$ 21000 galaxies with $M_{V}<-20$ at $0.2<z<1.0$, 
and fitted the distribution of the axial ratio with the triaxial 
ellipsoid models to statistically estimate their intrinsic 
3-dimensional shapes as a function of stellar mass, sSFR, and redshift.
Our main results are summarized as follows.
\begin{itemize}

\item We confirmed that star-forming galaxies on the main sequence show 
a thin disk shape with a intrinsic edge-on axial ratio of $C/A \sim $ 
0.2 -- 0.25, 
while passively evolving galaxies with a low sSFR have a more thick spheroidal 
shape with $C/A \sim $ 0.3 -- 0.5.
 
\item The transition from the thin disk to the thick spheroidal shape for 
galaxies with $M_{\rm star} = 10^{10}$--$10^{11} M_{\odot}$ occurs 
around $\Delta$MS $ \sim -1$ dex, i.e.,  an order of magnitude lower sSFR than 
that of the main sequence irrespective of redshift. 
The shape of galaxies with $M_{\rm star} = 10^{10.5}$--$10^{11} M_{\odot}$  
changes at slightly higher $\Delta$MS ($ \sim -0.75$ dex) than that of
 less massive ones with $10^{10}$--$10^{10.5} M_{\odot}$ ($\sim -1$ dex).

\item Passively evolving galaxies with $M_{\rm star}<10^{11} M_{\odot}$ 
show a significant evolution in the axial-ratio distribution and estimated 
intrinsic shape. 
The edge-on axial ratio $C/A$ decreases with time 
from  $\mu_{C/A}=0.500$ (0.400) at $0.6<z<1.0$ to $\mu_{C/A}=0.370$ (0.325) 
at $0.2<z<0.6$ for those galaxies with $M_{\rm star} = 10^{10}-10^{10.5} M_{\odot}$ 
($10^{10.5}$--$10^{11} M_{\odot}$). On the other hand, 
those massive galaxies with $M_{\rm star} > 10^{11} M_{\odot}$ have a thick shape 
with $\mu_{C/A} = $ 0.45 -- 0.49 and show no significant evolution in their shape
 at $0.2<z<1.0$.

\item The intrinsic shape of star-forming galaxies on the main sequence 
does not significantly evolve at $0.2<z<1.0$.
On the other hand, 
the intrinsic edge-on axial ratio $C/A$ (thickness) of 
the main-sequence galaxies 
decreases with increasing stellar mass from $\mu_{C/A} = 0.265$ 
(0.255) for galaxies with $M_{\rm star} = 10^{9.5}$--$10^{10} M_{\odot}$ at 
$0.2<z<0.6$ ($0.6<z<1.0$) to $\mu_{C/A} = 0.210$ (0.230) for those with 
$10^{10.5}$--$10^{11} M_{\odot}$, although the uncertainty is not negligible.  

\end{itemize}

We discussed that the quenching and migration to the passively evolving 
population of some main-sequence galaxies with a thin shape 
without violent morphological 
change can explain the shape transition at a nearly constant $\Delta$MS 
and the evolution of the fraction of passively evolving galaxies 
with a thin shape at $M_{\rm star} < 10^{11} M_{\odot}$. 
The scenario that 
the thin stellar disks of star-forming galaxies 
are formed by the gas supply through the hot-mode accretion 
could also explain the stellar mass dependence of the thickness 
of these galaxies and the increase of the fraction of passively evolving 
galaxies with a thin shape at $z<1$.
On the other hand, massive passively evolving galaxies with 
$M_{\rm star} > 10^{11} M_{\odot}$ are expected to be formed by dry major mergers 
at $z<1$, which leads to thick and spheroidal shapes.

The statistical analysis of the apparent axial ratio such as this study is 
a powerful tool to constrain the evolution in the intrinsic shape,   
especially thickness of galaxies over the cosmic time, 
but its advantage strongly depends on the sample size.
The future wide-field surveys with JWST and WFIRST 
will enable us to investigate the evolution more detailedly with high statistical 
accuracy.

\acknowledgments
We would like to thank the anonymous referee for the valuable suggestions and 
comments.
The HST COSMOS Treasury program was supported through NASA grant
HST-GO-09822. 
This work is based on observations taken by the CANDELS Multi-Cycle Treasury 
Program with the NASA/ESA HST, which is operated by the Association of Universities 
for Research in Astronomy, Inc., under NASA contract NAS5-26555.
Data analysis were in part carried out on common use data analysis computer
system at the Astronomy Data Center, ADC, of the National Astronomical
Observatory of Japan.  This work was financially supported by JSPS (17K05386).

%






\appendix

\section{Examples of sample galaxies as a function of axial ratio} \label{sec: montage}

\begin{figure}
\epsscale{1.0}
\plotone{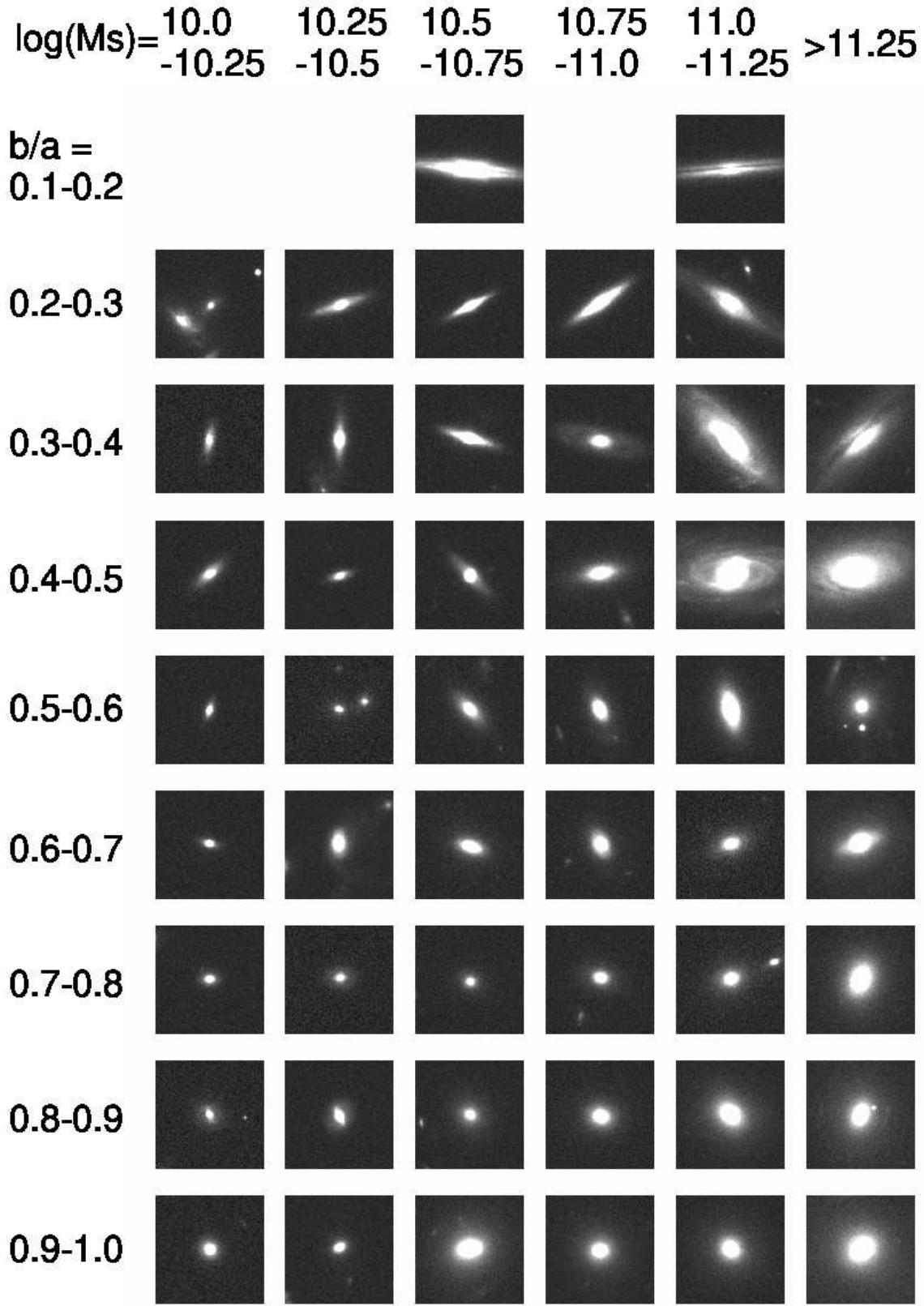}
\caption{
Examples of passively evolving galaxies with $\Delta$MS $ < -1.5$ dex 
at $0.2<z<0.6$ in our sample 
as a function of axial ratio and stellar mass.
The apparent axial ratio of the object increases from top to bottom row, 
and the stellar mass increases from left to right column.
Each panel is $6^{\prime\prime} \times 6^{\prime\prime}$ in size.  
These galaxies are randomly selected in the axial ratio and stellar mass bins.
Note that there is no sample galaxy in some of the bins at $b/a < 0.3$.
 \label{fig:monlps}}
\end{figure}

\begin{figure}
\epsscale{1.0}
\plotone{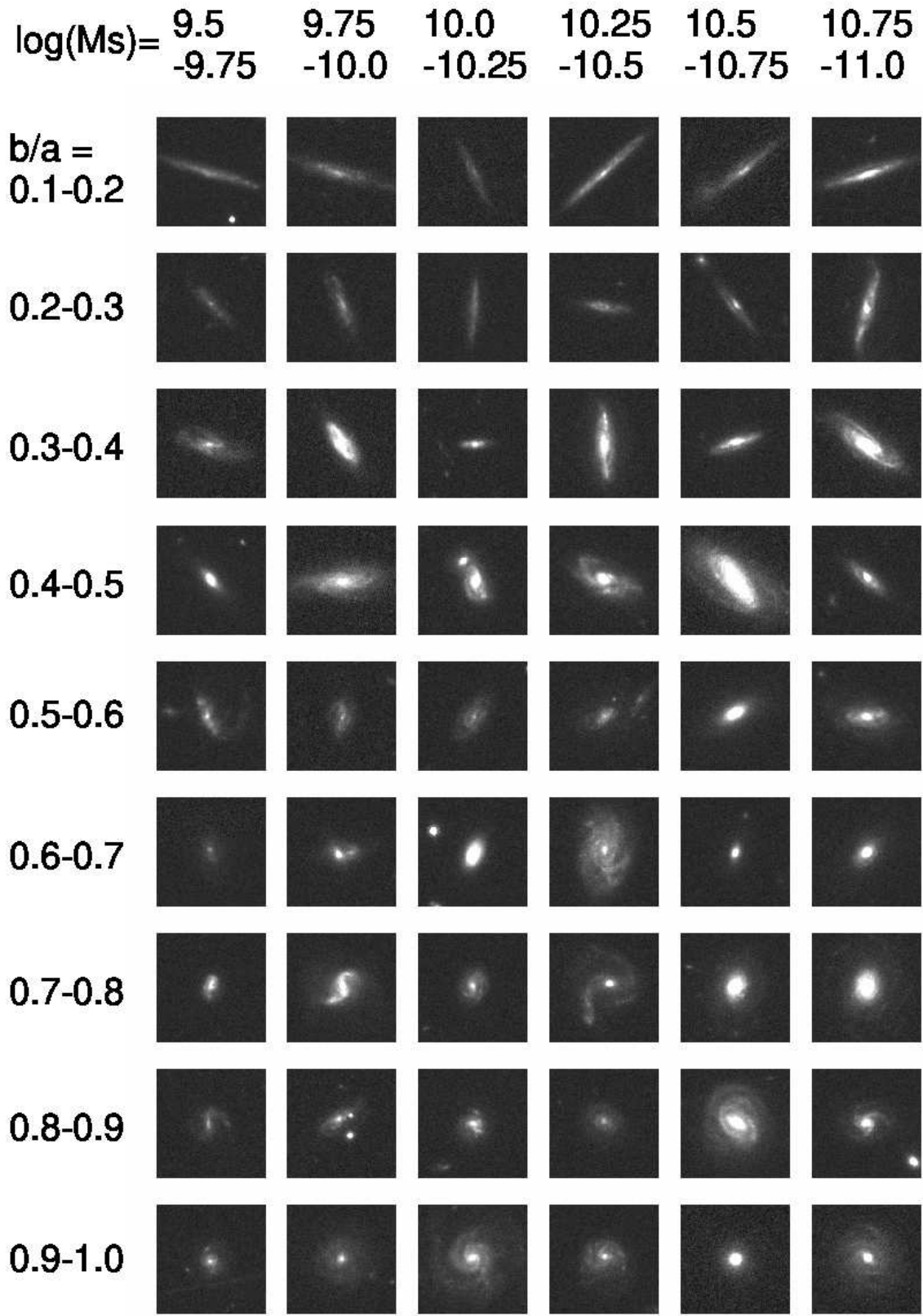}
\caption{ 
Same as Figure \ref{fig:monlps}, but for star-forming galaxies on the main sequence 
with $\Delta$MS $ = -0.5$ -- $+0.5$ at $0.2<z<0.6$.
\label{fig:monlms}}
\end{figure}

\begin{figure}
\epsscale{1.0}
\plotone{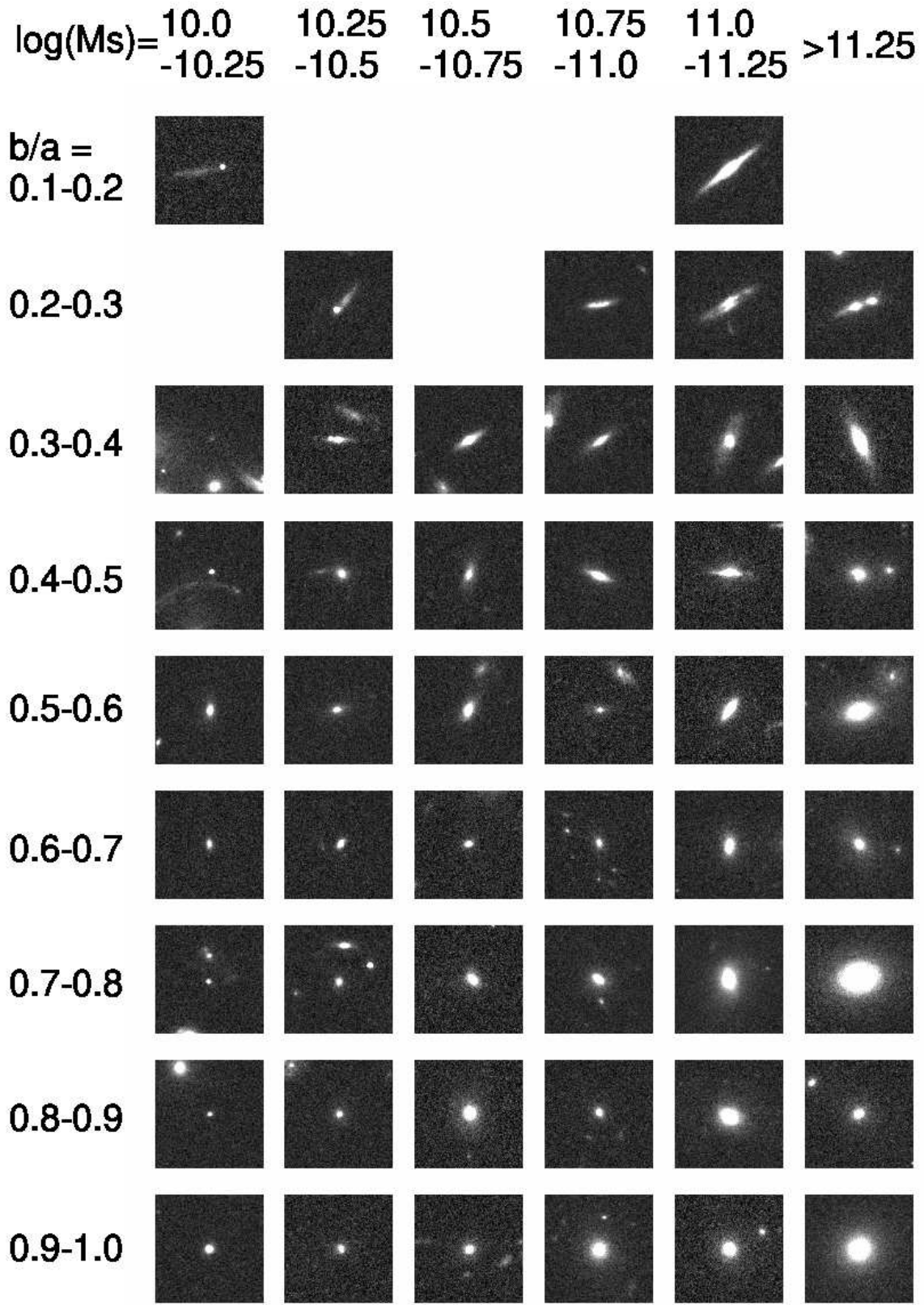}
\caption{ 
Same as Figure \ref{fig:monlps}, but for passively evolving galaxies with 
$\Delta$MS $ < -1.5$ dex at $0.6<z<1.0$.
\label{fig:monhps}}
\end{figure}

\begin{figure}
\epsscale{1.0}
\plotone{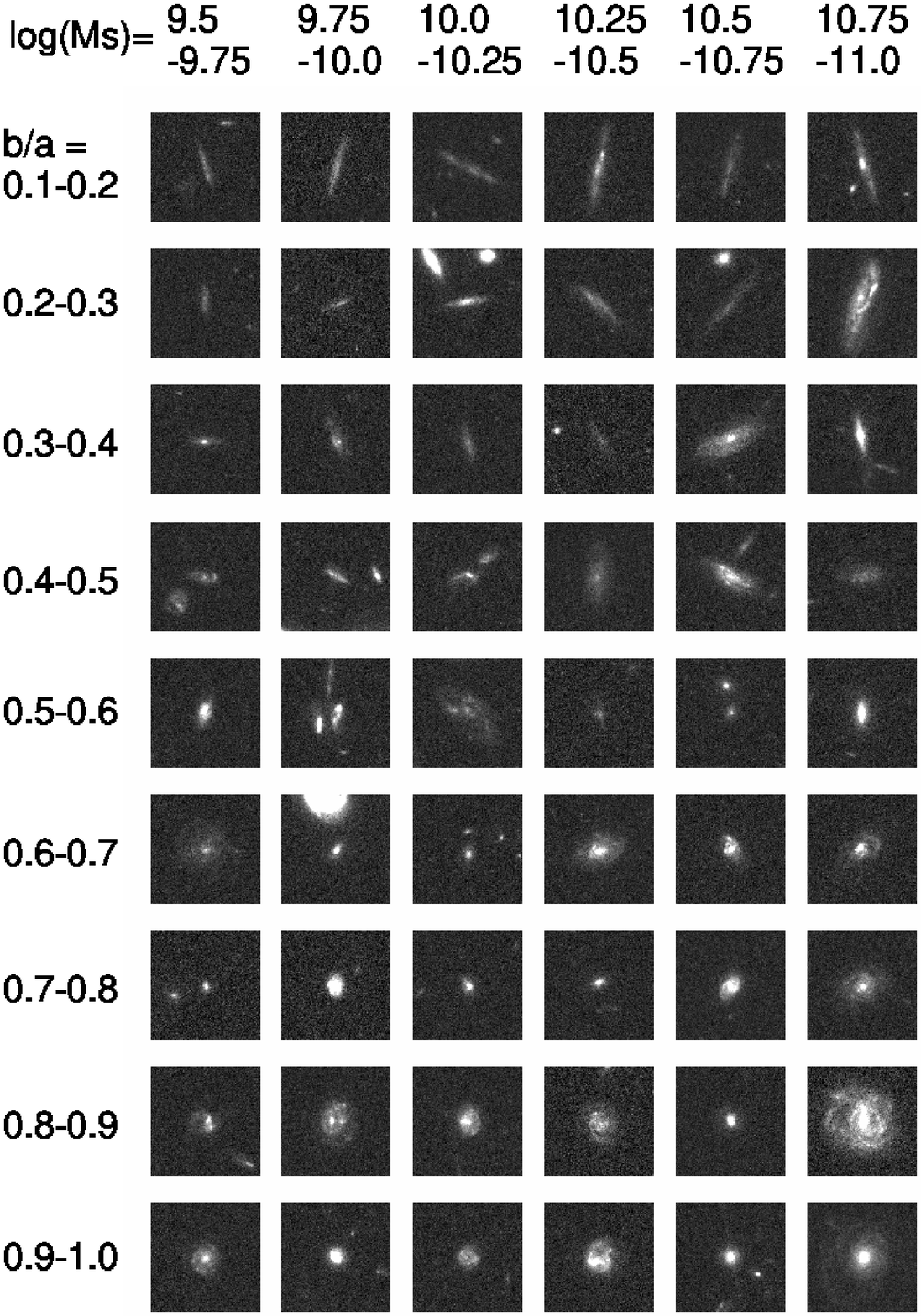}
\caption{ 
Same as Figure \ref{fig:monlps}, but for star-forming galaxies 
on the main sequence 
at $0.6<z<1.0$.
\label{fig:monhms}}
\end{figure}

We show HST/ACS $I_{\rm F814W}$-band images of some sample galaxies 
with different sSFRs and redshifts 
as a function of the measured axial ratio 
in Figures \ref{fig:monlps}--\ref{fig:monhms}.
Note that galaxies with a low sSFR of $\Delta$MS $< -1.5$ dex tend to show 
relatively large apparent axial ratios, and there are few those galaxies with 
a very small axial ratio of $b/a \lesssim 0.2$.

\section{The best-fit models for the subsamples} \label{sec:compmodel}

\begin{figure*}
\epsscale{1.1}
\plotone{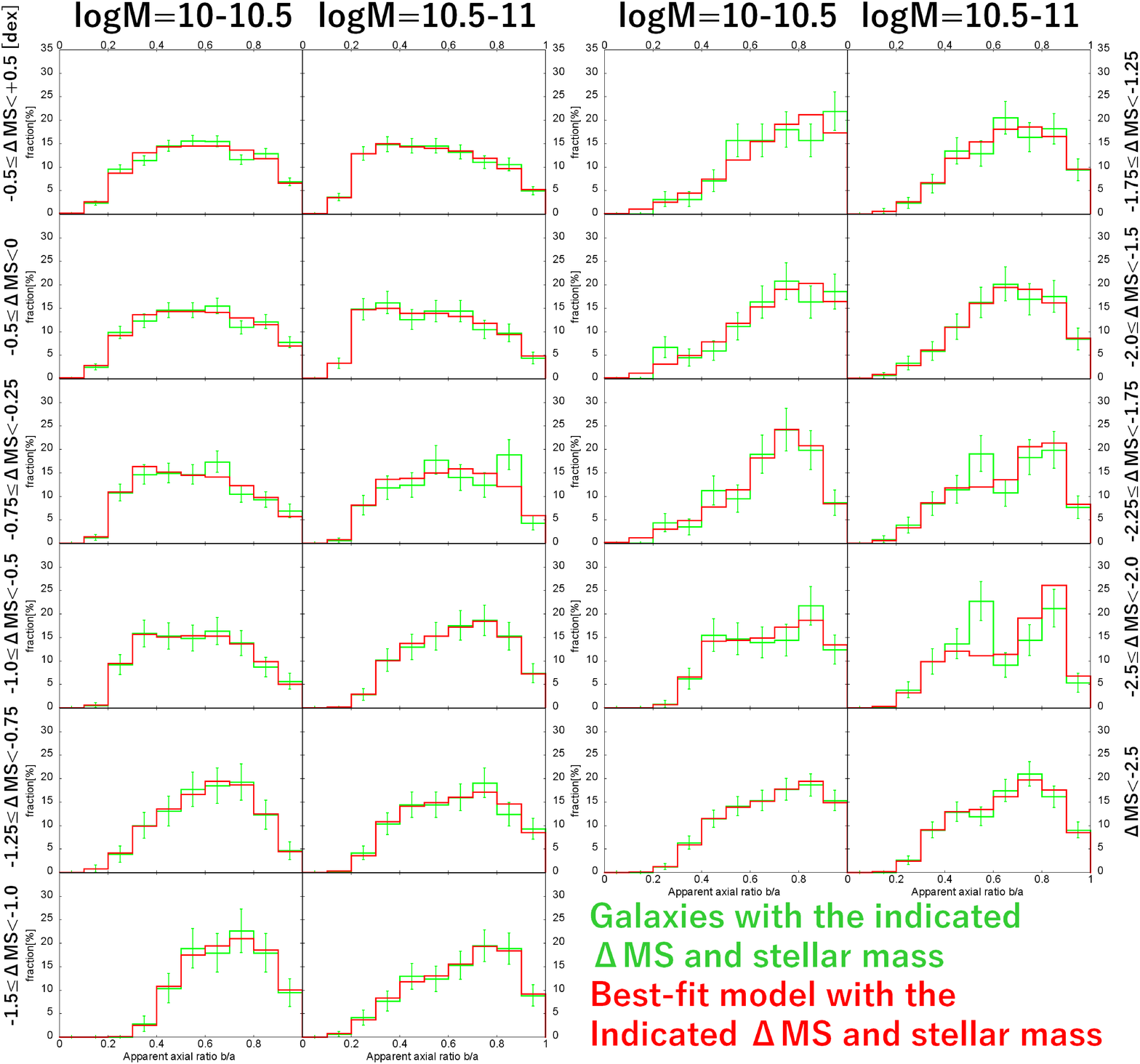}
\caption{ 
The observed distribution of the apparent axial ratio 
and the best-fit triaxial ellipsoid models described in 
Section \ref{subsec:intrinsic} for the subsamples 
 at $0.2<z<0.6$ as a function of $\Delta$MS and stellar mass.
The configuration of the panels are the same as Figure \ref{fig:dmslowz}.
\label{fig:fitdmslowz}}
\end{figure*}

\begin{figure*}
\epsscale{1.1}
\plotone{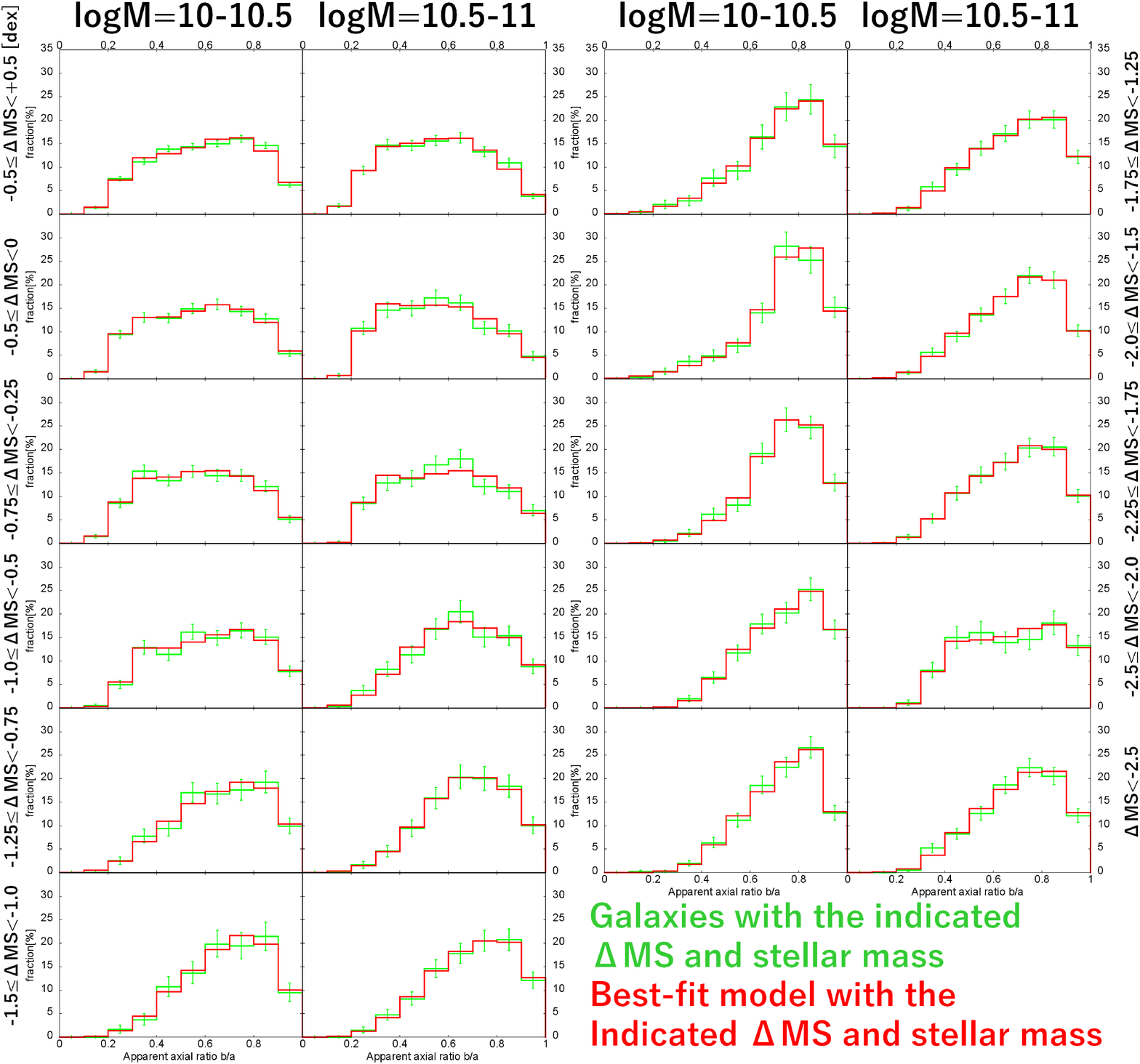}
\caption{ 
The same as Figure \ref{fig:fitdmslowz}, but for galaxies at $0.6<z<1.0$.
\label{fig:fitdmshighz}}
\end{figure*}

\begin{figure*}
\epsscale{1.05}
\plotone{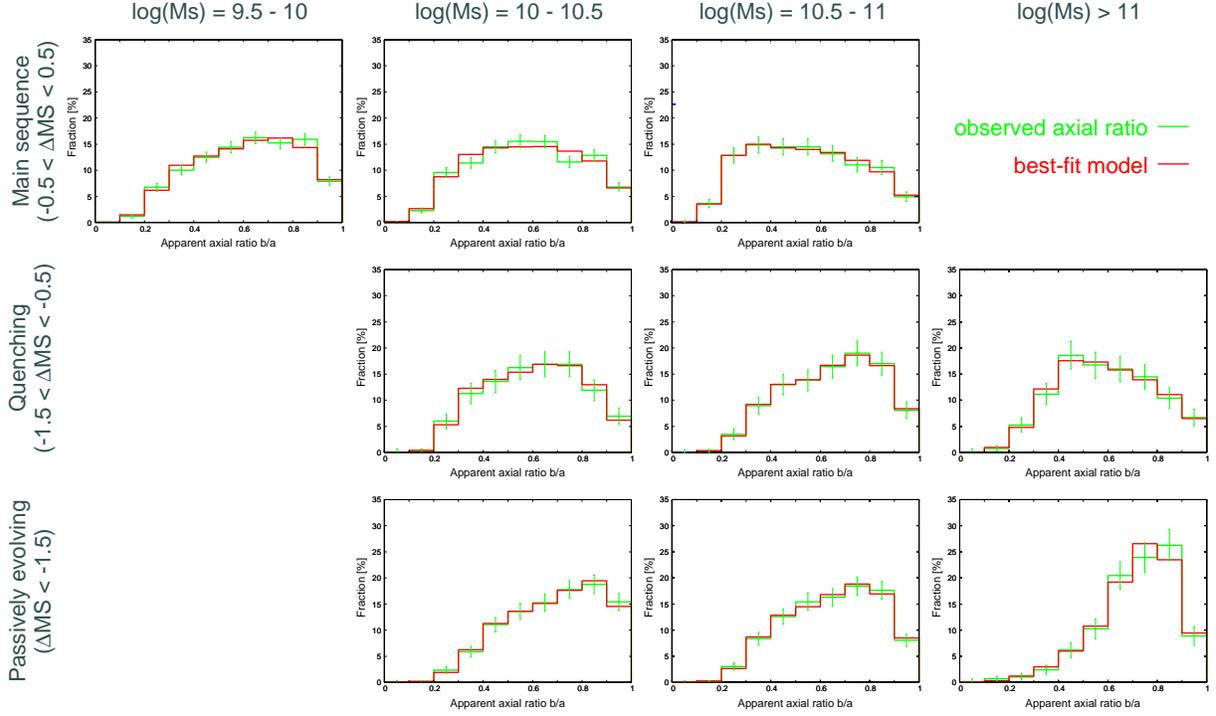}
\caption{ 
The observed distribution of the apparent axial ratio 
and the best-fit model for the subsamples 
 at $0.2<z<0.6$ as a function of $\Delta$MS and stellar mass.
The samples are divided by $\Delta$MS into the main-sequence, quenching, 
and passively evolving populations.
The configuration of the panels are the same as Figure \ref{fig:dms2evol}.
\label{fig:fit3lowz}}
\end{figure*}

\begin{figure*}
\epsscale{1.05}
\plotone{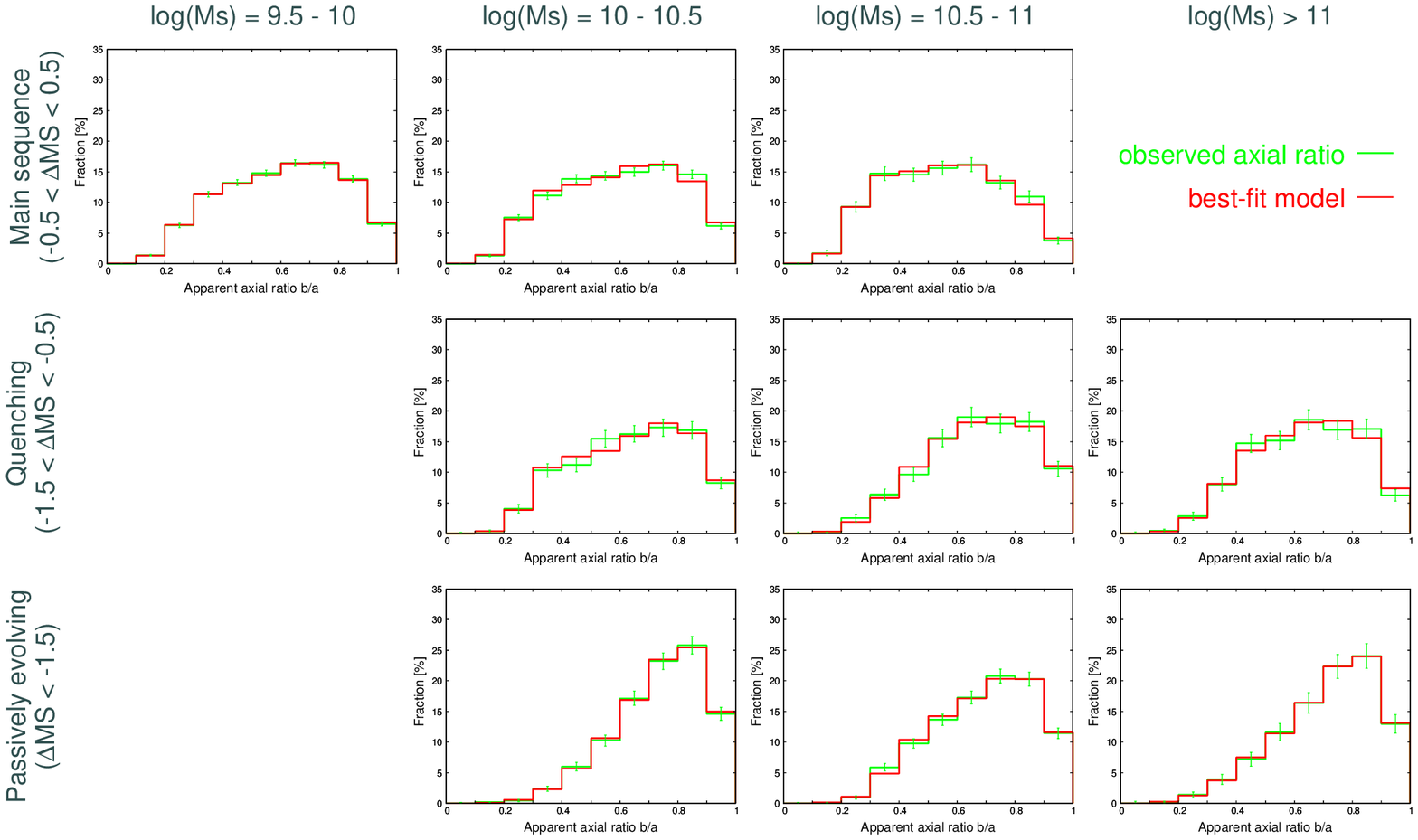}
\caption{ 
The same as Figure \ref{fig:fit3lowz}, but for galaxies at $0.6<z<1.0$.
\label{fig:fit3highz}}
\end{figure*}

In this study, we fitted the distribution of the apparent axial ratio for 
the subsamples with various stellar mass and $\Delta$MS ranges 
 with the triaxial ellipsoid models to estimate the intrinsic 3-dimensional 
shape. 
We show comparisons between the observed axial-ratio distributions and 
the best-fit models for the subsamples used in this study 
in Figures \ref{fig:fitdmslowz}--\ref{fig:fit3highz}, 
which enable to examine the goodness of fitting for each subsample and 
check whether the systematic difference between the observed distribution 
 and the best-fit model exists or not.
One can see that the observed distributions are well fitted 
by the models for all the subsamples and there seems to be
 no systematic difference.

\section{The model fitting for galaxies with $M_{\rm star} = 10^{10}$--$10^{10.5} M_{\odot}$ and $\Delta$MS $\sim -2.0$ -- $-1.5$ at $0.2<z<0.6$} \label{sec:largeerr}

\begin{figure}
\epsscale{1.05}
\plotone{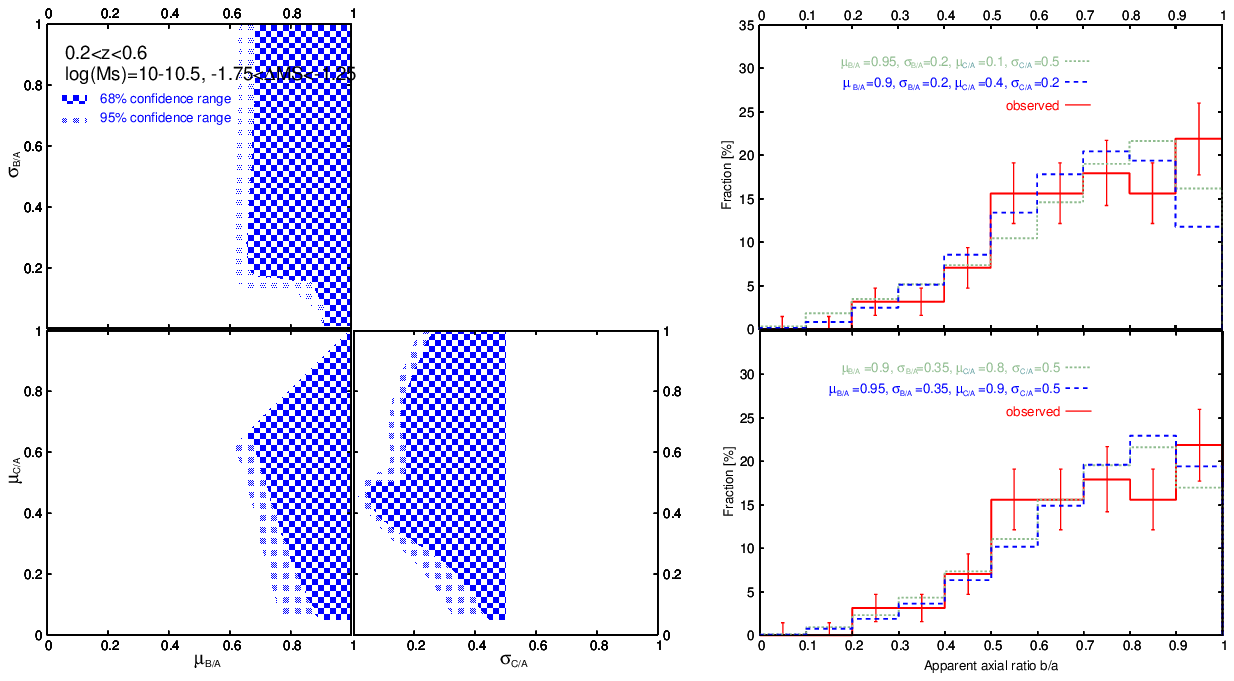}
\caption{ 
{\bf (left)} The confidence ranges of the intrinsic shape parameters, 
namely, $\mu_{B/A}$, $\sigma_{B/A}$, $\mu_{C/A}$, and $\sigma_{C/A}$,   
in the fitting of the observed axial-ratio distribution for galaxies 
with $M_{\rm star} = 10^{10}$--$10^{10.5} M_{\odot}$ and $\Delta$MS $= -1.75$ -- 
$-1.25$ dex at $0.2<z<0.6$. {\bf (right)} The observed axial-ratio distribution 
(solid line) 
and the acceptable models with various values of the fitting parameters 
within the 68\% confidence ranges for the same subsample (dashed lines).  
\label{fig:confhist1}}
\end{figure}

\begin{figure}
\epsscale{1.05}
\plotone{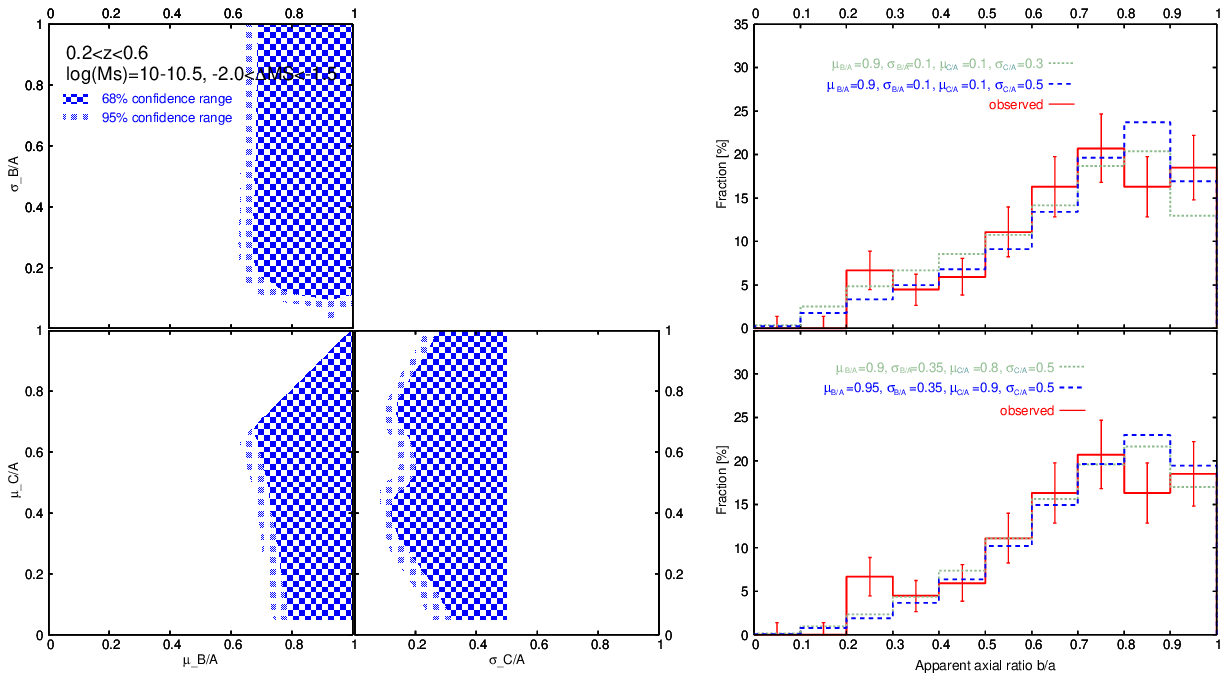}
\caption{ 
The same as Figure \ref{fig:confhist1} but for galaxies with 
$M_{\rm star} = 10^{10}$--$10^{10.5} M_{\odot}$ and $\Delta$MS $= -2.0$ -- 
$-1.5$ dex at $0.2<z<0.6$.
\label{fig:confhist2}}
\end{figure}

\begin{figure}
\epsscale{1.05}
\plotone{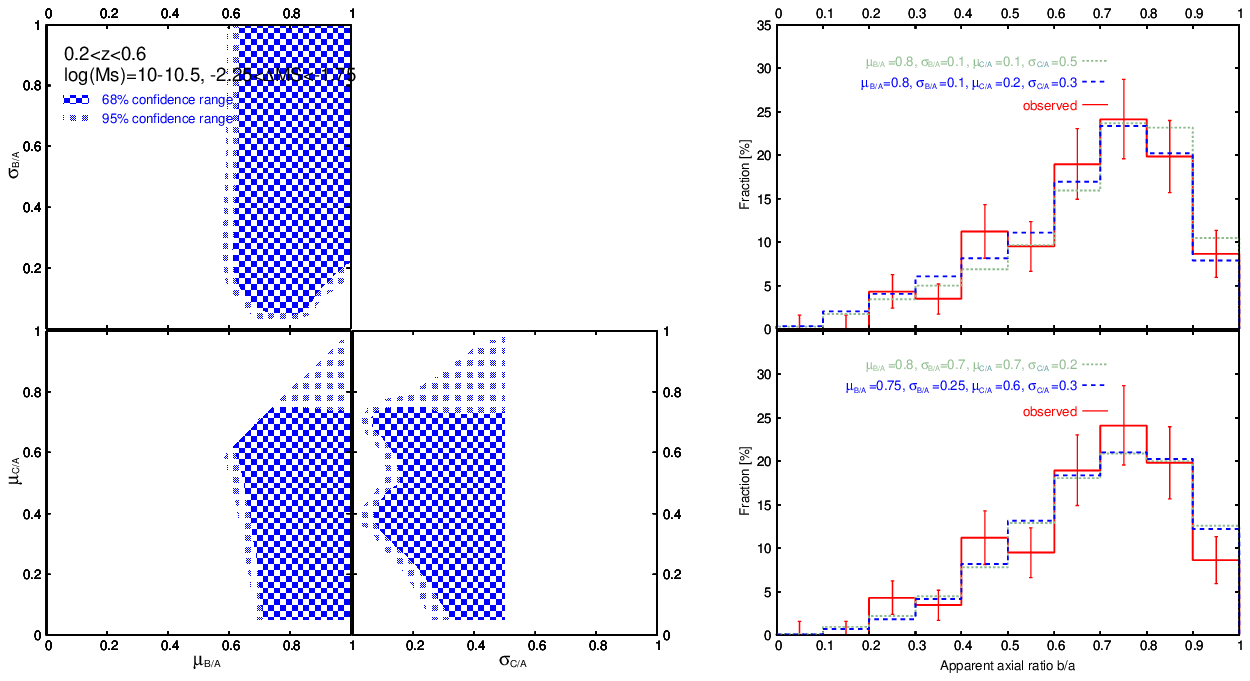}
\caption{ 
The same as Figure \ref{fig:confhist1} but for galaxies with 
$M_{\rm star} = 10^{10}$--$10^{10.5} M_{\odot}$ and $\Delta$MS $= -2.25$ -- 
$-1.75$ dex at $0.2<z<0.6$.
\label{fig:confhist3}}
\end{figure}

\begin{figure}
\epsscale{1.1}
\plotone{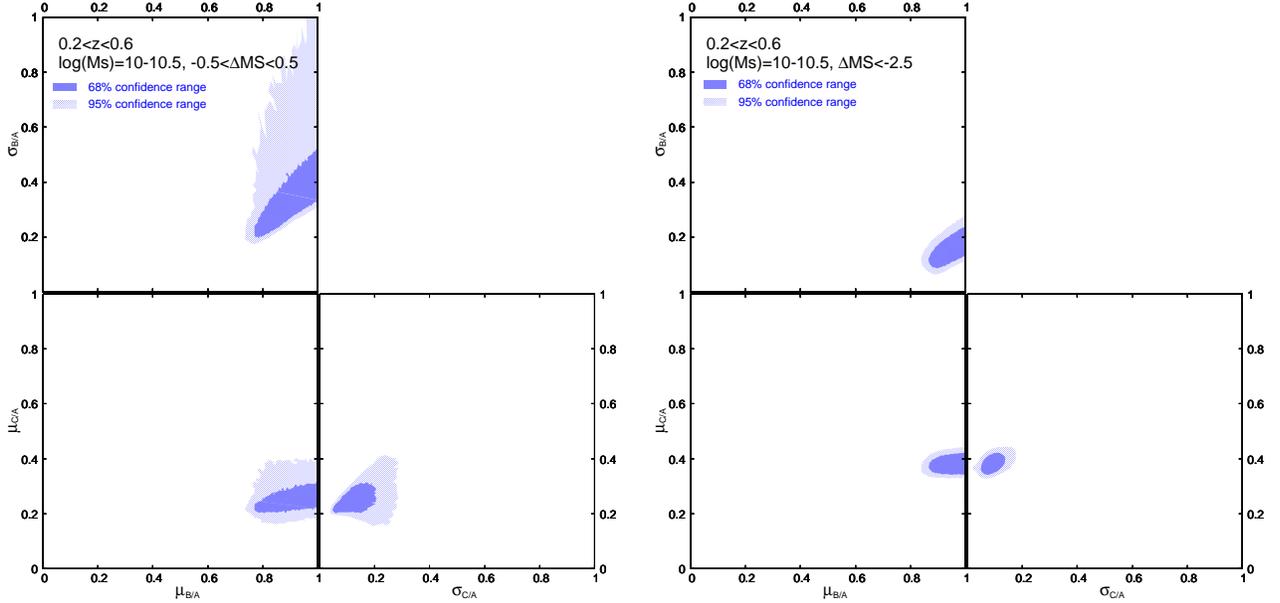}
\caption{ 
The confidence ranges of the intrinsic shape parameters   
in the fitting of the observed axial-ratio distribution for galaxies 
with $M_{\rm star} = 10^{10}$--$10^{10.5} M_{\odot}$ and $\Delta$MS $= -0.5$ -- 
$+0.5$ dex at $0.2<z<0.6$ (left) and for those with $\Delta$MS $< -2.5$ dex 
in the same stellar mass and redshift range (right).
\label{fig:confref}}
\end{figure}

\begin{figure}
\epsscale{1.0}
\plotone{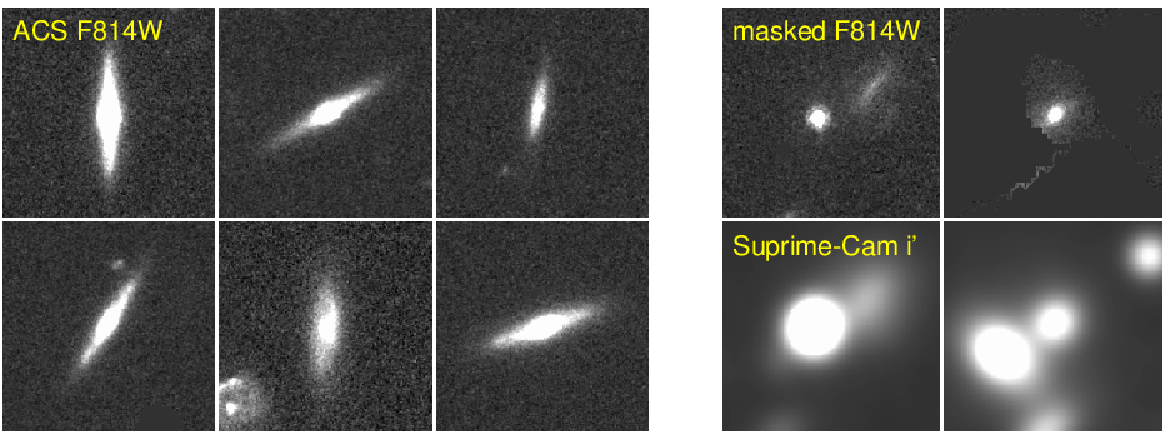}
\caption{ 
Examples of galaxies with $b/a = 0.2$ -- 0.3, 
$M_{\rm star} = 10^{10}$--$10^{10.5} M_{\odot}$, and $\Delta$MS $= -2.25$ -- $-1.25$ 
dex at $0.2<z<0.6$. The left panels show the ACS $I_{\rm F814W}$-band images of 
edge-on disk galaxies with a significant bulge. The right panels show 
examples of galaxies whose axial ratios are affected by nearby objects.
While the top panels show the $I_{\rm F814W}$-band images where nearby objects 
are masked out, the bottom panels represent the Suprime-Cam $i'$-band images 
of the same objects, where the source detection were performed.
The all images are $6^{\prime\prime} \times 6^{\prime\prime}$ in size.  
\label{fig:ba0203}}
\end{figure}

We here examine the model fitting for the axial-ratio distribution of 
galaxies with $M_{\rm star} = 10^{10}$--$10^{10.5} M_{\odot}$ and 
$\Delta$MS $\sim -2.0$ -- $-1.5$ dex at $0.2<z<0.6$, where 
the uncertainty of $\mu_{C/A}$ is very large.
The left panels of Figures \ref{fig:confhist1}, \ref{fig:confhist2}, 
and \ref{fig:confhist3} show the confidence ranges of the fitted 
intrinsic shape parameters, $\mu_{B/A}$, $\sigma_{B/A}$, $\mu_{C/A}$, and 
$\sigma_{C/A}$ for galaxies with $\Delta$MS $= -1.75$ -- $-1.25$, 
$-2.0$ -- $-1.5$, and $-2.25$ -- $-1.75$ dex, respectively.
One can see that the constraints on $\mu_{C/A}$ and $\sigma_{B/A}$ are 
very weak in the fittings for these galaxies.
Relatively higher values of $\sigma_{C/A} \gtrsim 0.2$ are preferred 
in the fitting for these galaxies, 
while $\sigma_{C/A}$ tends to be constrained to lower values  
in the fitting for the other subsamples, for example, 
those with $\Delta$MS $= -0.5$ -- $+0.5$ 
and $\Delta$MS $< -2.5$, whose all parameters are well constrained 
(Figure \ref{fig:confref}).
The right panels of Figures \ref{fig:confhist1}, \ref{fig:confhist2}, 
and \ref{fig:confhist3} show the observed axial-ratio distributions for 
these galaxies and those of acceptable models with various values of 
the fitting parameters within the 68\% confidence ranges.
The observed distributions for these galaxies have both a broad peak around 
$b/a \sim 0.8$ -- 1.0 and a non-negligible fraction of galaxies at 
$b/a = 0.2$ -- 0.3. 
Such distributions are difficult to be reproduced by the models with 
a small $\sigma_{C/A}$, because in such cases with small $\sigma_{C/A}$ 
the broad peak around $b/a \sim 0.8$ -- 1.0 requires a relatively high 
value of $\mu_{C/A}$ (and $\mu_{B/A}$), which leads to a very small 
fraction of those with $b/a < 0.3$. 
On the other hand, the models with a large $\sigma_{C/A}$ could roughly 
reproduce such distributions (the right panels of Figures 
\ref{fig:confhist1}, \ref{fig:confhist2}, and \ref{fig:confhist3}).
In the models with a large $\sigma_{C/A}$, $C/A$ are widely distributed 
irrespective of $\mu_{C/A}$  
and the effects of $\mu_{C/A}$ on the shape of the axial-ratio distribution 
tend to be relatively small.
This seems to be one of the reasons for the large acceptable range of 
$\mu_{C/A}$ in the fitting of these galaxies as well as the large statistical 
uncertainty of the observed distributions 
due to the small size of the subsamples.
We also note that 
there are some acceptable models with a relatively small value of 
$\sigma_{C/A} \lesssim 0.2$ in the left panels in Figures 
\ref{fig:confhist1}, \ref{fig:confhist2}, and \ref{fig:confhist3}.   
In such cases, a large value of $\sigma_{B/A}$ is needed 
to match with the observed distributions (the bottom right panel of 
Figure \ref{fig:confhist3}).

While most of those galaxies with $b/a = 0.2$ -- 0.3 and 
$\Delta$MS $\sim -2.25$ -- $-1.25$ dex are edge-on disk galaxies with a 
significant bulge, there are a few galaxies with a relatively round 
shape that affected by a nearby faint elongated companion or the outer 
part of nearby bright objects (Figure \ref{fig:ba0203}). 
For these objects, 
we could not completely mask out the nearby objects because 
the deblending is failed or not perfectly performed due to the lower 
resolution of the Suprime-Cam $i'$-band data.
Anyway, we cannot conclude 
whether the non-negligible fractions of those galaxies with $b/a = 0.2$ -- 0.3 
are caused by the statistical fluctuation or not 
with the small sizes of the subsamples.

\section{Monte Carlo simulation for the effect of the disk dimming} \label{sec:monsim}

\begin{figure}
\epsscale{1.0}
\plotone{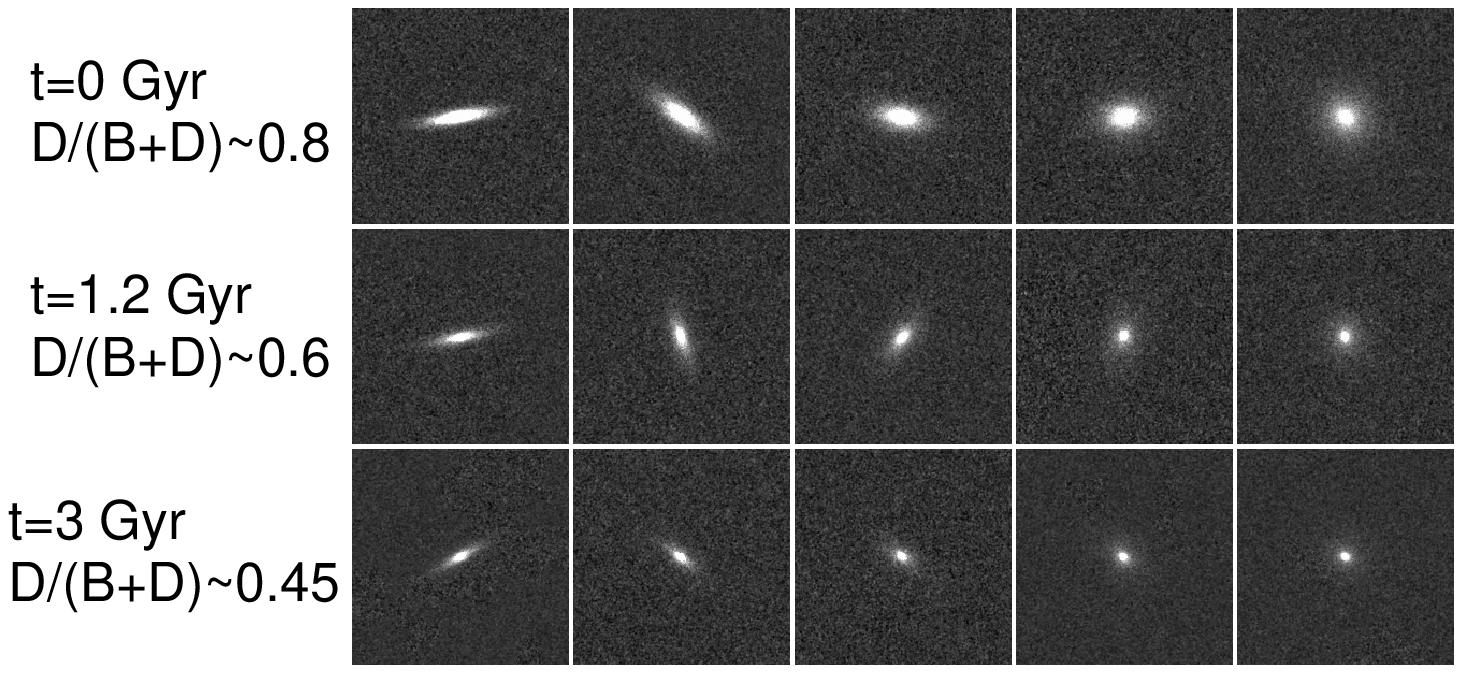}
\caption{ 
Examples of the artificial objects added to the $I_{\rm F814W}$-band images.
The age of the toy model with $\tau = 0.5$ Gyr 
(the disk fraction of the objects) 
increases (decreases) from $t=0$ Gyr (D/(B+D) $=$ 0.8) at the top row to 
$t=3$ Gyr (D/(B+D) $=$ 0.45) at the bottom row.
The inclination angle changes from nearly edge-on view at the left column 
to face-on view at the right column.
The three objects in the same column have the same intrinsic shape parameters 
for the disk and bulge components and inclination angles,  
but the different magnitudes of the disk and bulge components. 
The background sky and position angle on the image are 
randomly selected and different among the objects.
\label{fig:examsim}}
\end{figure}

We here describe the Monte Carlo simulation to check how the decrease 
of the disk fraction due to the quenching of star formation in the disk 
component affects the distribution of the apparent axial ratio.
Using the IRAF/ARTDATA package, we added artificial objects with 
disk and bulge components to sky regions at random positions in the HST/ACS 
$I_{\rm F814W}$-band images. 
Assuming these objects were located at $z=0.8$, 
we calculated the apparent $I_{\rm F814W}$-band magnitudes of the disk and 
bulge components from 
their absolute magnitudes of the toy model with $\tau =$ 0.5 Gyr at  
0, 1.2, and 3 Gyr after the start of the disk quenching, 
when the disk fraction was about 0.8, 0.6, and 0.45, respectively.
We assumed the exponential and de Vaucouleurs laws for 
surface brightness profiles of the disk and bulge components, respectively.
We performed the same Monte Carlo simulations as in the fitting of the 
axial-ratio distribution to generate the apparent axial ratios of the 
disk and bulge components.
In the simulations, we assumed the best-fit intrinsic shape 
parameters of the main-sequence and passively evolving galaxies 
with $M_{star} = 10^{10}$--$10^{10.5} M_{\odot}$ at $0.6<z<1.0$ for 
the disk and bulge components, respectively, except for 
$\mu_{C/A}$ of the disk component, for which we adopted $\mu_{C/A} = 0.18$ 
so that the thickness of artificial objects with the both disk and 
bulge components at the start of the quenching corresponds to that of 
 the observed main-sequence galaxies.
We also assumed that the major, middle, and minor axes of the disk and 
bulge  components are aligned in the same direction for simplicity.
The semi-major radii of the disk and bulge components were adjusted 
to match with those of the observed main-sequence and passively evolving
galaxies in Figure \ref{fig:arsizehighz}. 
While the assumption of the size of passively evolving galaxies may 
overestimate that of the bulge of normal star-forming galaxies, 
we confirmed that smaller/larger sizes of the bulge component do not 
significantly change the results. 
We performed 10000 such simulations for each age of the toy model mentioned 
above, and measured the apparent axial ratios of these artificial objects 
with the same way as for the observed galaxies.
Figure \ref{fig:examsim} shows examples of the artificial objects in the 
simulation.


\begin{thebibliography}{}

\bibitem[Abadi et al.(1999)]{aba99} Abadi, M.~G., Moore, B., \& Bower, R.~G.\ 1999, \mnras, 308, 947

\bibitem[Abraham, \& van den Bergh(2001)]{abr01} Abraham, R.~G., \& van den Bergh, S.\ 2001, Science, 293, 1273

\bibitem[Aguerri(2012)]{agu12} Aguerri, J.~A.~L.\ 2012, Advances in Astronomy, 2012, 382674

\bibitem[Barnes(1988)]{bar88} Barnes, J.~E.\ 1988, \apj, 331, 699

\bibitem[Balogh et al.(2000)]{bal00} Balogh, M.~L., Navarro, J.~F., \& Morris, S.~L.\ 2000, \apj, 540, 113

\bibitem[Bell et al.(2004)]{bel04} Bell, E.~F., Wolf, C., Meisenheimer, K., et al.\ 2004, \apj, 608, 752

\bibitem[Bell et al.(2007)]{bel07} Bell, E.~F., Zheng, X.~Z., Papovich, C., et al.\ 2007, \apj, 663, 834

\bibitem[Bell et al.(2012)]{bel12} Bell, E.~F., van der Wel, A., Papovich, C., et al.\ 2012, \apj, 753, 167

\bibitem[Bertin, \& Arnouts(1996)]{ber96} Bertin, E., \& Arnouts, S.\ 1996, \aaps, 117, 393

\bibitem[Bezanson et al.(2018)]{bez18} Bezanson, R., van der Wel, A., Pacifici, C., et al.\ 2018, \apj, 858, 60

\bibitem[Binggeli(1980)]{bin80} Binggeli, B.\ 1980, \aap, 82, 289

\bibitem[Binney, \& de Vaucouleurs(1981)]{bin81} Binney, J., \& de Vaucouleurs, G.\ 1981, \mnras, 194, 679

\bibitem[Binney(1985)]{bin85} Binney, J.\ 1985, \mnras, 212, 767


\bibitem[Birnboim, \& Dekel(2003)]{bir03} Birnboim, Y., \& Dekel, A.\ 2003, \mnras, 345, 349

\bibitem[Bluck et al.(2019)]{blu19} Bluck, A.~F.~L., Bottrell, C., Teimoorinia, H., et al.\ 2019, \mnras, 485, 666


\bibitem[Borch et al.(2006)]{bor06} Borch, A., Meisenheimer, K., Bell, E.~F., et al.\ 2006, \aap, 453, 869

\bibitem[Bremer et al.(2018)]{bre18} Bremer, M.~N., Phillipps, S., Kelvin, L.~S., et al.\ 2018, \mnras, 476, 12

\bibitem[Bruzual, \& Charlot(2003)]{bru03} Bruzual, G., \& Charlot, S.\ 2003, \mnras, 344, 1000

\bibitem[Carollo et al.(2013)]{car13} Carollo, C.~M., Bschorr, T.~J., Renzini, A., et al.\ 2013, \apj, 773, 112

\bibitem[Calzetti et al.(2000)]{cal00} Calzetti, D., Armus, L., Bohlin, R.~C., et al.\ 2000, \apj, 533, 682 

\bibitem[Capak et al.(2007)]{cap07} Capak, P., Aussel, H., Ajiki, M., et al.\ 2007, \apjs, 172, 99

\bibitem[Cervantes Sodi(2017)]{cer17} Cervantes Sodi, B.\ 2017, \apj, 835, 80

\bibitem[Ceverino et al.(2010)]{cev10} Ceverino, D., Dekel, A., \& Bournaud, F.\ 2010, \mnras, 404, 2151

\bibitem[Chabrier(2003)]{cha03} Chabrier, G.\ 2003, \pasp, 115, 763

\bibitem[Chang et al.(2013)]{cha13} Chang, Y.-Y., van der Wel, A., Rix, H.-W., et al.\ 2013, \apj, 773, 149

\bibitem[Conselice et al.(2005)]{con05} Conselice, C.~J., Blackburne, J.~A., \& Papovich, C.\ 2005, \apj, 620, 564

\bibitem[Darvish et al.(2017)]{dar17} Darvish, B., Mobasher, B., Martin, D.~C., et al.\ 2017, \apj, 837, 16

\bibitem[Dekel, \& Silk(1986)]{dek86} Dekel, A., \& Silk, J.\ 1986, \apj, 303, 39

\bibitem[Dekel et al.(2009a)]{dek09a} Dekel, A., Birnboim, Y., Engel, G., et al.\ 2009, \nat, 457, 451

\bibitem[Dekel et al.(2009b)]{dek09b} Dekel, A., Sari, R., \& Ceverino, D.\ 2009, \apj, 703, 785

\bibitem[Dekel, \& Burkert(2014)]{dek14} Dekel, A., \& Burkert, A.\ 2014, \mnras, 438, 1870


\bibitem[Duncan et al.(2019)]{dun19} Duncan, K., Conselice, C.~J., Mundy, C., et al.\ 2019, \apj, 876, 110

\bibitem[El-Badry et al.(2018)]{elb18} El-Badry, K., Quataert, E., Wetzel, A., et al.\ 2018, \mnras, 473, 1930

\bibitem[Elbaz et al.(2007)]{elb07} Elbaz, D., Daddi, E., Le Borgne, D., et al.\ 2007, \aap, 468, 33

\bibitem[Emsellem et al.(2011)]{ems11} Emsellem, E., Cappellari, M., Krajnovi{\'c}, D., et al.\ 2011, \mnras, 414, 888

\bibitem[Faber et al.(2007)]{fab07} Faber, S.~M., Willmer, C.~N.~A., Wolf, C., et al.\ 2007, \apj, 665, 265

\bibitem[Fabian(2012)]{fab12} Fabian, A.~C.\ 2012, \araa, 50, 455

\bibitem[Fiacconi et al.(2015)]{fia15} Fiacconi, D., Feldmann, R., \& Mayer, L.\ 2015, \mnras, 446, 1957

\bibitem[F{\"o}rster Schreiber et al.(2009)]{for09} F{\"o}rster Schreiber, N.~M., Genzel, R., Bouch{\'e}, N., et al.\ 2009, \apj, 706, 1364

\bibitem[Gehrels(1986)]{geh86} Gehrels, N.\ 1986, \apj, 303, 336

\bibitem[Governato et al.(2009)]{gov09} Governato, F., Brook, C.~B., Brooks, A.~M., et al.\ 2009, \mnras, 398, 312

\bibitem[Grogin et al.(2011)]{gro11} Grogin, N.~A., Kocevski, 
D.~D., Faber, S.~M., et al.\ 2011, \apjs, 197, 35 

\bibitem[Guedes et al.(2013)]{gue13} Guedes, J., Mayer, L., Carollo, M., et al.\ 2013, \apj, 772, 36

\bibitem[Hill et al.(2019)]{hil19} Hill, A.~R., van der Wel, A., Franx, M., et al.\ 2019, \apj, 871, 76

\bibitem[Holden et al.(2012)]{hol12} Holden, B.~P., van der Wel, A., Rix, H.-W., et al.\ 2012, \apj, 749, 96

\bibitem[Hopkins et al.(2009)]{hop09} Hopkins, P.~F., Somerville, R.~S., Cox, T.~J., et al.\ 2009, \mnras, 397, 802

\bibitem[Hubble(1926)]{hub26} Hubble, E.~P.\ 1926, \apj, 64, 321

\bibitem[Huertas-Company et al.(2009)]{hue09} Huertas-Company, M., Tasca, L., Rouan, D., et al.\ 2009, \aap, 497, 743

\bibitem[Ilbert et al.(2009)]{ilb09} Ilbert, O., Capak, P., 
Salvato, M., et al.\ 2009, \apj, 690, 1236 

\bibitem[Ilbert et al.(2010)]{ilb10} Ilbert, O., Salvato, M., Le Floc'h, E., et al.\ 2010, \apj, 709, 644

\bibitem[Ilbert et al.(2013)]{ilb13} Ilbert, O., McCracken, H.~J., Le F{\`e}vre, O., et al.\ 2013, \aap, 556, A55 

\bibitem[Ilbert et al.(2015)]{ilb15} Ilbert, O., Arnouts, S., Le Floc'h, E., et al.\ 2015, \aap, 579, A2 

\bibitem[Jesseit et al.(2009)]{jes09} Jesseit, R., Cappellari, M., Naab, T., et al.\ 2009, \mnras, 397, 1202

\bibitem[Kajisawa, \& Yamada(2001)]{kaj01} Kajisawa, M., \& Yamada, T.\ 2001, \pasj, 53, 833

\bibitem[Kajisawa et al.(2010)]{kaj10} Kajisawa, M., Ichikawa, T., Yamada, T., et al.\ 2010, \apj, 723, 129 

\bibitem[Kajisawa et al.(2011)]{kaj11} Kajisawa, M., Ichikawa, T., Yoshikawa, T., et al.\ 2011, \pasj, 63, 403

\bibitem[Kere{\v{s}} et al.(2005)]{ker05} Kere{\v{s}}, D., Katz, N., Weinberg, D.~H., et al.\ 2005, \mnras, 363, 2

\bibitem[Kere{\v{s}} et al.(2009)]{ker09} Kere{\v{s}}, D., Katz, N., Fardal, M., et al.\ 2009, \mnras, 395, 160

\bibitem[Koekemoer et al.(2007)]{koe07} Koekemoer,
		 A.~M., Aussel, H., Calzetti, D., et al.\ 2007, \apjs,
		 172, 196

\bibitem[Koekemoer et al.(2011)]{koe11} Koekemoer, A.~M., Faber, S.~M., Ferguson, H.~C., et al.\ 2011, \apjs, 197, 36

\bibitem[Kormendy, \& Kennicutt(2004)]{kor04} Kormendy, J., \& Kennicutt, R.~C.\ 2004, \araa, 42, 603

\bibitem[Laigle et al.(2016)]{lai16} Laigle, C., McCracken, H.~J., Ilbert, O., et al.\ 2016, \apjs, 224, 24

\bibitem[Lambas et al.(1992)]{lam92} Lambas, D.~G., Maddox, S.~J., \& Loveday, J.\ 1992, \mnras, 258, 404


\bibitem[Law et al.(2012)]{law12} Law, D.~R., Steidel, C.~C., Shapley, A.~E., et al.\ 2012, \apj, 745, 85

\bibitem[Larson et al.(1980)]{lar80} Larson, R.~B., Tinsley, B.~M., \& Caldwell, C.~N.\ 1980, \apj, 237, 692

\bibitem[Legrand et al.(2019)]{leg19} Legrand, L., McCracken, H.~J., Davidzon, I., et al.\ 2019, \mnras, 486, 5468

\bibitem[Lotz et al.(2008)]{lot08} Lotz, J.~M., Jonsson, P., Cox, T.~J., et al.\ 2008, \mnras, 391, 1137

\bibitem[Mager et al.(2018)]{mag18} Mager, V.~A., Conselice, C.~J., Seibert, M., et al.\ 2018, \apj, 864, 123

\bibitem[Mancini et al.(2019)]{man19} Mancini, C., Daddi, E., Juneau, S., et al.\ 2019, \mnras, 2068

\bibitem[Martig et al.(2009)]{mar09} Martig, M., Bournaud, F., Teyssier, R., et al.\ 2009, \apj, 707, 250

\bibitem[Moore et al.(1996)]{moo96} Moore, B., Katz, N., Lake, G., et al.\ 1996, \nat, 379, 613

\bibitem[Morselli et al.(2019)]{mor18} Morselli, L., Popesso, P., Cibinel, A., et al.\ 2019, \aap, 626, A61

\bibitem[Mundy et al.(2017)]{mun17} Mundy, C.~J., Conselice, C.~J., Duncan, K.~J., et al.\ 2017, \mnras, 470, 3507

\bibitem[Murata et al.(2014)]{mur14} Murata, K.~L., Kajisawa, M., Taniguchi, Y., et al.\ 2014, \apj, 786, 15

\bibitem[Muzzin et al.(2013)]{muz13} Muzzin, A., Marchesini, D., Stefanon, M., et al.\ 2013, \apj, 777, 18

\bibitem[Naab, \& Burkert(2003)]{naa03} Naab, T., \& Burkert, A.\ 2003, \apj, 597, 893

\bibitem[Noeske et al.(2007)]{noe07} Noeske, K.~G., Weiner, B.~J., Faber, S.~M., et al.\ 2007, \apj, 660, L43

\bibitem[Noguchi(2019)]{nog19} Noguchi, M.\ 2019, arXiv e-prints, arXiv:1905.08993

\bibitem[Oser et al.(2010)]{ose10} Oser, L., Ostriker, J.~P., Naab, T., et al.\ 2010, \apj, 725, 2312

\bibitem[Padilla, \& Strauss(2008)]{pad08} Padilla, N.~D., \& Strauss, M.~A.\ 2008, \mnras, 388, 1321

\bibitem[Peng et al.(2010)]{pen10} Peng, Y.-. jie ., Lilly, S.~J., Kova{\v{c}}, K., et al.\ 2010, \apj, 721, 193

\bibitem[Pillepich et al.(2019)]{pil19} Pillepich, A., Nelson, D., Springel, V., et al.\ 2019, arXiv e-prints, arXiv:1902.05553

\bibitem[Popesso et al.(2019)]{pop19} Popesso, P., Concas, A., Morselli, L., et al.\ 2019, \mnras, 483, 3213 

\bibitem[Qu et al.(2011)]{qu11} Qu, Y., Di Matteo, P., Lehnert, M.~D., et al.\ 2011, \aap, 535, A5

\bibitem[Quinn et al.(1993)]{qui93} Quinn, P.~J., Hernquist, L., \& Fullagar, D.~P.\ 1993, \apj, 403, 74

\bibitem[Ravindranath et al.(2006)]{rav06} Ravindranath, S., Giavalisco, M., Ferguson, H.~C., et al.\ 2006, \apj, 652, 963

\bibitem[Renaud et al.(2014)]{ren14} Renaud, F., Bournaud, F., Kraljic, K., et al.\ 2014, \mnras, 442, L33

\bibitem[Roberts, \& Haynes(1994)]{rob94} Roberts, M.~S., \& Haynes, M.~P.\ 1994, \araa, 32, 115

\bibitem[Rodriguez-Gomez et al.(2017)]{rod17} Rodriguez-Gomez, V., Sales, L.~V., Genel, S., et al.\ 2017, \mnras, 467, 3083

\bibitem[Ro{\v{s}}kar et al.(2008)]{ros08} Ro{\v{s}}kar, R., Debattista, V.~P., Stinson, G.~S., et al.\ 2008, \apj, 675, L65

\bibitem[Ryden(2004)]{ryd04} Ryden, B.~S.\ 2004, \apj, 601, 214

\bibitem[Sales et al.(2009)]{sal09} Sales, L.~V., Helmi, A., Abadi, M.~G., et al.\ 2009, \mnras, 400, L61

\bibitem[Sales et al.(2012)]{sal12} Sales, L.~V., Navarro, J.~F., Theuns, T., et al.\ 2012, \mnras, 423, 1544

\bibitem[Sandage et al.(1970)]{san70} Sandage, A., Freeman, K.~C., \& Stokes, N.~R.\ 1970, \apj, 160, 831

\bibitem[Sanders, \& Mirabel(1996)]{san96} Sanders, D.~B., \& Mirabel, I.~F.\ 1996, \araa, 34, 749

\bibitem[Shao et al.(2007)]{sha07} Shao, Z., Xiao, Q., Shen, S., et al.\ 2007, \apj, 659, 1159

\bibitem[Sparre, \& Springel(2016)]{spa16} Sparre, M., \& Springel, V.\ 2016, \mnras, 462, 2418

\bibitem[Springel, \& Hernquist(2005)]{spr05} Springel, V., \& Hernquist, L.\ 2005, \apj, 622, L9

\bibitem[Takeuchi et al.(2015)]{tak15} Takeuchi, T.~M., Ohta, K., Yuma, S., et al.\ 2015, \apj, 801, 2

\bibitem[Taniguchi et al.(2007)]{tan07} Taniguchi, Y., Scoville, N., Murayama, T., et al.\ 2007, \apjs, 172, 9 

\bibitem[Taniguchi et al.(2015)]{tan15} Taniguchi, Y., Kajisawa, M., Kobayashi, M.~A.~R., et al.\ 2015, \pasj, 67, 104

\bibitem[Teyssier et al.(2010)]{tey10} Teyssier, R., Chapon, D., \& Bournaud, F.\ 2010, \apj, 720, L149

\bibitem[Tinker et al.(2013)]{tin13} Tinker, J.~L., Leauthaud, A., Bundy, K., et al.\ 2013, \apj, 778, 93

\bibitem[van der Wel et al.(2009)]{van09} van der Wel, A., Rix, H.-W., Holden, B.~P., et al.\ 2009, \apj, 706, L120

\bibitem[van der Wel et al.(2014a)]{van14a} van der Wel, A., Chang, Y.-Y., Bell, E.~F., et al.\ 2014, \apj, 792, L6

\bibitem[van der Wel et al.(2014b)]{van14b} van der Wel, A., Franx, M., van Dokkum, P.~G., et al.\ 2014, \apj, 

\bibitem[Vika et al.(2013)]{vik13} Vika, M., Bamford, S.~P., H{\"a}u{\ss}ler, B., et al.\ 2013, \mnras, 435, 623

\bibitem[Villalobos, \& Helmi(2008)]{vil08} Villalobos, {\'A}., \& Helmi, A.\ 2008, \mnras, 391, 1806

\bibitem[Vincent, \& Ryden(2005)]{vin05} Vincent, R.~A., \& Ryden, B.~S.\ 2005, \apj, 623, 137

\bibitem[Whitaker et al.(2011)]{whi11} Whitaker, K.~E., Labb{\'e}, I., van Dokkum, P.~G., et 

\bibitem[White, \& Frenk(1991)]{whi91} White, S.~D.~M., \& Frenk, C.~S.\ 1991, \apj, 379, 52

\bibitem[Williams et al.(2009)]{wil09} Williams, R.~J., Quadri, R.~F., Franx, M., et al.\ 2009, \apj, 691, 1879

\bibitem[Windhorst et al.(2002)]{win02} Windhorst, R.~A., Taylor, V.~A., Jansen, R.~A., et al.\ 2002, \apjs, 143, 113

\bibitem[Wisnioski et al.(2015)]{wis15} Wisnioski, E., F{\"o}rster Schreiber, N.~M., Wuyts, S., et al.\ 2015, \apj, 799, 209

\bibitem[Wuyts et al.(2012)]{wuy12} Wuyts, S., F{\"o}rster Schreiber, N.~M., Genzel, R., et al.\ 2012, \apj, 753, 114

\bibitem[Yuma et al.(2011)]{yum11} Yuma, S., Ohta, K., Yabe, K., et al.\ 2011, \apj, 736, 92

\bibitem[Yuma et al.(2012)]{yum12} Yuma, S., Ohta, K., \& Yabe, K.\ 2012, \apj, 761, 19

\bibitem[Zhang et al.(2019)]{zha19} Zhang, H., Primack, J.~R., Faber, S.~M., et al.\ 2019, \mnras, 484, 5170

\end{thebibliography}
\end{document}